\documentstyle[12pt,epsf,epsfig,fleqn]{report}

\def\graphpath{}

\newcommand{\fips}[1]{\epsffile{#1.eps}}

\newcommand{\be}{\begin{equation}}
\newcommand{\ee}{\end{equation}}
\newcommand {\beaa}{\begin{eqnarray*}}   
\newcommand {\eeaa}{\end{eqnarray*}}
\newcommand{\lee}[1]{\label{#1} \end{equation}}
\newcommand{\bea}{\begin{eqnarray}}
\newcommand{\leea}[1]{\label{#1} \end{eqnarray}}
\newcommand{\eea}{\end{eqnarray}}
\newcommand{\nn}{\nonumber}
\newcommand{\eq}[1]{eq.~(\ref{#1})}
\newcommand{\Eq}[1]{Eq.~(\ref{#1})}
\newcommand{\eqs}[2]{eqs.~(\ref{#1}) and (\ref{#2})}

\newcommand{\fig}[1]{fig.~(\ref{#1})}
\newcommand{\loadeps}[1]{\epsfig{file=#1.eps,width=45mm}}
\newcommand{\floadeps}[1]{\epsfig{file=#1.eps,width=20mm}}
\newcommand{\diagform}[2]{\put(60,#1){\makebox(110,44.61)[l]{
  \begin{minipage}{110mm} #2 \end{minipage} }} }
\newcommand{\al}{\alpha}

\newcommand{\ga}{\gamma}
\newcommand{\de}{\delta}

\newcommand{\varep}{\varepsilon}

\newcommand{\et}{\eta}

\newcommand{\la}{\lambda}

\newcommand{\si}{\sigma}

\newcommand{\th}{\theta}

\newcommand{\ch}{\chi}

\newcommand{\om}{\omega}

\newcommand{\Ga}{\Gamma}
\newcommand{\De}{\Delta}
\newcommand{\La}{\Lambda}
\newcommand{\Si}{\Sigma}

\newcommand{\Om}{\Omega}

\def\bbbone{{\mathchoice {\rm 1\mskip-4mu l} {\rm 1\mskip-4mu l}
 {\rm 1\mskip-4.5mu l} {\rm 1\mskip-5mu l}}}
\begin{document}

\thispagestyle{empty}
\vspace{2cm}
\begin{center}
{\large\bf INAUGURAL-DISSERTATION}\\
\vskip0.1cm
{\bf zur}
\vskip0.1cm
{\bf Erlangung der Doktorw\"urde}\\
\vskip0.1cm
{\bf der}
\vskip0.1cm
{\bf Naturwissenschaftlich-Mathematischen}\\
\vskip0.1cm
{\bf  Gesamtfakult\"at}\\
\vskip0.1cm
{\bf der}\\
\vskip0.1cm
{\bf Ruprecht-Karls-Universit\"at}\\
\vskip0.1cm
{\bf Heidelberg}\\
\vspace{14cm}
vorgelegt von\\
\vskip0.2cm
{\large\bf  Dipl.-Phys. Elena Gubankova}\\
\vskip0.2cm
aus Moskau\\
\vskip0.5cm
Tag der m\"undlichen Pr\"ufung: 1. Juli 1998
\end{center}
\newpage
\thispagestyle{empty}
\cleardoublepage
\thispagestyle{empty}
\cleardoublepage
\begin{center}
{\Large \bf  FLOW EQUATIONS}\\ 
\vskip0.4cm
{\Large\bf FOR THE QUANTUM ELECTRODYNAMICS}\\
\vskip0.4cm
{\Large\bf ON THE LIGHT-FRONT}
\end{center}
\nopagebreak
\vspace{18cm}
\begin{center}
{\bf Gutachter: }{\bf Prof. Dr. Franz Wegner} 
\vskip0.1cm
\hspace{3.8cm}{\bf Prof. Dr.Hans-Christian Pauli}
\end{center}
\newpage

%
%



\thispagestyle{empty}
\cleardoublepage
 \begin{center}
{\bf Abstract}
\end{center}
 
The method of flow equations is applied to QED on  the light front.
Requiring that the particle number conserving terms in the Hamiltonian
are considered to be diagonal and the other terms off-diagonal
an effective Hamiltonian is obtained which reduces 
the positronium problem to a two-particle problem, since 
the particle number violating contributions are eliminated. 

Using an effective electron-positron Hamiltonian,
obtained in the second order in coupling,
we analyze the positronium bound state problem
analytically and numerically. The results obtained for Bohr
spectrum and hyperfine splitting coincide to a high accuracy
with experimental values. The rotational invariance,
that is not manifest symmetry on the light-front,
is recovered for positronium mass spectrum.

Except for the longitudinal infrared divergences,
that are special for the light-front gauge calculations,
no infrared divergences appear. 
The ultraviolet renormalization in the second order in coupling
constant is performed simultaneously. To preserve boost invariance
we take into account the diagrams arising from the normal ordering
of instantaneous interactions. Using flow equations
and coupling coherence we obtain the counterterms for 
electron and photon masses, which are free from longitudinal
infrared divergences.

\vspace{3.5cm}

\begin{center}
{\bf Abstrakt}
\end{center}

In dieser Arbeit wird die Methode von Flu{\ss}gleichungen
im Lichtkegel-Formalismus auf die QED angewandt.
Wir konstruieren einen effektiven, block-diagonalen
Hamiltonian, wobei wir fordern, da{\ss} diejenigen Terme,
welche die Teilchenzahl erhalten,
diagonal sind, w{\"a}hrend alle anderen Terme 
nichtdiagonal sind. Dieser effektive Hamiltonian
ver\-ein\-facht das Positronium-Problem auf ein
Zweik{\"o}rper-Problem, weil die Anteile,
die eine Teilchenzahl{\"a}nderung verursachen,
eliminiert sind.

Mittels des effektiven Elektron-Positron-Hamiltonians,
der in der elektromagnetischen Kopplung von zweiter Ordnung ist,
werden die Bindungszust{\"a}nde des Positroniums 
analytisch und numerisch untersucht. 
Unsere Resultate f{\"u}r das Bohr-Spektrum und 
die Hyperfein\-aufspaltung stimmen sehr gut mit
den experimentellen Werten {\"u}berein. Die Rotationsin\-varianz,
welche auf dem Lichtkegel nicht mehr gew{\"a}hrleistet ist,
kann f{\"u}r das Massenspektrum wieder hergestellt werden.

Au{\ss}er longitudinalen Infrarot-Divergenzen,
die f{\"u}r die Lichtkegel-Eichung spezifisch sind,
treten bei den Rechnungen keine Infrarot-Divergenzen auf.
Die Renormierung der Ultraviolet-Divergenzen 
bis zur zweiten Ordnung
in der Kopplungskonstanten erh{\"a}lt man mit der
Konstruktion des Hamiltonians automatisch. Um die Invarianz
des renormierten Hamiltonian unter Boosts sicherzustellen,
ber{\"u}cksichtigen wir auch die Diagramme, welche aus der
Normalordnung von instantanen Wechselwirkungen entstehen.
Die aus den Flu{\ss}gleichungen und der Kopplungskoh{\"a}renz
resultierenden Counterterme der Photon- und Elektronmasse
erhalten dann keine longitudinalen Infrarot-Divergenzen.


%
%



%
%


%
%

\tableofcontents

%
%

\chapter{Introduction}

Quantum chromodynamics (QCD) is widely accepted as the microscopic
theory of strong interaction, where quarks and gluons are
considered as the elementary degrees of freedom.
But we are still unable to solve this theory on the macroscopic level
(in the low energy domain) and to obtain an accurate description of the 
structure of hadrons, which are the strongly interacting particles
observed in nature.

A central goal of modern theoretical particle physics is to build a bridge
between microscopic (high energy) and macroscopic (low energy) 
domains of QCD.

At the microscopic level the theory is defined completely,
and no further development seems to be necessary. 
On the other side, when it comes to the macroscopic hadron
world, there is no rigorous way to calculate hadron properties
immediately from QCD.
One still has to resort to different
phenomenological methods and QCD-inspired models 
of covariant field theory to calculate
the hadron mass spectrum, form factors, wave functions, {\it etc}.
Also lattice calculations serving as a model-independent method in
covariant field theory improved our understanding of the hadronic
structure. However, besides different problems, this approach
is still strongly limited by the available computer power.

In this work we use the light-front formulation of field theory,
which is best suited for solving relativistic bound state 
problems in QCD \cite{LFref}, because of the simplified vacuum structure.  
We believe that light-front field theory 
together with the renormalization 
group approach for Hamiltonians \cite{We},\cite{GlWi} 
provides a good strategy to reach 
the above mentioned goal.
The physical idea behind this approach is to use
the renormalization concept for Hamiltonians 
on the light-front
to get an effective, low-energy Hamiltonian.

Recently, Glazek and Wilson suggested the renormalization
scheme for Hamiltonians, called similarity renormalization 
by the authors, 
where they developed a renormalization group and the basic elements
of renormalization group calculations for Hamiltonians 
on the light-front \cite{GlWi}. An alternative approach 
for Hamiltonian renormalization, the method of flow equations,
was proposed by Wegner \cite{We}.
It is common for both methods that they renormalize 
the canonical Hamiltonian of light-front field theory
to a given order of perturbation theory. A basic advantage
of the method of flow equations in comparison to the similarity
renormalization scheme is that one obtains an effective, 
renormalized Hamiltonian for a limited (truncated) Fock space.

The method of flow equations is based on the following idea:  
performing a set of infinitesimal unitary transformations,
with the condition that the particle number conserving terms 
in the Hamiltonian are considered to be diagonal and the other 
terms off-diagonal,
it is possible to get the block-diagonal effective Hamiltonian, 
where the particle number
in each block is conserved. This reduces the bound state
problem to a few-body problem, since the particle number 
violating contributions are eliminated.
This procedure is similar to the Tamm-Dancoff
space truncation \cite{TaDa} in the sense that
also in this truncation particle number changing interactions
are eliminated.
The effective Hamiltonian constructed by flow equations
is automatically renormalized to the given order in the coupling
constant, since an elimination of particle number changing sectors 
can not be
achieved in one step but rather sequently for transition 
amplitudes from large to small energy differences.

Low-energy QCD is challenging and explicit calculations 
become complicated due to the
nonperturbative nature of this theory. To test and to illustrate
our approach we consider $QED_{3+1}$ in the light-front dynamics and
investigate the corresponding positronium bound state problem.

This work is organized as follows:
In chapter \ref{chapter2} we outline the theoretical framework by
giving the key ingredients of the flow equation method and 
by enumerating their applications to problems of solid state physics and  
statistical mechanics done so far.
The approach of
similarity renormalization is also considered.
In chapter \ref{chapter3} we review the methods and results known 
for light-front $QED_{3+1}$. 
To second order in the coupling constant we obtain the effective,
renormalized QED Hamiltonian (chapter \ref{ch4}), which reduces 
the positronium problem 
to a two-particle problem to be analyzed further analytically 
(chapter \ref{ch5})
and numerically (chapter \ref{ch6}) for positronium bound states.
The renormalization issues of light-front QED are considered 
in chapter \ref{ch7}.

\chapter{Flow equations}
\label{chapter2}

In this section we give the key ingredients of flow equations method
and set up the framework to use flow equations for the problems 
of high-energy physics.

Flow equations are introduced in order to bring Hamiltonians closer
to diagonalization. The method is based on the numerical recipe
by Jacobi, consisting of unitary transformations between
two states which makes the connecting off-diagonal matrix elements vanishing.
If this is repeated for all off-diagonal matrix elements again and again
then the off-diagonal matrix elements will become arbitrarily small.
It is characteristic for these equations that matrix elements between
degenerate or almost degenerate states do not decay or decay very slowly.
In the next section we follow the original work of Wegner \cite{We}
to introduce flow equations for Hamiltonian matrices.

\section{Flow equations for Hamiltonian matrices}
\label{sec2.1}

The aim is to find continuous unitary transformation, that brings Hamiltonian 
to diagonalization. 
Continuous unitary transformation depend on the flow parameter $l$.
We call it $U(l)$. In what follows we assume that $U(0)=1$ is valid
and that $U(\infty)$ brings the given Hamiltonian operator $H$
or the given matrix $H$ to the diagonal form.
The operator $H$ acquires the $l$-dependence through the unitary transformation
$U(l)$
\bea
&& H(l)=U^+(l)HU(l)
\leea{1}
If the transformation $U(l)$ were known, one would find $H(\infty)$
and the problem would be solved. Generally, one does not know 
the transformation, which diagonalizes the given matrix. Therefore
the transformation is formulated in infinitesimal form
\bea
&& \frac{dH(l)}{dl}=[\eta(l),H(l)]
\leea{2}
Here $\eta$ is the generator of transformation; $\eta$ is antihermitian
$\eta^+=-\eta$. The connection between the unitary transformation
and its generator is given
\bea
&& U(l)=exp\left( \int_0^l \eta(l')dl' \right)
\leea{}
where the ordering along 'l-axis' is imposed. 
One has for the matrix elements
\bea
&& \frac{dh_{k,q}(l)}{dl}=\sum_p(\eta_{k,p}(l)h_{p,q}(l)
-h_{k,p}(l)\eta_{p,q}(l))
\leea{3} 
The generator $\eta$ must be chosen in a way, that the matrix $H$
is more and more diagonal as $l$ increases. We demand, that
$\sum_{k\neq q}h_{k,q}^2$ falls monotonously. We separate
$H=H_d+H_r$ and use that use that $TrH^2$ is invariant under the unitary
transformation
\bea
&& TrH_d^2+TrH_r^2=TrH^2=const
\leea{4}
$\sum_{k\neq q}h_{k,q}^2=TrH_r^2$ monoton falls when $TrH_d^2$
increases. One has
\bea
\frac{dTrH_d^2}{dl}=\frac{d}{dl}\sum_q h_{q,q}^2
=2\sum_q h_{q,q} \sum_p (\eta_{q,p}h_{p,q}-h_{q,p}\eta_{p,q})
=2\sum_{p,q}\eta_{p,q}h_{p,q}(h_{p,p}-h_{q,q})
\leea{5}
The right hand side must be negative. The possible choice for the generator
is $\eta_{p,q}=h_{p,q}(h_{p,p}-h_{q,q})$ or
\bea
&& \eta=[H_d,H_r]
\leea{6}
We have
\bea
&& \frac{dh_{k,q}(l)}{dl}=\sum_p (h_{k,k}(l)+h_{q,q}(l)-2h_{p,p}(l))
h_{k,p}(l)h_{p,q}(l)\nn\\
&& \frac{d}{dl}\sum_{k\neq q}h_{k,q}^2  = -\frac{d}{dl}\sum_k h_{k,k}^2
=-2\sum_{k,q} (h_{k,k}-h_{q,q})^2 h_{k,q}^2
=-2\sum_{k,q} \eta_{k,q}^2
\leea{7}
Since $\sum_{k\neq q}h_{k,q}^2$ falls monotonously and is restricted 
from below,
the derivative must vanish in the limit $l\rightarrow\infty$.
Therefore one has
\bea
&& \eta(l)=[H_d,H]\rightarrow_{l\rightarrow\infty}0
\leea{8}
Practically we have reached the aim, the matrix commute with its diagonal
part as $l\rightarrow\infty$. The \eqs{2}{8} are called flow equations
for Hamiltonians and are the basis for the work presented further.

\section{Similarity transformation}
\label{sec2.2}

Another method to diagonalize Hamiltonians continuously
was suggested independently by Glazek and Wilson \cite{GlWi},
which has been called similarity transformation by the authors.
In this subsection we follow the original work of Glazek and Wilson
\cite{GlWi} to give the key ingredients of this scheme.

We define the unitary transformation, that brings the Hamiltonian
operator in a form, where no transitions between the states
with energy difference larger than $\la$ are present.
The role of 'flow parameter' plays $\la$, which corresponds
to the ultraviolet (UV)-cutoff (see chapter \ref{ch7}) and 
is changed continuously.

The Hamiltonian operator $H_{\la}$ (and other quantities) depend then on the
continuous parameter $\la$. The Hamiltonian is given as a sum
$H_{\la}=H_{0\la}+H_{I\la}$, where $H_{0\la}$ is the free Hamiltonian
and $H_{I\la}$ contains (not renormalized) interactions and
counterterms.

We separate each matrix $M$ into two parts,
$M=D(M)+R(M)$. Let $E_{i\la}$ are the eigenvalues
of $H_{0\la}$, and
\bea
&& x_{ij\la}=\frac{|E_{i\la}-E_{j\la}|}{E_{i\la}+E_{j\la}+\la}\nn\\
&& u_{ij\la}=u(x_{ij\la}),~~r_{ij\la}=1-u_{ij\la}=r(x_{ij\la})
\leea{}
and $u(x)$ is the function, which is $1$ for small arguments and
vanishes for large arguments. Explicit one can choose for $u(x)$
\bea
&& u(x)=1~~for~~ 0\leq x\leq \frac{1}{3}\nn\\
&& u(x) ~falls~monotonously~from~1~to~0~for~ \frac{1}{3}\leq x\leq \frac{2}{3}\nn\\
&& u(x)=0~~for~~ \frac{2}{3}\leq x \leq 1
\leea{}
We define $D(M)_{ij}=u_{ij\la}M_{ij}$ and corresponding
$R(M)_{ij}=r_{ij\la}M_{ij}$. As we introduce $u_{ij\la}$ and
$r_{ij\la}$, we have the continuous transition between $D(M)$ and
$R(M)$. It is important in order to avoid the divergences in renormalization
equations. When one chooses $u(x)=\theta(x_0-x)$, then one has the original
definition for $D(M)$ and $R(M)$. The continuous transformation
for Hamiltonian operator can be
written as before in the form
\bea
&& \frac{dH_{\la}}{d\la}=[\eta_{\la},H_{\la}]
\leea{}
The generator $\eta_{\la}$ is chosen in a way that $D(H_{\la})=H_{\la}$.
For practical purpose to distinguish in the calculations
between the Hamiltonian operator $H_{\la}$ and its 'D'-part,
it is useful to introduce the operator $Q$, such that
\bea
&& H_{\la}=D(Q_{\la}) 
\leea{}
Then one has for the matrix elements 
\bea
&& \frac{du_{ij\la}}{d\la}Q_{ij\la}+u_{ij\la}\frac{dQ_{ij\la}}{d\la}
=\eta_{ij\la}(E_{j\la}-E_{i\la})+[\eta_{\la},H_{I\la}]_{ij}
\leea{}
One is not able to find $Q_{\la}$ and $\eta_{\la}$ from this equation.
The reason is, that for the given $\la$ 
the Hamiltonian operator $H_{\la}$ can be additionally
transformed unitary without breaking the condition $D(H_{\la})=H_{\la}$.
Thus one has one more condition. We rewrite the above equation in the form
\bea
&& u_{ij\la}\frac{dQ_{ij\la}}{d\la}-\eta_{ij\la}(E_{j\la}-E_{i\la})=G_{ij\la}
\leea{}
where we define
\bea
&&G_{ij\la}=[\eta_{\la},H_{I\la}]_{ij}-\frac{du_{ij\la}}{d\la}Q_{ij\la}
\leea{}
Now let
\bea
&& u_{ij\la}\frac{dQ_{ij\la}}{d\la}=D(G_{\la})_{ij}
\leea{}
and
\bea
&& \eta_{ij\la}(E_{j\la}-E_{i\la})=-R(G_{\la})_{ij}
\leea{}
Now all functions are defined. One has
\bea
&& \eta_{ij\la}=\frac{r_{ij\la}}{E_{i\la}-E_{j\la}}\left(
[\eta_{\la},H_{I\la}]_{ij}-\frac{du_{ij\la}}{d\la}\frac{H_{ij\la}}{u_{ij\la}}
\right)
\leea{}
and
\bea
&& \frac{dH_{ij\la}}{d\la}=u_{ij\la}[\eta_{\la},H_{I\la}]_{ij}
+r_{ij\la}\frac{du_{ij\la}}{d\la}\frac{H_{ij\la}}{u_{ij\la}}
\leea{}
There are no small energy denominators in these equations. These equations
should be solved iteratively.

\section{Flow equations for solid state physics}
\label{sec2.3}

 Flow equations were successfully applied to different systems in solid
state physics. Unitary transformation in the form of exact diaginalization
of Hamiltonian operator has been tested in\\  
$(1)$ model of impurity in the electron band \cite{Mi};\\
$(2)$ dissipative quantum systems, in particular for spin-boson problem \cite{KeMi1};\\
$(3)$ Lipkin model \cite{PiFr};\\
$(4)$ problem of interacting electrons and phonons in a solid
(referred to as BCS-theory) \cite{LeWe}.\\
In all these models a system of interest couples to its
environment Hamiltonian (for example, in $(1)$ single impurity couples to the
band of electrons; in $(2)$ a small quantum system couples to the
thermodynamical bath). The aim is then to decouple 'small' system from its
'large' environment to find a behavior of the system. It is different from
what is usually done: most theoretical work starts off by tracing out the
bath degrees of freedom and then using suitable approximation schemes for
the time evolution of the reduced density matrix of the small quantum
system. 
In general the approach of Hamiltonian diagonalization is particularly
suited for studying low-energy properties of the system, thereby being
complementary to most other approximation schemes.

In some cases the aim to get the diagonal Hamiltonian operator
for $l\rightarrow\infty$ can not be reached.
It was shown in the original work of Wegner \cite{We} for the model of
interacting fermions in one dimension, that a literal use of the concept
of Hamiltonian diagonalization can lead to convergency problems.
In the case discussed there the divergences appeared as $l\rightarrow\infty$.
The way out, as proposed by Wegner \cite{We}, is to bring Hamiltonian 
operator instead
of diagonal to the block-diagonal form, where the number of quasiparticles
is conserved in each block.

In many cases it is enough to transform the given Hamiltonian to the
block-diagonal form. In particular it is so, when the block, which describes
the states of interest, can be treated further with other methods. There are
many known transformations, that construct in this way from the initial
Hamiltonian operator an effective Hamiltonian operator, acting in a smaller
Hilbert space and which is simpler to consider. Flow equations, where
block-diagonalization is performed, have been compared with the following
transformations\\
$(1)$ the Schrieffer-Wolff transformation, which reduces the Anderson model
with single magnetic impurity to the Kondo problem \cite{KeMi2};\\
$(2)$ the Foldy-Wouthuysen transformation, which decouples the Dirac equation
into two two-component equations, one of which gives in the nonrelativistic
limit the known Pauli equation \cite{ByPi};\\
$(3)$ the Fr\"ohlich transformation, which constructs from the electron-phonon
interaction an effective electron-electron interaction \cite{LeWe}.
In all these cases flow equations re-examine the transformations used
before.

\section{Flow equations in field theory}
\label{sec2.4}

In this section we set up the framework to use flow equations for 
the problems of high-energy physics. We remind, that
flow equations perform the unitary transformation, which brings
the Hamiltonian to a block-diagonal form with the number of particles
(or Fock state) conserving in each block. In what follows we
distinguish between the 'diagonal' (here Fock state conserving)
and 'rest' (Fock state changing) sectors of the Hamiltonian.
We break the Hamiltonian as
\bea
&& H=H_{0d}+H_d+H_r
\leea{f1}
where $H_{0d}$ is the free Hamiltonian; and the indices 'd','r'
correspond to 'diagonal','rest' parts of the Hamiltonian, respectively.
The flow equation for the Hamiltonian \eq{2} and the generator
of unitary transformation \eq{6} are written \cite{We}
\bea
&& \frac{dH}{dl}=[\eta,H_d+H_r]+[[H_d,H_r],H_{0d}]
+[[H_{0d},H_r],H_{0d}]\nn\\
&& \eta=[H_{0d},H_r]+[H_d,H_r]
\leea{f2}
In the basis of the eigenfunctions of the free Hamiltonian
\bea
&& H_{0d}|i>=E_i|i>
\leea{f3}
one obtains for the matrix-elements between the many-particle states
\bea
&& \frac{dH_{ij}}{dl}=[\eta,H_d+H_r]_{ij}-(E_i-E_j)[H_d,H_r]_{ij}
-(E_i-E_j)^2H_{rij}\nn\\
&& \eta_{ij}=(E_i-E_j)H_{rij}+[H_d,H_r]_{ij}
\leea{f4}
The energy differences are given by
\bea
&& E_i-E_j=\sum_{k=1}^{n_2}E_{i,k}-\sum_{k=1}^{n_1}E_{j,k}
\leea{f5}
where $E_{i,k}$ and $E_{j,k}$ are the energies of the created and
annihilated particles, respectively.
The generator belongs to the 'rest' sector, i.e.
$\eta_{ij}=\eta_{rij}, \eta_{dij}=0$.
In what follows we use
\bea
&& [\hat{O}_r,\hat{H}_d]_d=0\nn\\
&& [\hat{O}_r,\hat{H}_d]_r\neq 0
\leea{}
where $\hat{O}_r$ is the operator from the 'rest' sector
(for example $\hat{H}_r$ or $\hat{\eta}_r$)
and $\hat{H}_d$ is the diagonal part of Hamiltonian.
 
For the 'diagonal' $(n_1=n_2)$
and 'rest' $(n_1\neq n_2)$
sectors of the Hamiltonian one has correspondingly
\bea
&& \frac{dH_{dij}}{dl}=[\eta,H_r]_{dij} \nn\\
&& \frac{dH_{rij}}{dl}=[\eta,H_d+H_r]_{rij}-(E_i-E_j)[H_d,H_r]_{rij}
+\frac{du_{ij}}{dl}\frac{H_{rij}}{u_{ij}}
\leea{f6b}
where we have introduced the cutoff function $u_{ij}(l)$
\bea
&& u_{ij}(l)={\rm e}^{-(E_i-E_j)^2l}
\leea{f7}
The energies $E_i(l)$ start to depend on the flow parameter $l$
in the second order of perturbation theory, that is taken into
account by the renormalization of single-particle 
electron and photon energies (see chapter \ref{ch7}).

The main difference between these two sectors is the presence
of the third term in the 'rest' sector
$\frac{du_{ij}}{dl}\frac{H_{rij}}{u_{ij}}$,
which insures the band-diagonal structure for the 'rest' part
\bea
&& H_{rij}=u_{ij}\tilde{H}_{rij}
\leea{f8}
i.e. in the 'rest' sector the matrix elements with the
energy differences larger than $1/\sqrt{l}$ are suppressed. 
In the similarity renormalization scheme \cite{GlWi}
the width of the band corresponds to the UV cutoff $\la$.
The connection between the two quantities is given
\bea
&& l=\frac{1}{\la^2}
\leea{f9}
The matrix elements of the interactions, which change
the Fock state, are strongly suppressed,if the energy difference
exceeds $\la$, while for the Fock state conserving part
of the Hamiltonian the matrix elements with all energy differences
are present.
As the flow parameter $l\rightarrow\infty$ (or $\la\rightarrow 0$)
the 'rest' part is completely eliminated, except maybe
for the matrix elements with $i=j$. One is left with the block-diagonal
effective Hamiltonian.
 
Generally, the flow equations are written
\bea
&& \frac{dH_{ij}}{dl}=[\eta,H_d+H_r]_{ij}-(E_i-E_j)[H_d,H_r]_{ij}
+\frac{du_{ij}}{dl}\frac{H_{ij}}{u_{ij}}\nn\\
&& \eta_{ij}=[H_d,H_r]_{ij}
+\frac{1}{E_i-E_j}\left(-\frac{du_{ij}}{dl}\frac{H_{ij}}{u_{ij}}\right)
\leea{f10}
where the following condition on the cutoff function
in 'diagonal' and 'rest' sectors, respectively, is imposed
\bea
&& u_{dij}=1\nn\\
&& u_{rij}=u_{ij}
\leea{f11}
One recovers with this condition the flow equations \eq{f6b}
for both sectors.
Other unitary transformations, which bring the Hamiltonian
to the block-diagonal form, with the Fock state conserving in each block
are used \cite{GlWi}
\bea
&& \frac{dH_{ij}}{d\la}=u_{ij}[\eta,H_d+H_r]_{ij}
+r_{ij}\frac{du_{ij}}{d\la}\frac{H_{ij}}{u_{ij}}\nn\\
&& \eta_{ij}=\frac{r_{ij}}{E_i-E_j}
\left([\eta,H_d+H_r]_{ij}-\frac{du_{ij}}{d\la}\frac{H_{ij}}{u_{ij}}\right)
\leea{f12}
and \cite{WiPe}
\bea
&& \frac{dH_{ij}}{d\la}=u_{ij}[\eta,H_d+H_r]_{ij}
+\frac{du_{ij}}{d\la}\frac{H_{ij}}{u_{ij}}\nn\\
&& \eta_{ij}=\frac{1}{E_i-E_j}
\left(r_{ij}[\eta,H_d+H_r]_{ij}
-\frac{du_{ij}}{d\la}\frac{H_{ij}}{u_{ij}}\right)
\leea{f13}
where $u_{ij}+r_{ij}=1$; and the constrain \eq{f11}
on the cutoff function in both sectors is implied.
One can choose the sharp cutoff function
$u_{ij}=\theta(\la-|\De_{ij}|)$.

\chapter{Light-front field theory}
\label{chapter3}

\section{Introduction}
\label{sec3.1}

The development of light-front field theory dates back to
the work of Dirac \cite{di:49}, where he introduced
the light-front coordinates (the coordinate vector
is $x=(x^+,x^-,x_{\perp})$ with $x^{\pm}=x^0\pm x^3$ 
and $x_{\perp}=(x_1,x_2)$) and the concept of front form
dynamics for Hamiltonians. Dirac suggested, that a Hamiltonian
operator can 'propagate' a physical system either in the usual
time $x^0$ (instant form dynamics) or in the light-front time 
$x^+$ (front form dynamics). The latter form of 
relativistic dynamics combines ``the restricted principle
of relativity with the Hamiltonian formulation of dynamics''
\cite{di:49}.

Later the rules for front form perturbation theory were
formulated \cite{KoSo}, and the equivalence of this theory
with the Feynman rules of covariant perturbation theory
was established \cite{ChRo},\cite{Li}.

Recent interest in light-front coordinates
is driven mainly by two topics: low-energy bound state
problem in QCD, where light-front coordinates offer
a scenario in which a constituent picture
of hadron structure can emerge from QCD, because
of the simplified vacuum on the light-front 
\cite{osu}, \cite{PaQCD}, \cite{BrPaPi};
and high-energy scattering processes, where 
light-front coordinates are the natural coordinates 
of the system \cite{Bu}, \cite{BrPaPi}.
For an extensive list of light-front references through
the early $1990's$ see \cite{LFref}, for the list of
recent reviews see \cite{lfreview}, \cite{BrPaPi}
and references within.

Below we give briefly an introduction to light-front field theory
(for introduction see also \cite{lfped}).
  
Light-Front (LF) quantization is very similar to canonical equal
time (ET) quantization (here we closely 
follow Ref. \cite{kent}). Both are Hamiltonian formulations of
field theory, where one specifies the fields on a
particular initial surface. The evolution of the fields
off the initial surface is determined by the
Lagrangian equations of motion. The main difference
is the choice of the initial surface, $x^0=0$ for
ET and $x^+=0$ for the LF respectively.
In both frameworks states are expanded in terms of fields
(and their derivatives) on this surface. Therefore,
the same physical state may have very different
wave functions\footnote{By ``wave function'' we mean here
the collection of all Fock space amplitudes.}
in the ET and LF approaches because fields at $x^0=0$
provide a different basis for expanding a state than
fields at $x^+=0$. The reason is that the microscopic
degrees of freedom --- field amplitudes at $x^0=0$
versus field amplitudes at $x^+=0$ --- are in general
quite different from each other in the two formalisms.

From the purely theoretical point of view, various advantages
of LF quantization derive from properties of the ten generators
of the Poincar\'e group (translations $P^\mu$,
rotations ${\vec L}$ and boosts ${\vec K}$) \cite{kent}.
Those generators which leave the initial surface
invariant (${\vec P}$ and ${\vec L}$ for ET and
$P_-$, ${\vec P}_\perp$, $L_3$ and ${\vec K}$ for LF)
are ``simple'' in the sense that they have very simple
representations in terms of the fields (typically just
sums of single particle operators). The other generators, 
which include the ``Hamiltonians'' ($P_0$, which is conjugate
to $x^0$ in ET and $P_+$, which is conjugate to the LF-time
$x^+$ in LF quantization) contain interactions among the
fields and are typically very complicated.
Generators which leave the initial surface invariant are also
called {\it kinematic} generators, while the others are called
{\it dynamic} generators. Obviously it is advantageous to have as
many of the ten generators kinematic as possible. There are
seven kinematic generators on the LF but only six in ET quantization.

The fact that $P_-$, the generator of $x^-$ translations, is
kinematic (obviously it leaves $x^+=0$ invariant!)
and positive has striking
consequences for the LF vacuum\cite{kent}. For free fields $p^2=m^2$ implies
for the LF energy $p_+ = \left(m^2 + {\vec p}_\perp \right)/p_-$.
Hence positive energy excitations have positive $p_-$. After the
usual re-interpretation of the negative energy states this implies
that $p_-$ for a single particle is non-negative [which makes sense,
considering that $p_- = p_0-p_3$].
$P_-$ being kinematic
means that it is given by the sum of single particle momenta $p_-$.
Combined with the non-negativity of $p_-$ this implies that, 
even in the presence of interactions,
the physical vacuum (ground state of the theory) differs
from the Fock vacuum (no particle excitations) only by
so-called {\it zero-mode} excitations, i.e. by excitations of modes
which are independent of the longitudinal LF-space coordinate
$x^-$. Due to this simplified vacuum structure, the LF-framework
seems to be the only framework, where a constituent quark picture
in a strongly interacting relativistic field theory has a chance
to make sense \cite{all:lftd,wi:mb,wi:dr,brodsky}.
This is the most attractive feature of LF-frame to approve
constituent quark model desription for QCD.

\section{Preceding work}
\label{sec3.2}

As far as progress is concerned, the light-front approach is not so far
along; most research effort has occurred since late $1980's$ \cite{LFref}.
Progress is currently limited by conceptual issues, mainly
by problems in renormalization program.

Further the methods available in the light-front field theory
to solve the bound state problem are discussed. Then the results 
for positronium problem on the light-front follow. 

\subsection{Models and methods in the light-front field theory}
\label{subsec3.2.1}

The ultimate goal of the light-front field theory is to start
with QCD Lagrangian and, with a minimum of approximation,
calculate the hadron spectrum. The basic idea behind this approach 
is to use Hamiltonian techniques in the coordinate system best suited 
for relativistic dynamics. For light-front field theory, physically 
interesting observables are quite easily calculated then from 
the bound state wave-functions.

Further we review several methods to solve the bound state problem
in the light-front field theory.

There are several attempts to solve for QCD bound states in 
the light-front approach, based on simplification of initial 
$QCD_{3+1}$ Lagrangian. Instead of real $QCD_{3+1}$ different models,
resembling QCD and having its main properties 
(confinement, asymptotic freedom as in transverse lattice model,
and also chiral symmetry breaking as in collinear model) are solved.

Collinear (tube) QCD model is a phenomenological model for light-front
QCD, where the transverse momenta of all constituents are neglected.
This reduces the $QCD_{3+1}$ Lagrangian to an effective $1+1$ dimensional
theory, which is then solved for the spectrum,
distribution amplitudes and form factors of mesons \cite{Coll}.

In the transverse lattice QCD approach one formulates an effective 
light-front Hamiltonian for $SU(N)$ Yang Mills theory in $2+1$ \cite{Trans1}
($(3+1)$ \cite{Trans2}) dimensions using two continuous space-time dimensions 
with the remaining transverse space dimension (two transverse dimensions)
discretized on a lattice. In the case of $(2+1)$ - dimensional theory
the reduction to a $(1+1)$ - dimensional theory takes place, that enables
to investigate the string tension and the glueball spectrum.

The goal of such toy model studies is to build intuition
which one can hopefully apply to $QCD_{3+1}$. However, while these
models have been very useful for studing nonperturbative
renormalization in $1+1$ dimesional LF field theories,
it is not clear to what extend these results can be generalized
to sufficiently nontrivial theories in $3+1$ dimensions.

In other approaches one reduces the initial Hamiltonian of
light-front field theory to an effective, low-energy Hamiltonian,
which can be solved for bound states. 
Different methods are used to construct such an effective Hamiltonian
on the light-front.

First group of methods, known as Tamm-Dancoff approach, 
is based on Tamm-Dancoff truncation \cite{TaDa} and 
projection technique in Fock space (Bloch-Fleshbach technique). 
By Tamm-Dancoff truncation one
simply restricts the full Fock space to several lowest Fock sectors
of interest. The problem of renormalization arises in this approach,
since in a limited Fock space the diagrams, needed for renormalization
but containing more Fock components in intermediate state as allowed,
inspite of it must be thrown away. 
The method of iterative resolvents
developed by Pauli, together with discretized light cone quantization
(DLCQ) approach is the synthesis
of the methods from the first group \cite{Itter}, \cite{DLCQ}.
In this approach one repeats projecting high Fock components upon
the lower in sequence several times. Thus constructed few-body effective
Hamiltonian describes most adequately low-lying bound states.
Application of this method to QED is discussed below.

Second group of methods uses renormalization group concept to construct
low-energy Hamiltonian. Similarity renormalization scheme \cite{GlWi}, 
formulated
by Glazek and Wilson, develops a renormalization group and the basic elements
of renormalization group calculations for Hamiltonians on the light-front.
In this method continuous unitary transformation, similarity transformation,
is performed to bring the Hamiltonian operator ${H_{ij}}$ to a band-diagonal
form $|E_i-E_j|\leq \la$. It turns out, that the width of the band $\la$
corresponds to the energy scale and plays the role of UV cutoff,
changing of which gives rise to renormalization group running of Hamiltonian.
Also new interactions, not present in canonical theory Hamiltonian, appear
through this scaling.

The alternative approach for Hamiltonian renormalization, 
the method of flow equations \cite{We}, was proposed independently by Wegner.
One performs a set of infinitesimal unitary transformations to get 
the block-diagonal effective Hamiltonian, with the particle number 
conserving in each block.  
This reduces the bound state problem to a few-body problem, since the particle
number violating contributions are eliminated.

There is a need for an effective formalism for gauge theories:
nobody ever has solved rigorously a relativistic many-body theory
in $3+1$ dimensions.

\subsection{Results for light front $QED_{3+1}$}
\label{subsec3.2.2}

In this section we review the results obtained by others on positronium
problem in the light front dynamics. First we discuss the first group of methods
based on Tamm-Dancoff approach (Tamm-Dancoff truncation and 
projection technique in Fock space).
The numerical recipes are well elaborated there.

Brodsky, Pauli, and Tang \cite{TaBrPa} showed how to set up the positronium 
bound state problem in DLCQ (discretized light cone quantization) with 
a Tamm-Dancoff truncation \cite{TaDa} to two sectors of Fock space:
(i) the electron-positron sector, and (ii) the electron-positron-photon sector.
In order to solve the associated eigenvalue problem, the diagonalization of the
discretized Hamiltonian, and variational methods were used. The leading results
for the triplet ground state of positronium were obtained not quit satisfactory.
To produce significant results by diagonalizing the Hamiltonian matrix, 
one would have had to include much more Fock states or to improve the numerical 
convergence of the method applied.

Krautg\"artner, Pauli, and W\"olz \cite{KrPaWo} derived the continuum limit
in the positronium model discussed above
for the calculation analytically, and with a prescription for handling an infrared
divergence they showed that to leading order the binding energy for positronium
results. They have used an effective interaction, obtained from a projecting
of the $|e\bar{e}\gamma>$-sector onto the $|e\bar{e}>$-sector. The corresponding
effective integral equation was solved numerically using the method of Coulomb
counterterms, introduced to improve numerical convergence. The results obtained
for Bohr spectrum and hyperfine splitting show an excellent convergence and coincide
with the expected values. For the first time the relativistic effects as fine structure
could be investigated (numerically) in the light-front Hamiltonian approach.

Kaluza and Pauli \cite{KaPa} showed that reasonably accurate numerical results 
can be obtained with proper renormalization of the Hamiltonian, although logarithmic
'divergences' remain in their calculations. These divergences are not a serious
problem if one uses a sufficiently small cutoff and weak coupling.
For calculation of fine structure the diagonalization technique was improved
to enlarge the Fock space feasible with the computer. But the convergence of spectra
was rather poor, since no proper counterterms for the Coulomb singularity
of the relativistic problem were taken.

Kaluza and Pirner extended the work in \cite{KrPaWo} through the fine structure \cite{KaPi}.

Pauli formulated the method of iterative resolvents \cite{Itter},
which enables, in general, 
to take into account
infinite many high Fock states by projecting in sequence high Fock sectors unto
the lower Fock sectors. The main advantage of the method is, that it requires
the inversion of the effective sector Hamiltonians, corresponding
to the given Fock sectors, at each step of iteration, instead of inversion
of the full Hamiltonian matrix, standing in the bound state equation. 
To avoid in the leading order the infrared divergences
(collinear singularities)
the full energy appeared in the interaction kernel for the electron-positron 
bound state was replaced with a free energy. In the physical amplitudes
the collinear singularities are exactly cancelled by the dynamical
terms \cite{BrLe}. 
Pauli and Trittmann
have solved numerically the corresponding light-front integral equation
for positronium bound state in the continuum limit \cite{TrPa}, including
the Coulomb counterterm technique in the computer code.
The results show excellent convergence and coincide to a high degree of accuracy
with the expected values.

Renormalization program for positronium in the light-front QED was performed
using similarity renormalization scheme (similarity transformation
and coupling coherence) by several authors.
Perry has shown \cite{Pe} that the resultant effective Hamiltonian leads
to standard lowest order bound state results, with the Coulomb interaction
emerging naturally.

For the first time Jones,Perry, and Glazek \cite{JoPeGl} presented
using similarity renormalization a systematic analytic calculation
in a light-front Hamiltonian approach of the singlet-triplet spin
splitting in the ground state of positronium through order $\alpha^4$.
The standard singlet-triplet splitting of positronium was obtained
with degeneracy of triplet state, that recovers rotational symmetry 
that is non-manifest symmetry in the light-front field theory.

Jones, and Perry have calculated also the Lamb shift in 
the same approach \cite{JoPe}. The dominant part of the splitting between the
$2S_{\frac{1}{2}}$ and $2P_{\frac{1}{2}}$ energy levels in hydrogen 
was obtained.

All calculations in similarity renormalization approach were done
within the perturbative theory and results for spectrum were obtained 
with several approximations analytically.

We do not mention here applications of both methods, Tamm-Dancoff approach
\cite{Wo} and similarity renormalization \cite{osu}, 
to bound state problem for light-front QCD.

\section{Canonical QED Hamiltonian}
\label{sec3.3}

In this section we write the canonical $QED_{3+1}$ Hamiltonian
in the light-front gauge and also give briefly its derivation
from Lagrangian density, following the original work of Zhang and Harindranath 
\cite{ZhHa}. The QED Hamiltonian in the secondary quantization is given 
further, which is used in the main part of the work 
(in the chapters \ref{ch4},\ref{ch5},\ref{ch7}). 

\subsection{Canonical light-front $QED_{3+1}$ Hamiltonian}
\label{subsec3.3.1}

Starting with the QED Lagrangian density
\bea
{\cal L}&=&-\frac{1}{4}F_{\mu \nu} F^{\mu \nu} +
\overline{\psi}(i \not\! {\partial } +e \not\!\! {A }- 
 m )\psi,\label{eq:L}
\eea
in a fixed gauge, $A^{+}=A^0+A^3=0$,\footnote{This derivation will not 
include a discussion of the gauge
field zero-modes. In this work we drop zero-modes. 
For a treatment that incorporates
these gauge field zero-modes from the start in QED, see \cite{KaRo} and
references within; (see also footnote at the end of this section).}  
the constrained degrees of freedom, $A^-$ and $\psi_-$
($\psi=\psi_++\psi_-,\psi_{\pm}=\Lambda_{\pm}\psi$; see all definitions
below), are removed explicitly, 
producing a canonical QED Hamiltonian, defined through the independent
physical fields $A_{\perp}$ and $\psi_+$. Details of derivation follow
below. The resulting canonical Hamiltonian $H_{can}$ is given as a sum
of the free Hamiltonian and the interacting term
\be
P^-=H_{can} = \int dx^-d^2x^{\bot}({\cal H}_0+{\cal H}_I)
\; . \lee{ch1}
where each term in \eq{ch1} is written \cite{ZhHa}
\be
{\cal H}_0=\frac{1}{2}(\partial ^iA^j)(\partial ^iA^j)+
\xi^+ \left( \frac{-\partial _{\bot}^2+m^2}{i\partial ^+} \right) \xi
\; , \lee{ch2}
\be
{\cal H}_I
 = {\cal H}_{ee \gamma} + {\cal H}_{ee \gamma \gamma}^{inst} + 
{\cal H}_{eeee}^{inst}
\lee{ch3}
and  
\be
{\cal H}_{ee\gamma}=e\xi^{\dag}\left[-2(\frac{\partial^{\perp}}
{\partial^+}\cdot A^{\perp})+\sigma \cdot A^{\perp} 
\frac{\sigma \cdot \partial^{\perp}+m}{\partial^+}+
\frac{\sigma \cdot \stackrel{\!\!\leftarrow}
{\partial^{\perp}}+m}{\stackrel{\!\!\leftarrow}{\partial^+}} 
\sigma \cdot A^{\perp}
\right]\xi
\;, \lee{ch4}
\be
{\cal H}_{ee \gamma \gamma}^{inst}= -ie^2 \left[ \xi ^+ \sigma\cdot A^{\bot}
\frac{1}{\partial ^+}(\sigma \cdot A^{\bot}\xi) \right]
\; , \lee{ch5}
\be
{\cal H}_{eeee}^{inst}=2e^2 \left[ \left( \frac{1}{\partial ^+}(\xi ^+ \xi) 
\right)
 \left( \frac{1}{\partial ^+}(\xi ^+ \xi) \right) \right]
\; , \lee{ch6a}       
where $\{\sigma ^i\}$ are the standard $2 \times 2$ Pauli matrices,
$i=1,2$ only, e.g.,  $\sigma \cdot \partial^{\perp}=
\sigma^i \partial^i = \sigma^1 (-\frac{\partial}{\partial x^1})+\sigma^2
(-\frac{\partial}{\partial x^2})$,
and
$\partial ^+=2\partial _- = 2\frac{\partial}{\partial x^-}$.
Formally one can write the last term \eq{ch6a} in the form
\be
{\cal H}_{eeee}^{inst}= 
-\frac{1}{2}J^+\frac{1}{\left(\partial^+\right)^2} J^+
\label{ch6ab}\;.
 \ee
where $J^+=-2 e\xi^\dagger \xi$.
We have used the two-component representation for fermion fields introduced
by Zhang and Harindranath \cite{ZhHa}\footnote{
In the two-component
representation the fermion fields are given
\bea
&& \psi_{+}=\left(
\begin{array}{c}
\xi\\
0
\end{array}
\right)\nn\\
&& \psi_{-}=\left(
\begin{array}{c}
0\\
\frac{1}{i \partial^+}\left\{\left[
\sigma^i\left(
i \partial^i+e A^i
\right)+i m
\right]\xi\right\}
\end{array}
\right)\;,
\leea{footnote1}
where $\xi$ is expressed through the two-component spinors
$\chi_s$ as follows
\bea
\xi(x) &=& \sum_{s\pm\frac{1}{2}} 
\chi_s \int\frac{dp^+d^2p^{\bot}}{2(2\pi)^3}\theta(p^+)
(b_{p,s}e^{-ipx}+d^{\dagger}_{p,\bar{s}}e^{ipx}) \nn \\
&& \chi(\uparrow) = {1\over\sqrt{2} } \, 
      \left(\begin{array} {r}
      1 \\  0  \end{array}\right)
       \qquad{\rm and}\qquad
       \chi(\downarrow) = {1\over\sqrt{2}} \, 
      \left(\begin{array} {r}
       0 \\ 1 \end{array}\right)
\leea{footnote2}
and $\bar{s}=-s$.

For completness we give the representation for the physical 
gauge field (transverse component) through the polarization vectors
$\varepsilon(\la)$
\bea
A^i(x) &=& \sum_{\lambda}\int \frac{dq^+ d^2q^{\bot}}{2(2\pi)^3}
\frac{\theta (q^+)}{\sqrt{q^+}}
(a_{q,\lambda} \varepsilon _{\lambda}^i e^{-iqx}+
a^{+}_{q,\lambda} \varepsilon _{\lambda}^{*i} e^{iqx})\nn\\
&& \varepsilon(\uparrow)=\frac{-1}{\sqrt{2}}(1,i)
\qquad{\rm and}\qquad
   \varepsilon(\downarrow)=\frac{1}{\sqrt{2}}(1,-i)
\; , \leea{footnote4}
}

Now the
details of the derivation of the canonical Hamiltonian are presented.
Given ${\cal L}$ of Eq.~(\ref{eq:L}), the equations of motion are
\bea
\partial_\mu F^{\mu \nu} &=&  J^{\nu}\;,\\
(i \not\! {\partial } +e \not\!\! {A }- 
 m )\psi&=&0\;,
\eea
where 
$J^\mu=-e{\overline \psi} \gamma^\mu \psi$. The physical
gauge $A^+=0$ is chosen
and the projection operators $\Lambda_+$ and $\Lambda_-$ 
are inserted into the equations
of motion. Note $\psi_-=\Lambda_- \psi$ and $\psi_+=\Lambda_+ \psi$.
Two of the equations are seen to be constraint equations:
\be
-\frac{1}{2} \partial^+ \partial^+ A^-+\partial^i\partial^+A^i=J^+\;,
\lee{con1}
\be
i\partial^+\psi_{-}=\left(
i\alpha^i \partial^i+e\alpha^iA^i+\frac{m \gamma^+}{2}
\right)\psi_{+}
\lee{con2}
The fact that these are constraints can be seen from the fact that
 no time derivatives $\partial^-$ 
appear. Note $\alpha^i=\gamma^0\gamma^i$.
Inverting the space
derivative $\partial^+$ gives
\be
A^-= \frac{-2}{( \partial^+)^2} J^+ + 2 \frac{ \partial^i}{ \partial^+} A^i
\;,\lee{con1a}
\be
\psi_{-}=\frac{1}{i\partial^+}\left[
\left(
i\alpha^i \partial^i+e\alpha^iA^i+\frac{m \gamma^+}{2}
\right)\psi_{+}
\right]
\; .\lee{con2a}
The gauge singularities in light-front QED (QCD) that arise
when one tries to eliminate the unphysical gauge degrees
of freedom by solving the constraint equations can be seen
clearly in momentum space. In momentum space,
the constraint \eq{con1} and \eq{con2} cannot determine
the dependent fields in terms of physical fields
for the single longitudinal momentum $k^+=0$. In coordinate
space, this implies that the $A^+=0$ gauge has a singularity
at longitudinal boundary. A careful treatment of the definition
of $1/\partial^+$ is necessary.
A suitable definition of $(1/\partial^+)$ 
(and consequently $(1/\partial^+)^2$), which determines uniquely
the initial value problem at $x^+=0$, is given \cite{ZhHa}  
\beaa
&&\Biggl{(} \frac{1}{\partial^+} \Biggr{)}f(x^-)=\frac{1}{4}
\int_{- \infty}^{+ \infty}dy^{-}\epsilon(x^{-} - y^{-})\,f(y^-)+g_1\;,\\
&&\Biggl{(} \frac{1}{\partial^+} \Biggr{)}^2f(x^-)=\frac{1}{8}
\int_{- \infty}^{+ \infty}dy^{-}|x^{-} - y^{-}|\,f(y^-)+g_2+x^-g_3\;,\\
 &&\partial^+=2\partial_-=2\frac{\partial}{\partial x^-}\;,\\
&&\partial_- \epsilon(x^- - y^-)=2 \delta(x^- - y^-)\;,\\
&&\epsilon(x)=\theta(x)-\theta(-x)\;.\label{foot1}
\eeaa
where $f(x)$ is an arbitrary field with 
$x^\perp$ and $x^+$ being implicitly in the argument of it. The functions
$g_1$, $g_2$ and $g_3$ are arbitrary fields independent of $x^-$. 
For a discussion on these boundary terms see \cite{ZhHa}.
Notice that this inverse longitudinal derivative
is non-local.

In practice,
we define the
 inverse longitudinal derivative in momentum space. We 
 explicitly put
the momentum representation of the field operators
into the respective terms of the Hamiltonian, multiply
the fields out explicitly, and then replace the inverse derivatives by
appropriate factors of longitudinal momentum  with
the restriction $|p^+|/{\cal P}^+\geq \epsilon=0_+$ 
[${\cal P}^+$ is the total longitudinal momentum of the physical state
of interest].
The absolute value sign on $\left|p^+\right|$ is required for the instantaneous interactions.
For example, a product of two fields gives
\be
\frac{1}{i\partial^+} \exp[-i(p-k)\cdot x]\longrightarrow
\frac{1}{p^+-k^+}\theta\left(|p^+-k^+|-\epsilon {\cal P}^+\right)
\exp[-i(p-k)\cdot x]
\;.\label{eq:2.67}
\ee
The dynamical degrees of freedom are $A^i$ and $\psi_{+}$. 
The canonical
Hamiltonian density is defined in terms of these dynamical degrees of freedom
\be
{\cal H}=\frac{\partial {\cal L}}{\partial(\partial^-A^i)}\partial^-A^i
+\frac{\partial {\cal L}}{\partial(\partial^-\psi_{+})}\partial^-\psi_{+}
-{\cal L}
\;.
\ee
Taking these derivatives of the Lagrangian density and combining terms,
the Hamiltonian density takes the following simple form
\be
{\cal H}=\frac{1}{2}(\partial^i A^j)^2+
\psi_{-}^\dagger i \partial^+ \psi_{-}
-2 \left(
\frac{1}{\partial^+}\frac{J^+}{2}
\right)^2
+\frac{J^+}{2}A^-
\;,
\ee
where the constraints of \eq{con1},\eq{con2} 
are assumed to be satisfied. In our $\gamma$-matrix representation,
only two of the components of the 4-spinor $\psi_{+}$  
are nonzero. Writing
these as the 2-spinor $\xi$,\footnote{
In other words, we are defining 
$\psi_{+}=\Lambda_+ \psi=(\psi_{1},\psi_{2},0,0)\equiv (\xi,0)$}
 and inserting the constraints
of  \eq{con1a},\eq{con2a},
$H_{can}$ takes on the form written earlier 
in Eqs.~(\ref{ch1})--(\ref{ch6a}), where surface terms such as in
\be
\int d^2 x^\perp dx^-\left(\frac{1}{\partial^+} J^+\right)
\left(\frac{1}{\partial^+} J^+\right)=
-\int d^2 x^\perp dx^-J^+\left(\frac{1}{\partial^+} \right)^2J^+
+{\rm surface ~terms}
\;,
\lee{foot2} 
are dropped \footnote{
A set of boundary integrals (as the first term
in \eq{foot2}), arising from elimination of unphysical
gauge degrees of freedom, is associated with the light-front
infrared singularity, $k^+=0$. Using the above definition \eq{foot1}
in the boundary integrals, the singularity at $k^+=0$ is removed;  
the surface terms (as the second term in \eq{foot2})
vanish from the LFQED (LFQCD) Hamiltonian and the light-front linear
infrared divergences cancel in all physical amplitudes \cite{ZhHa}.
The problem of 'zero modes' (the singularity of $A^+=0$ gauge at $k^+=0$)
is hidden now in the nontrivial asymptotic behavior of the transverse
(physical) gauge degrees of freedom at longitudinal
infinity, see the first reference of \cite{ZhHa}.
In this work we do not consider this problem.
}.

\subsection{QED Hamiltonian in second quantization}
\label{subsec3.3.2}

In this work we use the matrix elements of canonical Hamiltonian
$H_{can}$, given in Eqs.~(\ref{ch1})--(\ref{ch6a}), calculated in the free
basis of $H_0$. 
Below we write the canonical Hamiltonian $H_{can}$
in the form of second quantization.  

We use the following momentum-space representation 
for the field operators, \cite{PeWi}
and \cite{ZhHa}, (see footnote in the previous section)
\bea
\xi(x) &=& \sum_s \chi_s \int\frac{dp^+d^2p^{\bot}}{2(2\pi)^3}\theta(p^+)
(b_{p,s}e^{-ipx}+d^{\dagger}_{p,\bar{s}}e^{ipx}) \nn \\
A^i(x) &=& \sum_{\lambda}\int \frac{dq^+ d^2q^{\bot}}{2(2\pi)^3}
\frac{\theta (q^+)}{\sqrt{q^+}}(\varepsilon _{\lambda}^i a_{q,\lambda}
e^{-iqx}+h.c.)
\; , \leea{ch8}
where spinors are $\chi_{1/2}=(1,0)$, $\chi_{-1/2}=(0,1)$,
with $\bar{s}=-s$
and polarization vectors $\varepsilon_1^i=\frac{-1}{\sqrt{2}}(1,i)$,
$\varepsilon_{-1}^i=\frac{1}{\sqrt{2}}(1,-i)$;
the integration running over the $p^+\ge 0$ only these
states, that are allowed the light-front theory.  

The corresponding (anti)commutation relations are
\bea
& \{ b_{p,s},b_{p',s'}^+\}=\{d_{p,s},d_{p',s'}^+\}=\de_{p,p'}\delta_{ss'} & \nn \\
& [a_{q,\lambda},a_{q',\lambda '}^+]=\de_{q,q'} \delta_{\lambda,\lambda '} &
\; , \leea{ch10}
where 
\be
\de_{p,p'}\equiv 2(2\pi)^3\delta(p^+-p'^+)\delta^{(2)}
(p^{\bot}-p'^{\bot}) 
\; . \lee{ch11}
The light-front vacuum has trivial structure for both boson and fermion
sectors, namely $a_q|0>=0$; $b_p|0>=0$, simplifying the 
analytical calculations.
The normalization of states is according to
\be
<p_1,s_1|p_2,s_2>=\de_{p_1,p_2}\delta_{s_1,s_2}
\; , \lee{ch12}
where $b_{p,s}^+|0>=|p,s>$.

Making use of the field representation \eq{ch8},
we have the following Fourier transformed for

\noindent
the {\bf free} Hamiltonian 
\be
H_0=\sum _s\int\frac{dp^+ d^2p^{\bot}}{2(2\pi)^3}\theta (p^+)\\
\frac{p^{\bot 2}+m^2}{p^+} (b_{p,s}^+b_{p,s}+d_{p,s}^+d_{p,s})+\\
\sum_{\lambda}\int\frac{dq^+d^2q^{\bot}}{2(2\pi)^3}\theta(q^+)\\
\frac{q^{\bot 2}}{q^+}a_{q,\lambda}^+a_{q,\lambda}
\; , \lee{ch13}

\noindent
the leading order $O(e)$-{\bf the electron-photon coupling} 
\bea
H_{ee\gamma} &=& \sum_{\lambda s_1s_2}\int d^3p_1 d^3p_2 d^3q 
~ e~ [\varepsilon_{\lambda}^i\tilde{a}_{q}
 + \varepsilon_{\lambda}^{i *}\tilde{a}_{-q}^+]
(\tilde{b}_{p_2}^+\tilde{b}_{p_1} +\tilde{b}_{p_2}^+\tilde{d}_{-p_1}^+ +
\tilde{d}_{-p_2}\tilde{b}_{p_1} +\tilde{d}_{-p_2}\tilde{d}_{-p_1}^+)\nonumber\\
&&\times\chi_{s_2}^+\Gamma^i(p_1,p_2,-q)\chi_{s_1} \de_{q,p_2-p_1}
\; , \leea{ch14}
where
\be
\Gamma^i(p_1,p_2,q)=2\frac{q^i}{q^+}-
\frac{\sigma\cdot p_2^{\bot}-im}{p_2^+}\sigma^i-
\sigma^i\frac{\sigma\cdot p_1^{\bot}+im}{p_1^+}
\; , \lee{ch15}
We have for the {\bf instantaneous} interactions of the order $O(e^2)$
\bea
&& \hspace{-3cm}
H_{eeee}^{inst} = \sum_{s_1s_2s_3s_4}
\int d^3p_1 d^3p_2 d^3p_3 d^3p_4 
~ e^2~
(\tilde{b}_{p_3}^+ +\tilde{d}_{-p_3})(\tilde{b}_{p_4}^+ +\tilde{d}_{-p_4})
(\tilde{b}_{p_1} +\tilde{d}_{-p_1}^+)(\tilde{b}_{p_2} +\tilde{d}_{-p_2}^+)
\nonumber \\
& &\times\chi_{s_3}^+\chi_{s_4}^+\frac{4}{(p_1^+-p_3^+)^2}\chi_{s_1}\chi_{s_2}
\de_{p_3+p_4,p_1+p_2}\nn\\
\chi_{s_3}^+ \chi_{s_4}^+  & \bbbone & \chi_{s_1} \chi_{s_2} =
\delta_{s_1 s_3} \delta_{s_2 s_4} + \delta_{s_1 s_4} \delta_{s_2 s_3}
\leea{ch16}
and
\bea
&& \hspace{-3cm}
H_{ee\gamma\gamma}^{inst} = \sum_{s_1s_2\lambda_1\lambda_2}
\int d^3p_1 d^3p_2 d^3q_1 d^3q_2 
~ e^2~
(\varepsilon_{\lambda_1}^{i *}\tilde{a}_{q_1}^+ +
\varepsilon_{\lambda_1}^i\tilde{a}_{-q_1})
(\varepsilon_{\lambda_2}^j\tilde{a}_{q_2}+
\varepsilon_{\lambda_2}^{j *}\tilde{a}_{-q_2}^+)
(\tilde{b}_{p_2}^+ +\tilde{d}_{-p_2})(\tilde{b}_{p_1}+\tilde{d}_{-p_1}^+)
\nonumber\\
&& \times\chi_{s_2}^+\frac{\sigma^j\sigma^i}{(p_1^+-q_1^+)}\chi_{s_1}
\de_{p_1+q_2,q_1+p_2}\nn\\
\chi_{s_2}^+  & \bbbone & \chi_{s_1}  =
\delta_{s_1 s_2} 
\; ; \leea{ch17}
here 
\bea
& \tilde{a}_q\equiv a_{q,\lambda}\frac{\theta(q^+)}{\sqrt{q^+}}, \qquad
 \left[ \tilde{a}_{-q}\equiv a_{-q,\lambda}
\frac{\theta(-q^+)}{\sqrt{-q^+}} \right]
 \; , & \nn \\
& \tilde{b}_p\equiv b_{p,s}\theta(p^+), \qquad
 \tilde{d}_p\equiv d_{p,\bar{s}}\theta(p^+)
\; , \leea{ch19}
and the $\de$-symbol stands for the function defined in \eq{ch11}, the
short notation for the integral is
\be
\int d^3p \equiv\int\frac{dp^+d^2p^{\bot}}{2(2\pi)^3}
\; . \lee{ch20}

\chapter{Hamiltonian bound state problem on the light-front}
\label{ch4}

In this chapter we outline the program we follow to solve 
the bound state problem on the light-front. This includes
two issues: the derivation of effective Hamiltonian and solving
the corresponding bound state equation. The first point
is discussed here explicitly in application to positronium
bound state problem. The methods to solve bound state equation
(analytically and numerically) are discussed in the next chapters.

\section{Introduction}
\label{sec4.1}

We use the method of flow equations to construct an effective
Hamiltonian starting from the light-front formulation which can be
used to solve the bound state problem. The physical idea behind
this approach is to use renormalization concept for Hamiltonians
to get an effective, low-energy Hamiltonian for a limited
Fock space. The bound state problem is reduced then to a few-body
problem with the effective Hamiltonian acting on the energy scale
of bound state formation.

The key to renormalization of Hamiltonians is to diagonalize 
the Hamiltonian operator \cite{GlWi}. We have discussed briefly
in chapter $2$ the methods to diagonalize Hamiltonians continuously,
suggested by Wegner \cite{We} and Glazek and Wilson \cite{GlWi}, which have
been called flow equations and similarity renormalization by the authors, 
resp.  

It is common to both methods that they eliminate by means of a unitary
transformation initially the off-diagonal matrix elements between
states with large energy differences and continue with states closer and 
closer 
in energy, so that off-diagonal matrix elements between states of
energy difference larger than $\la$ are eliminated or strongly suppressed. 
The final aim is to eliminate them completely ($\la \rightarrow 0$)
and to obtain a diagonalized Hamiltonian.

We have mentioned in chapter $3$, that a literal
use of this concept can lead to convergence problems \cite{We}.
As was suggested by Wegner \cite{We}, one
may leave the idea of diagonalizing immediately in favor of 
block-diagonalizing.
If matrix-elements between states of equal particle number are considered
diagonal, then the procedure brings the Hamiltonian into a block-diagonal 
form. Application of flow equations to an $n$-orbital model has shown,
that procedure of block-diagonalizing works much better, where 
block-diagonalization with respect to the quasiparticle
number (number of electrons above the Fermi edge plus number of holes
below the Fermi edge) is performed \cite{We}. 

It becomes apparent from the calculations by Jones, Perry and Glazek 
\cite{JoPeGl} on the basis of the similarity transformation,
that this scheme works 
well down to energy differences of the order of Rydberg, but if one
goes below, then contributions in higher orders in the coupling
become important.

Indeed it seems to be rather difficult to obtain bound states from
plane waves by continuous unitary transformations. In eliminating only
the terms which do not conserve the number of particles one postpones
the diagonalization, but reduces the problem to one in the space
of fixed particle number \cite{HaOk}. Thus for the positronium problem it is 
sufficient
to determine the one- and two-particle contribution of the Hamiltonian
for electrons and positrons.

Basically the procedure is very similar to that of the elimination
of the electron-phonon interaction \cite{LeWe} which yields an effective
attractive interaction between electrons responsible for superconductivity.
Both the method of flow equations and the similarity remormalization \cite{LeWe}
(the ref. to A. Mielke)
yield results different from Fr\"ohlich's original ones \cite{Fr} but in very
good agreement with more sophisticated methods. In QED it is the interaction
of the electrons with the photons instead of the phonons which has to be 
eliminated.

A basic advantage of the methods of similarity renormalization and flow 
equations in comparison to conventional perturbation theory is, that one
obtains normally less singular effective interactions.

This procedure is similar to the Tamm-Dancoff Fock space truncation 
\cite{TaDa,TrPa} 
in the sense that also in this truncation particle number changing interactions
are eliminated.

We define an effective Hamiltonian 
\be
H^{eff}\rightarrow U H U^+
\lee{in1}
where $H$ is the bare Hamiltonian and the unitary transformation U 
is determined by the flow equations below.

\section{Flow equations in the perturbative frame}
\label{sec4.2}

In chapter \ref{chapter2} we have set up the framework to use flow equations
in the field theory. In this section we formulate
the equations, obtained in chapter \ref{chapter2}, 
for a canonical QED Hamiltonian
in the perturbative frame.

Flow equations read
\bea
&& \frac{dH(l)}{dl}=[\eta(l),H(l)]\nn\\
&& \eta(l)=[H_d(l),H_r(l)]
\leea{}
where the Hamiltonian is given $H=H_d+H_r$, with $H_d$ and $H_r$
including all the terms from the 'diagonal' and the 'rest' sectors, resp. 
 
Our goal is to transform the Hamiltonian into blocks with the 
same number of (quasi)-particles. 
This means, that we define the 'diagonal' part $H_d$ as the part 
of the interaction which conserves the number of particles
(electrons, positrons, photons),
and the 'rest' $H_r$ as the particle number changing part.
In the case of QED(QCD),
where the electron-photon (quark-gluon) coupling is present,
the number of photons (gluons) is conserved in each block of the 
final effective Hamiltonian.

As a result of the unitary transformation new interactions are induced
(see below). They are absent at $l=0$ and are generated as $l$ increases.
They also give rise to new terms 
in the generator of transformation $\eta(l)$. This in its turn
generates new interactions again.

To be able to perform the calculations analytically
we proceed in a perturbative frame and truncate
the series assuming the coupling constant is small.
In the case of QED on the light-front one has for any finite 
value of $l$

\be
H(l)=H_d^{(0)}+H_r^{(1)}+H_d^{(2)}+H_r^{(2)}+...
\ee
where the superscript denotes the order in the bare coupling constant,
$H^{(n)}\sim e^n$; the indices 'd' and 'r' indicate the diagonal and the
rest parts correspondingly.
The part $H_d^{(0)}$ is the free Hamiltonian, corresponding to the single
particle energies with the structure in secondary quantization
$a^+ a,b^+ b, d^+ d$, where $a,b,d$ are the annihilation operators
of the photons, electrons and positrons correspondingly; 
$H_r^{(1)}$ denotes the electron-photon coupling (of the type $a^+ b^+ b$);
$H_d^{(2)}$ is the second order diagonal part of the Hamiltonian,
having the structure $b^+ d^+ bd, b^+ b^+ bb, d^+ d^+ dd$
(in the light front
they correspond to the canonical instantaneous (seagull) and to newly 
generated interactions in the diagonal sector in second order). 
Note, that the diagonal part in the flow equations is not only the free
Hamiltonian but the full particle number conserving part of the 
effective Hamiltonian.
The choice of only $H_d^{(0)}$ as the diagonal part gives rise
to the band-diagonal structure of the effective Hamiltonian
in each 'particle number' sector in the similarity renormalization 
scheme \cite{GuWe}. However, this makes a difference for the diagonal
part only if one goes beyond third order in $e$.

The generator of the transformation is
\be
\eta(l)=[H_d,H_r]=[H_d^{(0)},H_r^{(1)}]+[H_d^{(0)},H_r^{(2)}]+...=
\eta^{(1)}+\eta^{(2)}+...
\ee
Up to second order the flow equation reads
\be
\frac{dH(l)}{dl}=[\eta,H]=
[[H_d^{(0)},H_r^{(1)}],H_d^{(0)}]+[[H_d^{(0)},H_r^{(1)}],H_r^{(1)}]+
[[H_d^{(0)},H_r^{(2)}],H_d^{(0)}]+...
\ee
Also terms of higher orders in $e$ are generated by the flow equations.

In the basis of the eigenfunctions of the free Hamiltonian
$H_d^{(0)}$
\be
H_d^{(0)}|i>=E_i|i>
\ee
one obtains for the matrix-elements between the many-particle states
\bea
&& \eta_{}=(E_i-E_j)H_{rij}^{(1)}+(E_i-E_j)H_{rij}^{(2)}+...\nn\\
&& \frac{dH_{ij}}{dl}=-(E_i-E_j)^2H_{rij}^{(1)}
+[\eta^{(1)},H_r^{(1)}]_{ij}-(E_i-E_j)^2H_{rij}^{(2)}+...
\leea{fe7}
The energy differences are given by
\bea
&& E_i-E_j = \sum_{k=1}^{n_2}E_{i,k}-\sum_{k=1}^{n_1}E_{j,k}
\leea{fe9}
where $E_{i,k}$ and $E_{j,k}$ are the energies of the created
and annihilated particles, respectively.

The energy $E_i$ depends on the flow parameter $l$ only in second
order in the coupling. Therefore one has
\bea
&& \frac{dH_{rij}^{(1)}}{dl}=-(E_i-E_j)^2H_{rij}^{(1)}\nn\\
&& H_{rij}^{(1)}(l)=H_{rij}^{(1)}(l=0){\rm e}^{-(E_i-E_j)^2l}
=H_{rij}^{(1)}(\la=\La\rightarrow\infty){\rm e}^{-\frac{(E_i-E_j)^2}{\la^2}}
\eea
Here we have used the physical meaning of the flow parameter $l$;
$l=\frac{1}{\la^2}$, where $\la$ is UV-cutoff 
(see chapter \ref{chapter2}, \eq{f9}).

In the flow equations $\la$ defines the smooth UV-cutoff.
This fact insures the analytical behavior of the effective
Hamiltonian with $\la$, that helps in numerical calculations.   

In second order one has to distinguish between the behavior of the 'diagonal'  
and the 'rest' term. For the 'rest' part one has
\be
\frac{dH_{rij}^{(2)}}{dl}=[\eta^{(1)},H^{(1)}]_{rij}
-(E_i-E_j)^2H_{rij}^{(2)},
\ee
where index 'r' by $[\eta^{(1)},H^{(1)}]_{r}$ defines
the particle number changing part of the commutator. Introduce
\be
H_{rij}^{(2)}(l)={\rm e}^{-(E_i-E_j)^2l}\tilde{H}_{rij}^{(2)}(l).
\ee
Then the solution reads
\be
\tilde{H}_{rij}^{(2)}(l)=\tilde{H}_{rij}^{(2)}(l=0)
+\int_0^l dl'{\rm e}^{(E_i-E_j)^2l'}
[\eta^{(1)}(l'),H^{(1)}(l')]_{rij}.
\ee
For the 'diagonal' part one has
\be
\frac{dH_{dij}^{(2)}}{dl}=[\eta^{(1)},H^{(1)}]_{dij}
\ee
and the solution is
\be
H_{dij}^{(2)}(l)=H_{dij}^{(2)}(l=0)
+\int_0^l dl'[\eta^{(1)}(l'),H^{(1)}(l')]_{dij}.
\lee{renormalization}
Note, that though in general the commutator $[[H_d^{(0)},H_d^{(2)}],H_d^{(0)}]$
is not zero, it is not present in the flow equation due to the definition 
of the diagonal part. The corresponding commutator 
$[[H_d^{(0)},H_r^{(2)}]H_d^{(0)}]$ in the 'non-diagonal' sector insures the 
band-diagonal
form for the 'rest' interaction and also gives rise to the different
structure of the generated interaction (the integral term) in the 'rest' 
and 'diagonal' sectors. 

The commutator $[\eta^{(1)},H^{(1)}]$ gives rise to new terms 
in second order in the bare coupling $e$. In the case of QED it induces
new types of interactions and generates the renormalization group corrections
to the electron (photon) masses. The coupling constant starts to run 
in third order in $e$.

\section{Effective low-energy Hamiltonian}
\label{sec4.3}

\subsection{Effective electron-positron interaction}
\label{subsec4.3.1}

In this section we follow mainly the work \cite{GuWe}.
We calculate by means of flow equations
the effective interaction between electron and positron,
generated in the second order in coupling $e$ 
by elimination of electron-photon vertex. 
We use the canonical QED Hamiltonian on the light-front
in second quantization, given in chapter \ref{chapter3}.

The term for electron-photon vertex by finite $l$ is given
\bea
&& \hspace{-2cm}
H_{ee\ga}=\sum_{\lambda s_1s_2}
\int d^3p_1 d^3p_2 d^3q
(g_{p_ip_f}^*(l)
\varepsilon_{\lambda}^i\tilde{a}_q+
g_{p_ip_f}(l)\varepsilon_{\lambda}^{i *}\tilde{a}_{-q}^+)
(\tilde{b}_{p_2}^+\tilde{b}_{p_1}+\tilde{b}_{p_2}^+\tilde{d}_{-p_1}^+ +
\tilde{d}_{-p_2}\tilde{b}_{p_1}+\tilde{d}_{-p_2}\tilde{d}_{-p_1}^+)\nonumber\\
& &\times\chi_{s_2}^+\Gamma_l^i(p_1,p_2,-q)\chi_{s_1} \de_{q,p_2-p_1}
\; , \leea{gi1a}
and
\bea
\Ga_l^i(p_1,p_2,q) &=& 2\frac{q^i}{q^+}
-\frac{\sigma\cdot p_2^{\perp}-im_{p_ip_f}(l)}{p_2^+}\sigma^i
-\sigma^i\frac{\sigma\cdot p_1^{\perp}+im_{p_ip_f}(l)}{p_1^+}
\; , \leea{gi1b}
where $l$-dependence comes from the unitary transformation performed.
Also we write explicitly the momentum dependence
of the coupling constant and the mass as long as $l\neq 0$,
here $p_i$ and $p_f$ stand for the set of initial and final momenta,resp. 
The initial conditions for the coupling constant and the mass
are defined at the value of bare cutoff  
$\La\rightarrow\infty~~ (l_{\La}=0)$
\bea
\lim_{\La\rightarrow\infty}g(l_{\La})=e\nn\\
\lim_{\La\rightarrow\infty}m(l_{\La})=m
\leea{}

Following the procedure outlined in the previous section, the leading order
generator of the unitary transformation is
\bea
&& \hspace{-2cm}
\eta^{(1)}(l)=\sum_{\lambda s_1s_2}
\int d^3p_1 d^3p_2 d^3q 
(\eta_{p_ip_f}^*(l)
\varepsilon_{\lambda}^i\tilde{a}_q+
\eta_{p_ip_f}(l)\varepsilon_{\lambda}^{i *}\tilde{a}_{-q}^+)
(\tilde{b}_{p_2}^+\tilde{b}_{p_1}+\tilde{b}_{p_2}^+\tilde{d}_{-p_1}^+ +
\tilde{d}_{-p_2}\tilde{b}_{p_1}+\tilde{d}_{-p_2}\tilde{d}_{-p_1}^+)\nonumber\\
& &\times\chi_{s_2}^+\Gamma_l^i(p_1,p_2,-q)\chi_{s_1} \de_{q,p_2-p_1}
\; , \leea{gi1}
\be
\eta_{p_ip_f}(l)=-\Delta_{p_ip_f}g_{p_ip_f}=
\frac{1}{\Delta_{p_ip_f}}\cdot\frac{dg_{p_ip_f}}{dl}
\; . \lee{gi2}
where we have introduced $\De_{p_ip_f}=\sum p_i^- -\sum p_f^-$,
and the light-front fermion energy is 
\mbox{$p^- = \frac{p^{\bot 2} + m^2}{p^+}$},
the photon one \mbox{$q^- = \frac{q^{\bot 2}}{q^+}$}.
Further we calculate the bound states of positronium.
In what follows we consider in the $|e\bar{e}>$ sector

\noindent
the {\bf generated interaction} to the first nonvanishing order
\be
H_{e\bar{e}e\bar{e}}^{gen}=\sum_{s_1\bar{s}_2s_3\bar{s}_4}
\int d^3p_1 d^3p_2 d^3p_3 d^3p_4
V_{p_ip_f}^{gen}(l)b_{p_3}^+d_{p_4}^+d_{p_2}b_{p_1}
\chi_{s_3}^+\chi_{\bar{s}_4}^+\chi_{\bar{s}_2}\chi_{s_1}
\de_{p_1+p_2,p_3+p_4}
\; , \lee{gi3a}
with the initial condition 
$\lim_{\La\rightarrow\infty}V_{p_ip_f}^{gen}(l_{\La})=0$,

\noindent
and the {\bf instantaneous interaction}
\be
H_{e\bar{e}e\bar{e}}^{inst}=\sum_{s_1\bar{s}_2s_3\bar{s}_4}
\int d^3p_1 d^3p_2 d^3p_3 d^3p_4
V_{p_ip_f}^{inst}(l)b_{p_3}^+d_{p_4}^+d_{p_2}b_{p_1}
\chi_{s_3}^+\chi_{\bar{s}_4}^+\chi_{\bar{s}_2}\chi_{s_1}
\de_{p_1+p_2,p_3+p_4}
\; , \lee{gi4a}
where 
\bea
&& V_{p_ip_f}^{inst}(l) = g_{p_ip_f}^{inst}(l) \, 
\frac{4}{(p_1^+-p_3^+)^2} ,\nn\\
&& \lim_{\La\rightarrow\infty}g_{p_ip_f}^{inst}(l_{\La})=e^2
\; . \leea{gi5a}
where the initial (bare) value of instantaneous
interaction (its matrix element in $|e\bar{e}>$ sector \eq{gi5a})
is defined by the instantaneous term ${\cal H}_{eeee}^{inst}$
\eq{ch6a},\eq{ch6ab} of the canonical light-front QED Hamiltonian.
The order of the field operators in both interactions
satisfies the prescription of standard Feynmann rules in the
$|e\bar{e}>$ sector.

We note, that generated interaction \eq{gi3a} is a new interaction,
induced by flow equations in the second order in coupling,
and corresponds to the dynamical photon exchange;
while the instantaneous term \eq{gi4a}  
enters the canonical light-front QED Hamiltonian 
and describes the instant photon exchange.
Instantaneous term is special 
for the light-front calculations \footnote{
The analogous problem, interacting electrons and phonons
in a solid, was considered by Wegner and Lenz \cite{LeWe}.
They used 'equal time' canonical electron-phonon Hamiltonian
known from BCS-theory. The elimination of electron-phonon interaction
by the flow equations generates a new interaction between
electrons, that defines an effective electron-electron interaction.
This effective interaction leads to the leading Coulomb behavior.
}.   

To the leading (second) order we neglect the $l$ dependence 
of the energies (which start to run in the second order
in coupling constant, see chapter \ref{ch7})
in the interactions, that
enables to write the flow equations for the corresponding couplings.
 
The flow equations to the first (for the electron-photon coupling) and
second (for the instantaneous and generated interactions) orders are 
\bea
\frac{dg_{p_ip_f}(l)}{dl}&=&-\Delta_{p_ip_f}^2g_{p_ip_f}(l)\nonumber\\
\frac{dg_{p_ip_f}^{inst}(l)}{dl}&=& 0\\
\frac{dV_{p_ip_f}^{gen}(l)}{dl}&=&<[\eta^{(1)}(l),H_{ee\gamma}]>_{|e\bar{e}>}
\nonumber
\; , \leea{gi6c}
where
\be
\Delta_{p_ip_f} = \sum p_i^- - \sum p_f^-
\lee{gi7}
The second and the third equations are written for 
the 'diagonal' sector. 
The matrix element \mbox{$<[\eta^{(1)}(l) , H_{ee\gamma}]>_{|e\bar{e}>}$}
is understood as the corresponding commutator between
the free electron-positron states, namely
\mbox{$<p_3 s_3, p_4 \bar{s}_4|...|p_1 s_1, p_2 \bar{s}_2>$}.
In the order $O(e^2)$ there is also the contribution of the commutator
$<[\eta^{(1)},H_{ee\gamma}]>$ to the free Hamiltonian in one-electron
and one-photon sectors, that defines the renormalization of electron
and photon masses, resp., (see chapter \ref{ch7}). 
Renormalization group running of both (instantaneous and generated)
interactions starts in the order $O(e^4)$, and the electron-photon
coupling starts to run in the order $O(e^3)$.

Neglecting the dependence of the light-front energies on 
the flow parameter $l$ (that is the corrections of higher orders), 
the solution of \eq{gi6c} reads 
\bea
g_{p_ip_f}(l)&=&f_{p_ip_f}\cdot e+O(e^3)\nonumber\\
g_{p_ip_f}^{inst}(l)&=&g_{p_ip_f}^{inst}(l_{\La}=0)=e^2+O(e^4)\nn\\
V_{p_ip_f}^{gen}(l)&=&\int_0^l dl'
<[\eta^{(1)}(l'),H_{ee\gamma}(l')]>_{|e\bar{e}>}+O(e^4)
\nonumber\\
f_{p_ip_f}&=&{\rm e}^{-\De^2_{p_ip_f}l}=
{\rm e}^{-\frac{\De^2_{p_ip_f}}{\la^2}}
\; , \leea{gi8}
where the subscript $|e\bar{e}>$ means, that the commutator is considered 
in the electron-positron sector.
The electron-photon interaction exists in the band of size $\la$
($|\De_{p_ip_f}|<\la$), whereas the matrix elements of instantaneous and
generated interactions in $|e\bar{e}>$ sector are defined for all
energy differences.

We give below the explicit expressions for the generated interaction,
and details of calculations can be found in Appendix \ref{appA}.
In what follows we use the notations of this Appendix.
 
The matrix elements  of the commutator $[\eta^{(1)},H_{ee\gamma}]$
in the exchange and annihilation channels are 
\be
<[\eta^{(1)},H_{ee\gamma}]> = 
\left\{ \begin{array}{l}
 M_{2ii}^{(ex)}  \frac{1}{(p_1^+-p_3^+)}(\eta_{p_1,p_3}g_{p_4,p_2}+
\eta_{p_4,p_2}g_{p_1,p_3}) \; , \\
\\
-M_{2ii}^{(an)}  \frac{1}{(p_1^++p_2^+)}(\eta_{p_1,-p_2}g_{p_4,-p_3}+
\eta_{p_4,-p_3}g_{p_1,-p_2}) \; ,
\end{array} \right. 
\lee{gi9}
\noindent
where
\bea
&& \eta_{p_1,p_2}(l) = e\cdot\frac{1}{\Delta_{p_1p_2}}
 \frac{df_{p_1,p_2}(l)}{dl} \nn \\
&& g_{p_1,p_2}(l) = e\cdot f_{p_1,p_2}(l) \nn\\
&& \Delta_{p_1,p_2} = p_1^- - p_2^- - (p_1-p_2)^-
\leea{gi10}
and the conservation of ``$+$'' and ``$\perp$'' components 
of the total momentum
is implied, i.e. $p_1^+ +p_2^+=p_3^+ +p_4^+$ and
$p_1^{\perp}+p_2^{\perp}=p_3^{\perp}+p_4^{\perp}$.
The matrix elements $M_{2ii}$ between the corresponding spinors
in both channels are given
\bea
M_{2ij}^{(ex)}&=&[\chi_{s_3}^+\Gamma^i(p_1,p_3,p_1-p_3)\chi_{s_1}]\,
[\chi_{\bar{s}_2}^+\Gamma^j(-p_4,-p_2,-(p_1-p_3))\chi_{\bar{s}_4}]
\nonumber\\
\\
M_{2ij}^{(an)}&=&[\chi_{s_3}^+\Gamma^i(-p_4,p_3,-(p_1+p_2))\chi_{\bar{s}_4}] \,
[\chi_{\bar{s}_2}^+\Gamma^j(p_1,-p_2,p_1+p_2)\chi_{s_1}]\nonumber
\leea{gi11}
where
\be
\Gamma^i(p_1,p_2,q) = 2\frac{q^i}{q^+} -
\frac{\sigma\cdot p_2^{\bot} - im}{p_2^+}\sigma^i -
\sigma^i\frac{\sigma\cdot p_1^{\bot} + im}{p_1^+}
\lee{b4c}
This equation, \eq{gi11}, defines the spin structure 
of the generated interaction.

We combine the formulas for commutator $[\eta^{(1)},H_{ee\ga}]$ together
with the generator $\eta(l)$ and coupling constant $g(l)$, expressed
through the similarity function $f(l)$.
The generated interactions \eq{gi8} are given then in both channels 
\bea
V_{gen}^{(ex)}(\la)&=&-e^2M_{2ii}^{(ex)}\frac{1}{(p_1^+-p_3^+)}
\left(\frac{\int_{\la}^{\infty}\frac{df_{p_1,p_3,\la'}}{d\la'}f_{p_4,p_2,\la'}d\la'}
{\De_{p_1,p_3}}+
\frac{\int_{\la}^{\infty}\frac{df_{p_4,p_2,\la'}}{d\la'}f_{p_1,p_3,\la'}d\la'}
{\De_{p_4,p_2}}\right)
\nonumber\\ 
V_{gen}^{(an)}(\la)&=&e^2M_{2ii}^{(an)}\frac{1}{(p_1^++p_2^+)}
\left(\frac{\int_{\la}^{\infty}\frac{df_{p_1,-p_2,\la'}}{d\la'}f_{p_4,-p_3,\la'}d\la'}
{\De_{p_1,-p_2}}+
\frac{\int_{\la}^{\infty}\frac{df_{p_4,-p_3,\la'}}{d\la'}f_{p_1,-p_2,\la'}d\la'}
{\De_{p_4,-p_3}}\right)
\nonumber\\ 
\; . \leea{gi13a}         
where in the integral we have neglected the dependence of light-front energies
on the cutoff $\la$ (that is the correction of the order $O(e^2)$),
and the connection between flow parameter and cutoff, $l=1/\la^2$, is used. 
We use the explicit form for similarity function  
\be
f_{p_1,p_2,\la}={\rm e}^{-\frac{\De_{p_1p_2}^2}{\la^2}}
\; . \lee{gi14}  
that gives for the generated interaction in both channels
fig.($2$), fig.($3$)
\bea
V_{gen}^{(ex)}(\la) &\hspace{-5em}=\hspace{-5em}& 
-e^2M_{2ii}^{(ex)}\frac{1}{(p_1^+-p_3^+)} \,
\frac{\De_{p_1,p_3}+\De_{p_4,p_2}}{\De_{p_1,p_3}^2+\De_{p_4,p_2}^2}
\cdot(1-f_{p_1,p_3,\la}f_{p_4,p_2,\la})\nn \\
\\
V_{gen}^{(an)}(\la) &\hspace{-5em}=\hspace{-5em}& 
e^2M_{2ii}^{(an)}\frac{1}{(p_1^++p_2^+)} \,
\frac{\De_{p_1,-p_2}+\De_{p_4,-p_3}}{\De_{p_1,-p_2}^2+\De_{p_4,-p_3}^2}
\cdot(1-f_{p_1,-p_2,\la}f_{p_4,-p_3,\la})\nn 
\; , \leea{gi14}

For finite values of $l$ the solutions of the
flow equations have no divergences in the form of small energy 
denominators, present in perturbation theory.
The divergent contribution in the generated interaction 
as $\De_{p_1,p_3}\sim\De_{p_4,p_2}\sim 0$
is effectively cancelled  
by the factor in bracket containing similarity functions 
$(1-f_{p_1,p_3,\la}f_{p_4,p_2,\la})$
(the same for annihilation channel).
One has for the generated interaction in the exchange channel
\be
\frac{\De_{p_1,p_3}+\De_{p_4,p_2}}{\De_{p_1,p_3}^2+\De_{p_4,p_2}^2}
=\frac{\De_{p_1,p_3}+\De_{p_4,p_2}}{\De_{p_ip_f}^2+
2\De_{p_1,p_3}\De_{p_4,p_2}}
\lee{gi14a}
where  
\be
\Delta_{p_ip_f}\equiv p_1^-+p_2^--p_3^--p_4^-=
\Delta_{p_1,p_3}-\Delta_{p_4,p_2}
=\Delta_{p_1,-p_2}-\Delta_{p_4,-p_3}
\lee{gi14b}
due to the total momentum conservation in $'+'$ and 'transversal' 
directions. The matrix elements with any energy differences
(i.e. $\forall \De_{p_ip_f}$) are present in 'diagonal' sector.   

The effective Hamiltonian is defined in the limit $\la\rightarrow 0$. 
In this limit the electron-photon coupling,
present in generated interaction through the similarity functions $f_{p_ip_f}$,
is completely eliminated ~~$f_{p_ip_f}(\la\rightarrow 0)=0$ for 
$\De_{p_ip_f}\neq 0$, and generated interaction is given by 
the expression that does not depend explicitly on the cutoff
$\la$. 

The resultant {\bf generated interactions} in both channels are given
fig.($2$),fig.($3$)
\bea
\tilde{V}_{gen}^{(ex)} &=& -e^2 N_{1,\la} \,
\frac{\tilde{\De}_1+\tilde{\De}_2}{\tilde{\De}_1^2+\tilde{\De}_2^2}\nn\\
\tilde{V}_{gen}^{(an)} &=& e^2 N_{2,\la} \,
\frac{M_0^2+M_0^{'2}}{M_0^4+M_0^{'4}}
\; , \leea{fgi15}
where we have introduced
\bea
& P^{+ 2}M_{2ii,\la}^{(ex)} = - N_1 \quad;\qquad 
P^{+ 2}M_{2ii,\la}^{(an)} = N_2 & \nn \\
& \De_{p_1 p_3} = \frac{\De_1}{P^+} = \frac{\widetilde{\De}_1}
{(x' - x) P^+} \quad;\qquad
 \De_{p_4 p_2} = \frac{\De_2}{P^+} = \frac{\widetilde{\De}_2}
{(x' - x) P^+}; & \nn \\
& \De_{p_1,-p_2} = \frac{M^2_0}{P^+} \quad;\qquad
 \De_{p_4,-p_3} = \frac{{M'}^2_0}{P^+} & 
\leea{gi16}
(see Appendix \ref{appA} for the explicit definition of these quantities
in the light-front frame).

The expression \eq{fgi15} is written for the rescaled value
of the potential, i.e. $V_{\la}=\tilde{V}_{\la}/P^{+2}$,
and the cutoff is defined in units of the total momentum $P^+$,
i.e. $\la\rightarrow\frac{\la^2}{P^+}$, with $l=1/\la^2$.
The spin structure of the interaction is carried by
the matrix elements $M_{2ii}$, defined in Appendix \ref{appA}.

We summarize the {\bf instantaneous interactions} in both channels
in the order ${\bf O(e^2)}$ fig.($2$),fig.($3$)
\bea
&& V_{inst}^{(ex)} = -\frac{4e^2}{(p_1^+-p_3^+)^2} \;
\delta_{s_1s_3} \delta_{s_2s_4}  \nn \\
&& V_{inst}^{(an)} = \frac{4e^2}{(p_1^++p_2^+)^2} \;
\delta_{s_1\bar{s}_2} \delta_{s_3\bar{s}_4} 
\; , \leea{gi17}
where we have used
\mbox{$\chi_{s_3}^+ \chi_{\bar{s}_2}^+ \bbbone \chi_{s_1} \chi_{\bar{s}_4} =
\delta_{s_1 s_3} \delta_{s_2 s_4} + \delta_{s_1 \bar{s}_2} \delta_{s_3 \bar{s}_4}$}.
For the rescaled potential in the light-front frame Appendix \ref{appA}
\eq{b14} we have 
\bea
&& \tilde{V}_{inst}^{(ex)} = -\frac{4e^2}{(x-x')^2} \;
\delta_{s_1s_3} \delta_{s_2s_4} \nn \\
&& \tilde{V}_{inst}^{(an)} = 4e^2 \;
\delta_{s_1\bar{s}_2} \delta_{s_3\bar{s}_4} 
\; . \leea{gi18}
In the second order in coupling the effective electron-positron interaction,
calculated in the light-front dynamics, is given as a sum of interactions,
generated by the flow equations \eq{fgi15}, and instantaneous interactions
\eq{gi18}, that are present in the light-front gauge calculations;i.e.
\bea
&& V^{eff}=\tilde{V}^{gen}+\tilde{V}^{inst}=
\sum_{(i)=ex,ann}(\tilde{V}_{gen}^{(i)}+\tilde{V}_{inst}^{(i)})
\leea{gi18a}
where we sum over exchange and annihilation channels.

In the chapters \ref{ch5} and \ref{ch6} we use the effective electron-positron
interaction \eq{gi18a} to calculate positronium mass spectrum. 

In the next section we outline the general computational strategy
to solve for bound states of positronium using an effective Hamiltonian.

\subsection{Positronium model (general computational strategy)}
\label{subsec4.3.2}

In this section we give the program to solve positronium bound state 
problem in the light-front dynamics. This approach can be applied
also to the other systems. 

Required that the particle number (Fock state) conserving terms 
in the Hamiltonian were considered to be diagonal and the other terms 
off-diagonal an effective Hamiltonian was obtained which is block-diagonal
in particle number (Fock) space. This means, that the 'diagonal' sectors
are decoupled, since the particle number (Fock state) violating
contributions are eliminated, and one is able then to truncate
the full Fock space in the effective Hamiltonian to the lowest 
Fock sectors of interest. The bound state problem is reduced
then to a few-particle Hamiltonian problem. Sure, this is true
to the given order of perturbation theory. 
This is the idea of the approach.

Below we illustrate schematically this procedure for  
the case of light-front QED, where the positronium 
bound state problem is reduced to a two-particle problem.

We start with the light front Schr\"odinger equation 
for the positronium model 
\be
H_{LC}|\psi_n>=M_n^2|\psi_n>
\lee{re1}
where $H_{LC}=P^{\mu}P_{\mu}$ is the invariant mass (squared) operator, 
referred for convenience to as the light front Hamiltonian of positronium 
and $|\psi_n>$ being the corresponding eigenfunction; $n$ labels all
the quantum numbers of the state. 

The canonical Hamiltonian of the system
$H_{LC}$ contains infinitely many Fock sectors (i.e. one has for 
the positronium  wave function $|\psi_n>=c_{e\bar{e}}|(e\bar{e})_n>
+c_{e\bar{e}\ga}|(e\bar{e}\ga)_n>
+c_{e\bar{e}\ga\ga}|(e\bar{e}\ga\ga)_n>+...$) and each Fock sector contains
states with arbitrarily large energies. We now

$(1)$ introduce the bare cutoff (regularization) with the result
$H^B(\La)$ - the bare Hamiltonian;

$(2)$ perform the unitary transformation by means of flow equations
with the result $H^{eff}$ - the effective renormalized
Hamiltonian (table $1$ for finite value of $\la$);

$(3)$ truncate the Fock space to the lowest Fock sector ($|e\bar{e}>$)
with the result $\tilde{H}^{eff}$ - the effective, 
renormalized Hamiltonian acting in the electron-positron sector.

Then the eigenvalue equation reads
\be
\tilde{H}^{eff}|(e\bar{e})_n>=M_n^2|(e\bar{e})_n>.
\lee{re1a}
The effective light front Hamiltonian consists of
the free (noninteracting) part and the effective 
electron-positron interaction
\be
\tilde{H}^{eff}=H^{(0)}+V^{eff}
\lee{re2}
The light front equation \eq{re1a} 
is then expressed by the integral equation (the coordinates are given in
fig.($4$)
\bea
&& \hspace{-2cm} \left(\frac{m^2+\vec{k}_{\perp}^{'2}}{x'(1-x')}-M_n^2\right)
\psi_n(x',\vec{k}_{\perp}^{'};s_3,s_4)\nn\\
&+& \sum_{s_1,s_2}\int_D \frac{dx d^2 k_{\perp}}{2(2\pi)^3}
<x',\vec{k}_{\perp}^{'};s_3,s_4|V^{eff}|x,\vec{k}_{\perp};s_1,s_2>
\psi_n(x,\vec{k}_{\perp};s_1,s_2)=0\nn\\
\leea{re3ab}
The kernel of this equation, the effective electron-positron interaction
$V^{eff}$, was obtained in the previous section, \eq{gi18a}.
Note, that the effective interaction $V^{eff}$, present in \eq{re3ab},
is boost invariant (i.e. does not depend on $P^+$).  
The integration domain $D$ in \eq{re3ab} is restricted by the covariant cutoff
condition \cite{BrLe}
\be
\frac{m^2+\vec{k}_{\perp}^{2}}{x(1-x)}\leq \La^2+4m^2
\lee{re3a}
which allows for states which have a kinetic energy below the 
cutoff $\La$.

In the chapter \ref{ch5} we solve eigenvalue equation \eq{re1a} analytically,
using the bound state perturbation theory.
In the chapter \ref{ch6} we solve the corresponding integral equation
\eq{re3ab} numerically, using the numerical methods 
elaborated in Tamm-Dancoff approach and discussed
in the section \ref{subsec3.2.2}.
 
We have considered here $|e\bar{e}>$-sector. In the next section we consider
extended Fock space to give diagrammatic representation
of the effective Hamiltonian, obtained in the second order in $e$.

\subsection{Effective, renormalized QED Hamiltonian}
\label{subsec4.3.3}

In this section we review diagrammatic rules for the effective Hamiltonian,
obtained by flow equations in the second order in $e$. It is useful
to represent Hamiltonian matrix elements in the form of table,
consisting of different Fock blocks, each of them has infinite many
matrix elements describing the transition between corresponding Fock states
and with infinite many energy differences (Table $1$). 
In this representation the canonical field theory Hamiltonian
on the light-front has definite structure, that is simpler
than that in equal-time formalism. On the light-front
the vacuum 'does not fluctuate'. This means, that only one- 
and two-particle transitions
are possible, three- and more-particle transitions are absent.
In the case of QED on the light-front one-particle transitions are
defined by the electron-photon interaction and two-particle transitions
are given by the instantaneous interactions.
Kinetic energy term is diagonal in Fock space representation,
also there are instantaneous terms that does not change particle number.
The canonical QED Hamiltonian has therefore pentadiagonal structure 
in this representation \cite{Itter}.

Now consider the effective QED Hamiltonian. We do not include 
instantaneous diagrams,
since the flow equations do not change them
to the second order in coupling.
We start with the situation, when the 'rest' (Fock state changing) 
sector consists of matrix elements of electron-photon vertex, and 
the 'diagonal' (Fock state conserving)
sector has matrix elements of kinetic terms for electron and photon.
We perform the unitary transformation to eliminate the 'rest' sector.
In the second order in $e$ the elimination of electron-photon vertex
generates one- and two-particle operators. The elimination of 'rest'
sector in the next orders in $e$ generates many-particle operators.

The matrix elements of the effective Hamiltonian, obtained by flow equations
in the second order in $e$, namely the diagrams in different Fock sectors
are depicted in Table $1$. Dot denote the zero in the second order matrix elements. 
Corresponding analytic expressions for 
the matrix elements in 'diagonal' and 'rest' sectors are listed in fig.($2$).
The diagrammatic rules are obtained by direct calculation of matrix elements
between free particle states.
We consider the situation with finite $\la$. The matrix elements
of the 'rest' sectors are squeezed in the energy band
$\De_{p_ip_f}=|\sum p_i^- -\sum p_f^-|<\la$; there are matrix elements 
with all energy differences in 'diagonal' sector. As $\la\rightarrow 0$
the 'rest' sector is completely eliminated to the given order 
of perturbation theory. One ends up with the block-diagonal effective
Hamiltonian, where in each block the Fock state is conserved.     

For completenes we give below the matrix elements
of effective interactions in different Fock sectors,
generated by the flow equations
in the second order in coupling constant.
The transitions between Fock states and corresponding matrix
elements are given, resp.,\\ 
in the {\bf 'diagonal' sectors} 
\bea
&& |e\bar{e}>\rightarrow |e\bar{e}>,
|e\bar{e}e\bar{e}>\rightarrow |e\bar{e}e\bar{e}>, ... \nn\\
&& -e_{\la}^2M_{2ij,\la}\de^{ij}\frac{1}{[p_1^+-p_3^+]}\left(
\frac{\int_{\la}^{\infty}\frac{df_{p_1p_3\la'}}{d\la'}f_{p_4p_2\la'}d\la'}
{\De_{p_1p_3\la}}+
\frac{\int_{\la}^{\infty}\frac{df_{p_4p_2\la'}}{d\la'}f_{p_1p_3\la'}d\la'}
{\De_{p_4p_2\la}}\right)\nn\\
&& |e\bar{e}\ga>\rightarrow |e\bar{e}\ga>,
|e\bar{e}\ga\ga>\rightarrow |e\bar{e}\ga\ga>, ...\nn\\
&& e_{\la}^2\tilde{M}_{2ij,\la}\varepsilon^{i*}\varepsilon^j\left(
\frac{\int_{\la}^{\infty}\frac{df_{p_1k_1\la'}}{d\la'}f_{p_2k_2\la'}d\la'}
{\De_{p_1k_1\la}}+
\frac{\int_{\la}^{\infty}\frac{df_{p_2k_2\la'}}{d\la'}f_{p_1k_1\la'}d\la'}
{\De_{p_2k_2\la}}\right)
\; , \leea{re1c}
in the {\bf 'rest' sectors}   
\bea
&& |e\bar{e}>\rightarrow |e\bar{e}e\bar{e}>,
|e\bar{e}e\bar{e}>\rightarrow |e\bar{e}>, ... \nn\\
&& -e_{\la}^2f_{p_ip_f\la}M_{2ij,\la}\de^{ij}\frac{1}{[p_1^+-p_3^+]}\left(
\frac{\int_{\la}^{\infty}\frac{1}{f_{p_ip_f\la'}}
\frac{df_{p_1p_3\la'}}{d\la'}f_{p_4p_2\la'}d\la'}{\De_{p_1p_3\la}}+
\frac{\int_{\la}^{\infty}\frac{1}{f_{p_ip_f\la'}}
\frac{df_{p_4p_2\la'}}{d\la'}f_{p_1p_3\la'}d\la'}{\De_{p_4p_2\la}}\right)\nn\\
&& |e\bar{e}>\rightarrow |\ga\ga>,
|\ga\ga>\rightarrow |e\bar{e}>, ...\nn\\
&& e_{\la}^2f_{p_ip_f\la}\tilde{M}_{2ij,\la}\varepsilon^{i*}\varepsilon^j\left(
\frac{\int_{\la}^{\infty}\frac{1}{f_{p_ip_f\la'}}
\frac{df_{p_1k_1\la'}}{d\la'}f_{p_2k_2\la'}d\la'}{\De_{p_1k_1\la}}+
\frac{\int_{\la}^{\infty}\frac{1}{f_{p_ip_f\la'}}
\frac{df_{p_2k_2\la'}}{d\la'}f_{p_1k_1\la'}d\la'}{\De_{p_2k_2\la}}\right)
\; , \leea{re2c}
where 'dots' denote the higher Fock sectors;
one obtains the next higher Fock sector, when in the given Fock state
an additional $(e\bar{e})$-pair or $\ga$-photon are created.
Note, that in \eq{re1c},\eq{re2c} the order of momenta is given
for $|e\bar{e}>$ channel, fig.($2$).
We use the similarity function
\bea
&& f_{p_ip_f\la}={\rm e}^{-\frac{\De_{p_ip_f}^2}{\la^2}} \nn\\
&& \De_{p_ip_f}=\sum p_i^--\sum p_f^-
\; , \leea{re3}
in \eq{re1c},\eq{re2c} to get the explicit form of interactions in both
sectors listed in fig.($2$). 'Rest' diagrams are drawn schematically 
to show the difference between the interactions in 'diagonal' and 'rest'
sectors. Namely, for the 'rest' sectors we imply, that the 
corresponding momentum 
exchange must be done in the diagrams of fig.($2$) to get 
analytical expressions for diagrams depicted in Table $1$.

In the second order flow equations give rise to the mass corrections
in one-particle sector. It turns out that the electron, photon mass
corrections are equal, but have the opposite sign, to the standard 
electron, photon self energy terms, 
obtained in the light-front perturbation theory.  
We consider this explicitly in the chapter \ref{ch7}.
In order to make the diagrammatic rules complete we include mass diagram
for the photon in the table (sector $|\ga>\rightarrow |\ga>$).
After the unitary transformation is performed the electron (photon)
mass depends on the cutoff
\be
m_{\la}^2=m_0^2-\de\Sigma_{\la}
\; , \lee{re4}
where $\de\Sigma_{\la}$ is the self energy term ($m_0=0$ for a photon),
and subscript $0$ denotes the bare mass. 
In the third order the coupling constant gets the cutoff dependence, 
i.e. $e_{\la}=e_0(1+O(e_{\la}^2))$, that is beyond our consideration.

The two-component LF theory, introduced by Zhang and Harindranath \cite{ZhHa},
as compared to four-component formalism of Brodsky and Lepage,
is formulated purely in terms of physical degrees of freedom;
so that each term in the effective, renormalized Hamiltonian 
corresponds to a real dynamical process (or give renormalization term).
Different 'diagonal' sectors of the effective Hamiltonian 
contribute to: in $|\ga>$ sector - to self energy photon operator,
in $|e\bar{e}>$ sector - to electron-positron bound state (or corresponding
scattering process),
in $|\ga\ga>$ sector - to light-light scattering, 
in $|e\bar{e}\ga>$ sector - to Compton scattering, {\it et. cetera} 
(see Table $1$).

In the previous section we considered this formalism in application 
to the positronium bound state problem.

\chapter{Positronium spectrum (analytically)}
\label{ch5}

In this chapter we solve an effective eigenvalue equation
for positronium, obtained in the previous chapter, analytically.
We perform the bound state calculations perturbatively. 
The idea of calculations is the following.
We split an effective Hamiltonian $H^{eff}$ into $H^{(0)}$, 
a part which
is solved nonperturbatively, and $\de V$, the difference
between the original Hamiltonian and $H^{(0)}$. The effects of $\de V$ 
are to be computed using bound state perturbation theory. 
The criteria for choosing 
$H^{(0)}$ is that it approximates the physics relevant for the given
bound states (positronium in our case) as closely as possible.
As a consequence, $H^{(0)}$ contributes the dominant term to the mass spectrum 
and the bound state perturbation theory converges 
with respect to $\de V/H^{(0)}$. For QED, where the analytic answer is known,
the leading order solution $H^{(0)}$ is simply given by a sum of 
kinetic terms and the Coulomb potential. This result arise straightforward
from the form of the effective electron-positron Hamiltonian 
in the nonrelativistic limit. 
For more complicated theories as QCD the hint to choose $H^{(0)}$
comes from phenomenological models.

In the next section we define explicitly bound state perturbation theory
for positronium system. The calculation of Bohr spectrum and the ground state
spin-splittings is given further. In analytical calculations of 
singlet-triplet spin-splitting we follow the work \cite{JoPeGl}.

\section{Bound state perturbative theory (BSPT)}
\label{sec5.1}

First introduce instead of the light front parameterization
fig.($3$), used before for the single-particle momenta, 
the instant form. We express the variable
$(x,\vec{\kappa}_{\perp})$ in terms of the equal-time variable 
$\vec{p}=(p_z,\vec{\kappa}_{\perp})$ as
\bea
&& x=\frac{1}{2}\left(1 + \frac{p_z}{\sqrt{\vec{p}^2 + m^2}} \right)\\
&& \vec{p}^2=p_z^2+\vec{\kappa}_{\perp}^2
\leea{bspt1}
and similarly for $x'$ and $\vec{p}^{'2}$ as function of $p_z^{'}$.

The Jacobian of this transformation, $J(p)$, is
\be
J(p)=\frac{dx}{d p_z}=\frac{\kappa_{\bot}^2+m^2}{2(\vec{p}^2+m^2)^{3/2}}
\; . \lee{bspt2}

The instant form is used for practical purposes: it is simpler
to recover the rotational symmetry there, the symmetry that is not manifest
in the light-front frame.

We choose the leading order Hamiltonian operator for positronium
\be
H^{(0)} = h + \hat{V}_{coul}
\; , \lee{bspt5}
where $h$ is the free part (sum of corresponding kinetic terms)
and $\hat{V}_{coul}$ is the Coulomb interaction 
\be
V_{coul} = -\frac{16 e^2 m^2}{(\kappa_{\bot} - \kappa'_{\bot})^2 + \\
(p_z - p'_z)^2} = -\frac{16 e^2 m^2}{(\vec{p}-\vec{p'})^2}
\; . \lee{bspt4}
Let us solve the corresponding Schr\"odinger equation
on the light-front
\be
H^{(0)} |\psi_N(P)> = E_N |\psi_N(P)>
\; , \lee{bspt6}
where $P$ is the positronium momentum. The eigenvalues and 
eigenfunctions for positronium bound state on the light-front
are defined in a standard way
\bea
& E_N = \frac{P_{\bot}^2 + M_N^2}{P^+} & \nn \\
& |\psi_N(P)> = \sum_{s_1 s_2} \, \int_{p_1 p_2} \, \sqrt{p_1^+ p_2^+} \, 
2(2\pi)^3 \, \de^{(3)}(P - p_1- p_2) \, 
\tilde{\Phi}_N (x \kappa_{\bot} s_1 s_2) \,
b^+_{s_1}(p_1) \, d^+_{s_2}(p_2)|0> & \nn \\
& \sum_{s_1s_2} \, \frac{\int d^2\kappa_{\bot} \, \int_0^1 dx}{2(2\pi)^3} \,
\tilde{\Phi}_N^*(x \kappa_{\bot} s_1 s_2) \,
\tilde{\Phi'}_N(x \kappa_{\bot} s_1 s_2) = \de_{NN'} &
\leea{bspt7}
$M_N$ stands here for the leading order mass of positronium.
Combining the definitions for the wave function and the energy
together with the light-front Schr\"odinger equation, one has
\be
\biggl[ M_N^2-\frac{\kappa_{\bot}^{'2}+m^2}{x'(1-x')} \biggr]
\tilde{\Phi}_N (x'\kappa'_{\bot}s_3s_4) = \sum_{s_1 s_2}
\frac{\int d^2\kappa_{\bot} \int_0^1 dx}{2(2\pi)^3} \,V_{coul} \, 
\tilde{\Phi}_N(x\kappa_{\bot}s_1s_2)
\; , \lee{bspt8}
or, after change of coordinates according to \eq{bspt1},
\be
\left( M_N^2 - 4 (\vec{p'}^2 + m^2) \right) \Phi_N(\vec{p'} s_3 s_4) =
\sum_{s_1 s_2} \frac{\int d^3p \sqrt{J(p) J(p')}}{2(2\pi)^3} \,
V_{coul}(\vec{p},\vec{p'}) \, \Phi_N(\vec{p} s_1 s_2)
\; , \lee{bspt9}
where the wave functions are redefined to have the norm
\be
\sum_{s_1s_2} \, \int d^3p \, \Phi_N^*(\vec{p} s_1 s_2) \,
\Phi_N'(\vec{p} s_1 s_2) = \de_{NN'} 
\; . \lee{bspt10}
Our aim is to obtain the nonrelativistic Schr\"odinger equation for positronium.
Note, that in the nonrelativistic limit \mbox{$\vec{p}^2/m^2 <\!\!< 1$}
we have 
\bea
& \sqrt{J(p) J(p')} \; \approx \;
\frac{1}{2m} \left( 1 - \frac{\vec{p}^2 + (p_z^2 + {p'_z}^2)} {2m^2} \right) & \nn \\
& M_N \; = \; (2m + B_N)^2 \; \approx \; 4m^2 + 4m B_N^{(0)} &
\; , \leea{bspt11}
where the leading order binding energy $B_N^{(0)}$ is introduced.
Then in the leading order the bound state equation for positronium is
\bea
&& \left( \frac{\vec{p'}^2}{m} - B_N \right) \Phi_N(\vec{p'} s_3 s_4)
\; = \;- \sum_{s_1 s_2} 
\int d^3p \left( \frac{1}{2m} \frac{1}{2(2\pi)^3} 
\frac{1}{4m} \, V_{coul} \right) \,
\Phi_N(\vec{p}s_1s_2)\nn\\ 
\; . \leea{bspt12}

Using the explicit form for the Coulomb potential, \eq{bspt4}, we get
the equation that determines the leading order bound state wave function:
\be
\left( \frac{\vec{p'}^2}{m} - B_N \right) \, \Phi_{\mu}(\vec{p'}) \; = \;
\frac{\al}{2\pi^2} \, \int \, \frac{d^3p}{(\vec{p}-\vec{p'})^2} \, \Phi_{\mu}(\vec{p})
\lee{bspt13}
with
\be
\Phi_N \; = \; \Phi_{\mu, s_e, s_{\bar{e}}}(\vec{p'} s_3 s_4) \; = \;
\Phi_{\mu}(\vec{p'}) \, \de_{s_e s_3} \, \de_{s_{\bar{e}} s_4}
\; . \lee{bspt14}

This is the standard nonrelativistic Schr\"odinger equation 
for positronium. The solution is characterized by
\mbox{$\mu \! = \!(n,l,m)$}, the usual principal and angular 
momentum quantum numbers.
The wave functions are given through the hyperspherical harmonics
\bea
&& Y_{\mu}(\Om) \; = \;
\frac{(e_n^2 + \vec{p}^2)^2}{4 \, e_n^{5\!/\!2}} \, \Phi_{\mu} \nn \\
&& Y_{\mu} \; = \; Y_{n,l,m} \; = \; f_{n,l}(\om) \, Y_{l,m}(\theta,\phi) \nn \\
&& B_N \; = \; -\frac{m \alpha^2}{4n^2},\qquad e_n \; = \; \frac{m\alpha}{2n} 
\leea{bspt15}
and for the binding energy one has the standard 
nonrelativistic expression for positronium bound state to $O(e^2)$.
We write for completeness the coordinates used in the solution
\bea
&& (e_n^2 = -m B_N, \vec{p}) \; \longrightarrow \; (u_0, \vec{u}) \nn \\
&& u_0 \; = \; \cos\om \; = \; \frac{e_n^2 - \vec{p}^2}{e_n^2 + \vec{p}^2} \nn \\
&& \vec{u} \; = \; \frac{\vec{p}}{|\vec{p}|}
\sin\omega \; = \; \frac{2e_n \vec{p}}{e_n^2 + \vec{p}^2} 
\; , \leea{bspt16}
but, for details, refer to \cite{JoPeGl}.

Now we define BSPT to solve positronium bound state problem. 
We introduce the potential, arising in the
nonrelativistic Schr\"odinger equation, \eq{bspt13},
\bea
&& V'(\vec{p'}s_3s_4;\vec{p}s_1s_2) =
\lim_{\frac{\vec{p}^2}{m^2}<<1} \frac{\sqrt{J(p)J(p')}}{2(2\pi)^3}\frac{1}{4m}
 V^{eff}\nn\\
&& \approx \frac{1}{2(2\pi)^3}\frac{1}{2m}\frac{1}{4m}
\lim_{\frac{\vec{p}^2}{m^2}<<1} (\tilde{V}_{exch}+\tilde{V}_{ann})
\; . \leea{bspt17}
where \mbox{$\tilde{V}_{LC}^{eff}=\tilde{V}_{exch}+\tilde{V}_{ann}$} is the effective
electron-positron interaction, obtained in the previous chapter.
The leading order solution is given in \eq{bspt15}. 
We perform perturbative bound state calculations 
with respect to the difference
\be
\de V=V'(\vec{p'}s_3s_4;\vec{p}s_1s_2)
-(-\frac{\alpha}{2\pi^2})\frac{1}{(\vec{p}-\vec{p'})^2}
\de_{s_1s_3}\de_{s_2s_4}
\; , \lee{bspt18}
Note, that, in order to define the Coulomb potential, i.e.
the $e\bar{e}$ interaction in the leading order of BSPT,
we take only the first
term of nonrelativistic expansion of the Jacobian $J(p)$. 

Further we use the matrix elements of $\de V$, that are defined as 
\bea
&& <\Phi_{nlm}|\de V|\Phi_{nlm}>=\int d^3pd^3p'
\Phi_{nlm}^*(\vec{p})\de V\Phi_{nlm}(\vec{p'})
\; , \leea{bspt19}
where $\Phi_{nlm}$ are the Coulomb wave functions.

\section{Effective electron-positron interaction
in light-front and instant frames}
\label{sec5.2}

We summarize together all the terms defining the electron-positron
interaction $V^{eff}$, \eq{gi18a} 
\be
 V^{eff} = \tilde{V}_{exch}+\tilde{V}_{ann}=
\sum_{channel}
(\tilde{V}^{gen}+\tilde{V}^{inst})
\lee{re4}
we remind that {\it tilde} above the interaction terms denotes the rescaled
potential ($V=\tilde{V}/P^{+2}$), 
that does not depend on the total momentum $P^+$, i.e.
is invariant under the light-front boosts.

In the {\bf light-front frame}
the generated interaction and instantaneous term
are given resp. 
in the {\bf exchange channel}
\bea
\tilde{V}^{gen} &=& -e^2 N_1\left(
\frac{\tilde{\De}_1+\tilde{\De}_2}{\tilde{\De}_1^2+\tilde{\De}_2^2}\right) \nn\\
\tilde{V}^{inst} &=& -\frac{4 e^2}{(x - x')^2} \: \de_{s_1 s_3} \: 
\de_{s_2 s_4} 
\leea{repi2}
in the {\bf annihilation channel}
\bea
\tilde{V}^{gen} &=& e^2 N_2\left(
\frac{M_0^2+M_0^{'2}}{M_0^4+M_0^{'4}}\right) \nn\\
\tilde{V}^{inst} &=& 4 e^2
 \: \de_{s_1 \bar{s}_3} \: \de_{s_2 \bar{s}_4} 
\; , \leea{repi3}
The functions $N_1, N_2$ (current-current terms) and 
the energy denominators $\tilde{\De}_i, i=1,2,3$, $M_0^2,M_0^{'2},M_n^2$ 
are defined in the light-front dynamics \cite{ZhHa} 
as follows (see Appendix \ref{appA})  fig.($3$)
\bea
N_1&=&\delta_{s_1s_3}\delta_{s_2s_4}
 T_1^{\bot}\cdot T_2^{\bot}
 -\delta_{s_1\bar{s}_2}\delta_{s_1\bar{s}_3}\delta_{s_2\bar{s}_4}
 2m^2\frac{(x-x')^2}{xx'(1-x)(1-x')}\nonumber\\
&&+im\sqrt{2}(x'-x) \left[ \delta_{s_1\bar{s}_3}\delta_{s_2s_4}
 \frac{s_1}{xx'}T_1^{\bot}\cdot \varepsilon_{s_1}^{\bot}
 +\delta_{s_1s_3}\delta_{s_2\bar{s}_4}
 \frac{s_2}{(1-x)(1-x')}T_2^{\bot}\cdot \varepsilon_{s_2}^{\bot} \right]
\nn\\
N_2&=&\delta_{s_1\bar{s}_2}\delta_{s_3\bar{s}_4}
 T_3^{\bot}\cdot T_4^{\bot}
 +\delta_{s_1s_2}\delta_{s_3s_4}\delta_{s_1s_3}
 2m^2\frac{1}{xx'(1-x)(1-x')}\nonumber\\
&&+im\sqrt{2} \left[ \delta_{s_3\bar{s}_4}\delta_{s_1s_2}
 \frac{s_1}{x(1-x)}T_3^{\bot}\cdot \varepsilon_{s_1}^{\bot}
 -\delta_{s_3s_4}\delta_{s_1\bar{s}_2}
 \frac{s_3}{x'(1-x')}T_4^{\bot}\cdot \varepsilon_{s_4}^{\bot *} \right] \nn \\
&& \varepsilon_s^i = -\frac{1}{\sqrt{2}}(s, i)
\leea{re7}
and
\bea
T_1^i&=&- \left[ 2\frac{(\kappa_{\bot}-\kappa'_{\bot})^i}{(x-x')}
+\frac{\kappa_{\bot}^i(s_2)}{(1-x)}+
 \frac{\kappa_{\bot}^{'i}(\bar{s}_2)}{(1-x')} \right] \; ; \qquad
T_2^i=2\frac{(\kappa_{\bot}-\kappa'_{\bot})^i}{(x-x')}
-\frac{\kappa_{\bot}^i(s_1)}{x}-
 \frac{\kappa_{\bot}^{'i}(\bar{s}_1)}{x'} \nonumber \\
T_3^i&=&-\frac{\kappa_{\bot}^{'i}(\bar{s}_3)}{x'}
 +\frac{\kappa_{\bot}^{'i}(s_3)}{(1-x')} \; ; \qquad
T_4^i=\frac{\kappa_{\bot}^i(\bar{s}_1)}{(1-x)}
 -\frac{\kappa_{\bot}^i(s_1)}{x}\nonumber\\
&& \kappa_{\bot}^i(s) = \kappa_{\bot}^i+is\varepsilon_{ij}\kappa_{\bot}^j \; ;
 \qquad
 \varepsilon_{ij}=\varepsilon_{ij3} \; ; \qquad
 \bar{s} = -s \nn 
\leea{re8}
with the definitions
\bea
\tilde{\De}_1 &=& \frac{(x\kappa'_{\bot}-x'\kappa_{\bot})^2+m^2(x-x')^2}
{xx'}\; ; \qquad
 \tilde{\De}_2=\tilde{\De}_1|_{x\rightarrow(1-x),x'\rightarrow(1-x')} \nn \\
&&\De_1=\frac{\tilde{\De}_1}{x'-x} \; ; \qquad
 \De_2=\frac{\tilde{\De}_2}{x'-x} \nn \\
&&M_0^2=\frac{\kappa_{\bot}^2+m^2}{x(1-x)} \; ; \qquad
 M_0^{'2}=\frac{\kappa_{\bot}^{'2}+m^2}{x'(1-x')} \nn \\
&&P^-=\frac{(P^{\bot})^2+M_N^2}{P^+} \; ; \qquad P=(P^+,P^{\bot}) \; ; \qquad
 M_N = 2m + B_N
\; . \leea{re9}
Here $x$ is the light front fraction of the electron momentum,
$P$ is the total momentum of positronium and $B_N$ the binding energy of the
positronium.
The effective interaction \eq{re4}, generated by the flow equations,
is defined in the whole parameter region, (except maybe for 
the Coulomb singularity point $\vec{q}=\vec{p}-\vec{p'}=0$, where 
we are not able to eliminate the electron-photon vertex) 
as follows 
\bea
 V^{eff} &=& \tilde{V}_{exch}+\tilde{V}_{ann} \nn \\
&=&-e^2N_1 
\left(\frac{\tilde{\De}_1+\tilde{\De}_2}
{\tilde{\De}_1^2+\tilde{\De}_2^2}\right) 
+\left( -\frac{4 e^2}{(x - x')^2} \: \de_{s_1 s_3} \: 
 \de_{s_2 s_4} \right)\nn\\
&&+e^2N_2 
\left(\frac{M_0^2+M_0^{'2}}{M_0^4+M_0^{'4}}\right)
+\left( 4 e^2
 \: \de_{s_1 \bar{s}_3} \: \de_{s_2 \bar{s}_4} \right)
\; , \leea{re18}
To get the effective interaction in the {\bf instant frame}
the substitutions $x(p_z), x'(p_z^{'})$ \eq{bspt1}
are to be done, also in the instant frame holds
\bea
&& M_0^2=4(\vec{p}^2+m^2) \; ;\qquad
M_0^{'2}=4(\vec{p'}^{2}+m^2)
\; ; \leea{repi12}
At the end of this section we illustrate, that the effective
interaction \eq{re18} gives in the leading order of nonrelativistic
expansion $|\vec{p}|/m \ll 1$ the Coulomb interaction.
In the leading order 
one has (for the exchange channel, that gives the dominant contribution)
\bea
&& \tilde{\De}_1\sim\tilde{\De}_2=\tilde{\De}=
(\vec{p}-\vec{p'})^2\nn\\
&& V^{gen}\approx -e^2\frac{N_1}{(\vec{p}-\vec{p'})^2}\nn\\
&& \De=\frac{(\vec{p}-\vec{p'})^2}{x'-x}
\leea{re14}
and the electron-positron interaction is 
\be
V^{|e\bar{e}>}\approx -e^2\frac{N_1}{(\vec{p}-\vec{p'})^2}
- \frac{4e^2}{(x-x')^2}\de_{s_1s_3}\de_{s_2s_4}
\lee{re15}
Using the following expressions
\bea
&& N_1^{diag}\approx -4\frac{(\vec{\kappa}_{\perp}-\vec{\kappa}_{\perp}^{'})^2}
{(x-x')^2}\de_{s_1s_3}\de_{s_2s_4}\nn\\
&& (\vec{p}-\vec{p'})^2=(\vec{\kappa}_{\perp}-\vec{\kappa}_{\perp}^{'})^2+
(p_z-p_z^{'})^2\approx (\vec{\kappa}_{\perp}-\vec{\kappa}_{\perp}^{'})^2+
4m^2(x-x')^2
\leea{re16}
one obtains in leading order of the nonrelativistic approximation
the $3$-dimensional Coulomb interaction $(e^2=4\pi\al)$
\be
V^{|e\bar{e}>}\approx 16m^2\left( 
 - \frac{e^2}{(\vec{p}-\vec{p'})^2}\right)
\de_{s_1s_3}\de_{s_2s_4}
\lee{re17}
Hence the rotational invariance is restored in this order.
This result \eq{re17} 
does not depend on the details of the similarity function $f_{\la}(\De)$.

\section{Positronium's ground state spin splitting}
\label{sec5.3}

In this section we use the effective electron-positron interaction
\eq{re18} to calculate the ground state singlet-triplet
splitting for positronium. We follow the work \cite{JoPeGl},
where the similarity renormalization scheme was used
to get an effective electron-positron interaction.

There is an important difference between the two approaches,
flow equations and similarity transformation.
Flow equations eliminate the matrix elements between states
with large energy differences ($|E_i-E_j|>\la$, where $\la$ is
an effective UV-cutoff), but only for those blocks that
change the number of quasiparticles. The value of
$\la=\La\rightarrow\infty$ corresponds to the initial bare Hamiltonian,
a finite $\la$ determines the effective Hamiltonian at an intermediate
stage, for $\la=0$ the elimination of the particle number changing
sectors is complete and the effective Hamiltonian is 
block-diagonal in particle number. One can work then
in a few (or even one) lowest Fock sectors.

The effective Hamiltonian, obtained by the similarity transformation,
is band-diagonal in the 'energy space'.
\footnote{Under the 'energy space' we understand the basis of the free
Hamiltonian $H_0$.}
The width of the band $\la$ (namely $\la^2/P^+$)
introduces the artificial parameter in the procedure,
which must be adjusted from the physical reasoning.
The effective UV-cutoff must be low enough to neglect
the contribution of high Fock states, but is restricted from below
to stay in perturbation theory region.
In the case of positronium the ``window of opportunity''
is simple to choose, since there are two dynamical energy scales
in QED, $\frac{m^2\alpha^2}{P^+}$ and $\frac{m^2\alpha}{P^+}$,
that is one of the reasons why QED calculations
have been always so successful. Namely one chooses
$m^2\alpha^2<<\la^2<<m^2\alpha$, and the effective 
electron-positron interaction does not depend on $\la$ 
and is defined on the energy scale of positronium bound state 
formation \cite{JoPeGl}. In the case of QCD, unfortunately,
there is no such 'window', and the procedure of fitting
the cutoff seems to be nontrivial \cite{osu}.
In the case of flow equations elimination of particle number changing
sectors can be performed completely, so that there is
no artificial parameter (as effective UV-cutoff) 
left in the effective Hamiltonian.

We use the potential \eq{bspt17}, that appears in the nonrelativistic
Schr\"odinger equation for positronium \eq{bspt13}, and defines
the nonrelativistic binding energy $B_N$ 
$(M_N^2\approx 4m^2+4mB_N)$. 
The effective electron-positron interaction \eq{re18}
has the following form in the nonrelativistic limit
\bea
\hspace{-6em} V' &=& \frac{1}{2(2\pi)^3} \frac{1}{4m} 
\frac{1}{2m} \,
 \biggl(1 - \frac{\vec{p}^2}{2m^2}\biggr) \, V^{eff} \nn \\
\hspace{-6em} V^{eff} &=& \tilde{V}_{exch} + 
\tilde{V}_{ann} \nn \\
&=& - \frac{e^2N_1}{(\vec{p} - \vec{p'})^2}
    - \: \frac{4 e^2}{(x - x')^2} \: \de_{s_1 s_3} \: 
      \de_{s_2 s_4} \nn\\  
 && + \: \frac{e^2N_2}{4m^2}
    + \: 4 e^2
      \: \de_{s_1 \bar{s}_3} \, \de_{s_2 \bar{s}_4}
\; , \leea{nra1}
where the energy denominators were simplified as 
\bea
&& x - x' = \frac{p_z-p_z^{'}}{2m} \left[ 1 + \frac{\vec{p}^2}{2m^2} \right]
 + O \left( m^2 \left(\frac{p}{m}\right)^5 \right) \nn \\
&& \tilde{\De}_1 = \tilde{\De}_2 = (\vec{p} -\vec{p'})^2
 + O \left(m^2 \left(\frac{p}{m}\right)^5 \right) \nn \\
&& \De_1 = \De_2 = \frac{2m(\vec{p'} - \vec{p})^2}{(p_z^{'}-p_z)}
 \left[ 1 + O \left( \left(\frac{p}{m}\right)^2 \right) \right] \; ;\qquad
 \De = \frac{2m (\vec{p'} - \vec{p})^2}{(p_z' - p_z)} \nn \\
&& M_0^2 = M_0^{'2} = 4m^2 + O \left(m^2 \left(\frac{p}{m} \right)^2 \right) 
\; , \leea{nra2}
The expression for Jacobian of the coordinate change is given
\be
\sqrt{J(p)J(p')} = \frac{1}{2m} \left[ 1 - \frac{\vec{p}^{\,2}}{2m^2}
 + O \left( \frac{p_z^2}{m^2}, \frac{p_z^{'2}}{m^2} \right) \right]
\; , \lee{nra3}
Combining all together, we have
\bea
\hspace{-1em} V'(\vec{p},\vec{p'})
 &=& \frac{1}{2(2\pi)^3} \frac{1}{4m}\frac{1}{2m} \,
 \biggl( 1 - \frac{\vec{p}^2}{2m^2} \biggr) \\
&&\hspace{1em} \times \biggl[
 - \frac{e^2 N_1}{(\vec{p} - \vec{p'})^2} \;  
 - \frac{16 e^2 m^2}{(p_z - p'_z)^2} \left( 1 + \frac{\vec{p}^2}{m^2} \right) \;
  \de_{s_1 s_3} \: \de_{s_2 s_4} \nn \\
&&\hspace{3.8em} + \; \frac{e^2 N_2}{4m^2} \; 
 \hspace{8.2em} + \; 4e^2 \; \de_{s_1 \bar{s}_2} \: \de_{s_3 \bar{s}_4}
 \biggr] \nn
\; . \leea{nra6}
We expand the factors $N_1$ and $N_2$,
appearing in the interaction, in the nonrelativistic limit.

\noindent
The term $N_1$ contributes in $V'$ in the order

\noindent
\mbox{$\underline{O(1),O\left( \left(\frac{p}{m} \right)^2 \right) }$}:
\bea
-T_1^{\bot}T_2^{\bot} &=& 16m^2 \frac{q_{\bot}^2}{q_z^2}
 \left( 1 + \frac{\vec{p}^2}{m^2} \right) + 16 \frac{q_{\bot}^i}{q_z}
 \left( \kappa_{\bot}^ip_z + \kappa_{\bot}^{'i} p_z^{'} \right) \nn \\
&& -16i(s_1+s_2)[\kappa_{\bot}^{'}\kappa_{\bot}]
-4(\kappa_{\bot}+\kappa_{\bot}^{'})^2
+4s_1s_2q_{\bot}^2\nn
\; , \leea{nra7}
 
\noindent
\mbox{$\underline{O\left(\frac{p}{m}\right),
 O\left( \left(\frac{p}{m} \right)^2 \right) }$} :
\bea
&&\hspace{-1em} im\sqrt{2}(x'-x)
 \left(\frac{s_1}{xx'} \: \varep^\perp_{s_1} \cdot T^\perp_1 \;
  \de_{\bar{s}_1 s_3} \: \de_{s_2 s_4} \;
  + \frac{s_2}{(1-x)(1-x')} \: \varep^\perp_{s_2} \cdot T^\perp_2 \;
  \de_{\bar{s}_4 s_2} \: \de_{s_1 s_3} \right) \nn \\
&&\hspace{0em} = 8 \, \de_{\bar{s}_1 s_3}  \: \de_{s_2 s_4} \,
 \left[m \, (iq_{\bot}^x - s_1q_{\bot}^y) \left( 1 - \frac{p_z+p_z^{'}}{m} 
\right) \,
  + q_z \, (i\tilde{p}_{\bot}^{x}-s_1\tilde{p}_{\bot}^{y})
  + \frac{1}{2} s_2 q_z(q_{\bot}^y - is_1q_{\bot}^x) \right] \nn \\
&&\hspace{1em} - \; \de_{\bar{s}_4 s_2} \: \de_{s_1 s_3}
 \left[m \, (iq_{\bot}^x - s_2q_{\bot}^y) \, \left( 1 + \frac{p_z + p_z'}{m} 
\right) \,
  - q_z \, (i\tilde{p}_{\bot}^{x}-s_2\tilde{p}_{\bot}^{y})
  - \frac{1}{2} s_1 q_z(q_{\bot}^y - is_2q_{\bot}^x) \right] \nn
\; , \leea{nra8}

\noindent
\mbox{\underline{$O\left( \left(\frac{p}{m} \right)^2 \right)$}} :
\bea
&& 2m^2\frac{(x-x')^2}{xx'(1-x)(1-x')} = 8 q_z^2 \nn
\; . \leea{nra9}

\noindent
The term $N_2$ contributes to $V'$ in the order
\\
\noindent
\mbox{\underline{$O\left( \left(\frac{p}{m} \right)^2 \right)$}} :
\be
2m^2\frac{1}{xx'(1-x)(1-x')} = 32m^2 \nn
\; . \lee{nra10}

In these formulas we have used
\mbox{$[\kappa_{\bot}^{'},\kappa_{\bot}] = 
\varepsilon_{ij}\kappa_{\bot}^{'i}\kappa_{\bot}^{j}$},
$\varepsilon_{ij}=\varepsilon_{ij3}$ and
\mbox{$\varepsilon_{s}^{i} = -\frac{1}{\sqrt{2}}(s,i)$}; also
the following variables have been introduced
\bea
&& q_{\bot} = \kappa_{\bot}^{'}-\kappa_{\bot} \; ,\qquad (\bot=x,y)~,~
   q_z = p_z^{'}-p_z \nn \\
&& \tilde{p}_{\bot} = \frac{\kappa_{\bot} + \kappa'_{\bot}}{2}
\; . \leea{nra11}
We do not analyse here the
expressions for $N_1$ and $N_2$, where also in this form some terms
can be identified as spin-orbit and spin-spin interactions
in the transverse plane and in longitudinal (z) direction.

Instead we follow \cite{JoPeGl}, where an analogous calculation
of singlet-triplet ground state mass splitting of positronium was performed 
in the similarity renormalization scheme. 
We therefore can drop in $N_1$, except for the leading
order term $O(1)$, the diagonal part in spin space. Also the
terms of the type 
$f=\kappa_{\bot}^{x,y}p_z~,~\kappa_{\bot}^{x,y}p_z^{'}~,~
\kappa_{\bot}^x \kappa_{\bot}^y$
do not contribute to the ground state mass splitting, since
\bea
&& \int d^3pd^3p' \, \Phi_{100}^*(\vec{p}) \, 
\frac{f}{\vec{q}^2} \, \Phi_{100}(\vec{p'})
\; , \leea{nra12}
average over directions, gives zero.

In the {\bf leading order $O(1)$}
of nonrelativistic expansion we obtain the following $e\bar{e}$-potential
\bea
V^{'(0)}(\vec{p'}, \vec{p})
 &=& \frac{1}{2(2\pi)^3} \frac{1}{4m}\frac{1}{2m} \,
 \left( 1 - \frac{\vec{p}^2}{2m^2} \right) \nn \\
&& \times \left[ \frac{16e^2m^2}{\vec{q}^2} \frac{q_{\bot}^2}{q_z^2}
 \left( 1 + \frac{\vec{p}^2}{m^2} \right) \; 
 - \frac{16e^2m^2}{q_z^2} \, \left( 1 + \frac{\vec{p}^2}{m^2} \right) \; \right]
 \de_{s_1s_3} \: \de_{s_2s_4} \nn \\
&=& - \frac{\alpha}{2\pi^2} \frac{1}{\vec{q}^2} 
 \left( 1 + \frac{\vec{p}^2}{2m^2} \right) \,
 \de_{s_1s_3} \: \de_{s_2s_4} \\
&&\hspace{-2em}\longrightarrow \quad V(r) \, 
 \left( 1 + \frac{\vec{p}^2}{2m^2} \right) \nn   
\; . \leea{nra13}
where \mbox{$\vec{q} = \vec{p'}-\vec{p}$}, and we have done in the last expression 
the Fourier transformation with respect to $\vec{q}$
to the coordinate space. 
In the leading order of nonrelativistic expansion
we have reproduced the Coulomb potential, defined before as 
the leading order term in BSPT. 

We combine this expression with the kinetic term
from the Schr\"odinger equation, \eq{bspt13}, and write it in the form
\be
\frac{1}{m} \left( 1 + \frac{V(r)}{2m} \right) \vec{p}^{\,2} + V(r)
\; . \lee{nra14}
Here the potential $V(r)$ plays a different role in the two terms.
In the first term, corresponding to kinetic energy, it generates
an effective mass of the electron, which depends on the relative
position and manifests the non-locality of the interaction.
The second term is the usual potential energy, in our case, the Coulomb
interaction.

The energy of the Coulomb level for positronium with quantum numbers $(nlm)$ is 
the standard one
\be
B^{(0)} = <\Phi_{nlm}|V^{'(0)}|\Phi_{nlm}>
= \int d^3p d^3p'\Phi^{*}_{nlm}(\vec{p}) \, 
\left(- \frac{\alpha}{2\pi^2}\frac{1}{(\vec{p}-\vec{p}')^2}\right) 
\, \Phi_{nlm}(\vec{p'})
= -\frac{m\alpha^2}{4n^2}
\; , \lee{nra15}
where the Coulomb wave function $\Phi_{nlm}$ is defined in \eq{bspt14},\eq{bspt15}. 
We have used in \eq{nra15} the following representation
\bea
&& (\vec{p} - \vec{p'})^2 = \frac{(e_n^2 + \vec{p}^2) \,
(e_n^2 + \vec{p'}^2)}{4e_n^2} \, (u-u')^2 \nn \\
&& \frac{1}{(u-u')^2} = \sum_{\mu} \frac{2\pi^2}{n} \,
Y_{\mu}(\Omega_{p}) \, Y_{\mu}^{*}(\Omega_{p'}) \nn \\
&& d^3p = d\Omega_p \left( \frac{e_n^2 + \vec{p}^2}{2e_n} \right)^3
\leea{nra16}
and also orthogonality of the hyperspherical harmonics
\be
\int d\Omega \, Y_{\mu}^{*} \, Y_{\mu^{'}} = \de_{\mu\mu'}
\; . \lee{nra17}
More details can be found in \cite{JoPeGl}.

Now the result of the first and the second order bound state perturbation
theory for positronium ground state $(n=1)$ is presented.
These are the corrections to the leading order binding energy 
$B^{(0)}_n$ \eq{nra15}, defined by the nonrelativistic Schr\"odinger equation
for positronium \eq{bspt13}. 

{\bf The next to leading order \mbox{$O\left(\frac{p}{m}\right)$}}
\bea
\de V^{(1)} &=& \frac{1}{2(2\pi)^3} \frac{1}{4m} \frac{1}{2m}
 \left(-\frac{e^2}{\vec{q}^{\,2}}\right) \nn \\
&& \times \left(
  8m(iq_{\bot}^x-s_1q_{\bot}^y) \de_{s_1\bar{s}_3} \de_{s_2s_4}
 -8m(iq_{\bot}^x-s_2q_{\bot}^y) \de_{s_1s_3} \de_{s_2\bar{s}_4} \right)
\leea{nra18}
contributes (because of the spin structure) to the second order of BSPT:
\be
\de^{(2)} B = \sum_{\mu\neq(1,0,0),s_i}
 \frac{< \Phi_{100} |\de V^{(1)}| \Phi_{\mu,s_i} > \, 
< \Phi_{\mu,s_i} |\de V^{(1)}| \Phi_{100} >}
      {B^{(0)}_1-B^{(0)}_n}
\; . \lee{nra19}
Recall, that $\mu=(n,l,m)$, the usual principal and angular momentum
quantum numbers of nonrelativistic positronium.
{\bf The order $O\left( \left(\frac{p}{m} \right)^2\right)$} 
(cf. remark after \eq{nra11}) is 
\be
\de V^{(2)} = \frac{1}{2(2\pi)^3} \frac{1}{4m} \frac{1}{2m} \,
 \left( 8e^2 \frac{q_z^2}{\vec{q}^{\,2}} \de_{s_1\bar{s}_2} \de_{s_1\bar{s}_3}
 \de_{s_2\bar{s}_4} + 8e^2 \de_{s_1s_2} \de_{s_3s_4} \de_{s_1s_3}
 +4e^2 \de_{s_1\bar{s}_3} \de_{s_2\bar{s}_4} \right)
\lee{nra20}
and contributes to the first order of BSPT:
\be
\de^{(1)} B = < \Phi_{100} |\de V^{(2)}| \Phi_{100} >
\; . \lee{nra21}
Both contributions were calculated in \cite{JoPeGl} with the result
\bea
\de B &=& \de^{(1)} B + \de^{(2)} B \nn \\
<1|\de B|1> &=& -\frac{5}{12} m \alpha^4 \nn \\
<2|\de B|2> &=& <3|\de B|3>= <4|\de B|4> = \frac{1}{6} m \alpha^4
\; , \leea{nra22}
where the eigenvectors in spin space are defined as follows:
\bea
&& |1> = \frac{1}{\sqrt{2}} \, (|\!\!+-\!\!> - |\!\!-+\!\!>) \; , \nn \\
&& |2> = \frac{1}{\sqrt{2}} \, (|\!\!+-\!\!> + |\!\!-+\!\!>) \; ,\qquad
   |3> = |\!\!--\!\!> \; ,\qquad
   |4> = |\!\!++\!\!>
\; . \leea{nra23}
Using the relation between Coulomb energy units
and ${\rm Ryd}=\frac{1}{2}m\alpha^2$, we get the standard result
for the singlet-triplet mass splitting of positronium
\bea
B_{triplet}-B_{singlet}=\frac{7}{6}\alpha^2 {\rm Ryd}+O(m\alpha^5)
\leea{}
The degeneracy of the triplet ground state
$n=1$ signals the rotational invariance, that is not-manifest symmetry
on the light-front.

Brisudova and Perry \cite{BrPerotation} tried to bring the effective light-front
Hamiltonian of positronium, obtained in the second order
in coupling by similarity transformation,
to the rotational invariant form. They succeeded to get the correct
spin-spin interactions, namely they obtained the familiar
Breit-Fermi spin-spin and tensor terms. They failed to reproduce
the standard (equal-time) spin-orbit interaction using the effective
Hamiltonian in the order $O(e^2)$. 
The reason can be the following.

Each spin enters the electron-positron interaction with a factor
of order $q/m$ as compared to the leading Coulomb interaction,
that is of order $q^{-2}$. Thus the two-spin interaction
enters only in the order $q^{0}$ or higher.
The only contribution to fine structure splitting,
$\alpha^4$, comes from order $e^2$ (from the terms $e^2q^0$).
The same holds for the spin-triplet splitting
(which is quadratic in the spin), discussed above.
The contributions to the spin-orbit coupling are of the order
$q^{-1}$. In the order $\alpha^4$ also term $e^4q^{-2}$
is important, its contribution to the mass spectrum in order
$\alpha^4$ must be considered together with the contribution
of $e^2q^{-1}$ term. To get the correct spin-orbit splittings
one has to derive the effective interaction to the order
$e^4$. 

Spin independent interaction is of order $q^{-2}$
and contain also subleading terms in $q$ ($q^{-1},q^{0}$).
The spin-independent term $e^2q^0$ contribute to the fine structure
splitting, in the order $\alpha^4$. In this order also terms 
of order $e^4q^{-1}$ and $e^6q^{-2}$ are important.
Thus to obtain all contributions of order $\alpha^4$
one has to consider the effective interaction in order
$e^2$, $e^4$ and $e^6$.

\chapter{Positronium spectrum (numerically)}
\label{ch6}
This part of the work is done in collaboration with
Dr. U.Trittmann.
In this chapter we suggest the form of the effective electron-positron
interaction, that preserves the rotational symmetry
(at least on the level of the mass spectrum).
Using this interaction the light-front integral equation
for positronium bound states is solved numerically\footnote{
All numerical calculations were performed
by Dr.U.Trittmann. The calculations are preliminary.
}.

\section{Light-front bound state equation}
\label{sec6.1}

The effective light-front eigenvalue equation for positronium
bound states reads (see subsection \ref{subsec4.3.2})
\be
\tilde{H}^{eff}|(e\bar{e})_n>=M_n^2|(e\bar{e})_n>.
\lee{re1}
where $n$ labels all quantum numbers, and
the effective light-front Hamiltonian consists of
the free (noninteracting) part and the effective 
electron-positron interaction
\be
\tilde{H}^{eff}=H^{(0)}+V^{eff}
\lee{re2}
Note, that the rescaled value
(that does not depend on $P^+$) of the effective electron-positron
interaction, obtained in the chapter \ref{ch4}, stands in \eq{re1},\eq{re2}.
The integral light-front equation in the momentum space, corresponding 
to \eq{re1}, is given
\bea
&& \hspace{-2cm} \left(M_n^2-\frac{m^2+\vec{k}_{\perp}^{2}}{x(1-x)}\right)
\psi_n(x,\vec{k}_{\perp};\la_1,\la_2)\nn\\
&& \hspace{1.5cm} =\sum_{\la^{'}_1,\la^{'}_2}\int_D 
\frac{dx' d^2 \vec{k}_{\perp}^{'}}{2(2\pi)^3}
<x,\vec{k}_{\perp};\la_1,\la_2|
V^{eff}|x',\vec{k}_{\perp}^{'};\la_1^{'},\la_2^{'}>
\psi_n(x',\vec{k}_{\perp}^{'};\la_1^{'},\la_2^{'})
\leea{r4}
where the wave function is normalized
\bea
&& \sum_{\la_1,\la_2} 
\frac{dx d^2 \vec{k}_{\perp}}{2(2\pi)^3}
\psi_n^{*}(x,\vec{k}_{\perp};\la_1,\la_2)
\psi_{n'}(x,\vec{k}_{\perp};\la_1,\la_2)=\de_{nn'}
\leea{r5}
For practical purposes we have chosen Jacobi momenta
as depicted on fig.($4$).
The integration domain $D$ is restricted by the covariant cutoff
condition \cite{BrLe}
\bea
&& \frac{m^2+\vec{k}_{\perp}^{2}}{x(1-x)}\leq \La^2+4m^2
\leea{r6}
which allows for states to have the kinetic energy below the bare 
cutoff $\La$.

In order to introduce the spectroscopic notation for positronium
mass spectrum we integrate out the angular degree of freedom
$\varphi$, where $\vec{k}_{\perp}=k_{\perp}(\cos\varphi,\sin\varphi)$, 
by substituting it with the discrete quantum number
$J_z=n$, $n\in {\bf Z}$
(actually for the annihilation channel only $|J_z|\leq 1$ is possible)
\bea
&& \hspace{-2cm} <x,k_{\perp};J_z,\la_1,\la_2|\tilde{V}^{eff}|
x',k'_{\perp};J'_z,\la'_1,\la'_2>\nn\\
&=& \frac{1}{2\pi}\int_0^{2\pi}d\varphi{\rm e}^{-iL_z\varphi}
\int_0^{2\pi}d\varphi'{\rm e}^{iL'_z\varphi'}(-\frac{1}{2(2\pi)^3})
<x,k_{\perp},\varphi;\la_1,\la_2|V^{eff}|
x',k'_{\perp},\varphi';\la'_1,\la'_2>\nn\\
&&
\leea{r43}
where $L_z=J_z-S_z$; $S_z=\frac{\la_1}{2}+\frac{\la_2}{2}$ and the states
can be classified (strictly speaking only for rotationally invariant
systems, that is the case in the nonrelativistic limit considered in
the chapter \ref{ch5}) 
according to their quantum numbers of total angular momentum $J$,
orbit angular momentum $L$, and total spin $S$. 
It should be noted
that the definition of angular momentum operators in light-front
dynamics is problematic because they include the interaction
(see introduction in the chapter \ref{chapter3}).

We proceed now to solve for the positronium spectrum 
in all sectors of $J_z$. We formulate therefore 
the light-front integral equation \eq{r4} in the form
where the integral kernel is given by the effective interaction
for the total momentum $J_z$ \eq{r43}. After the change 
of variables \eq{bspt1}
$(\vec{k}_{\perp};x)=(k_{\perp},\varphi;x)\rightarrow
\vec{p}=(\vec{k}_{\perp},k_z)=
(\mu\sin\theta\cos\varphi,\mu\sin\theta\sin\varphi,\mu\cos\theta)$
\bea
&& x=\frac{1}{2}\left(1+\frac{\mu\cos\theta}{\sqrt{\mu^2+m^2}}\right)
\leea{r46}
where the Jacobian reads
\bea
&& J=\frac{\mu^2}{2}
\frac{m^2+\mu^2(1-\cos^2\theta)}{(m^2+\mu^2)^{3/2}}\sin\theta
\leea{r47}
one has the following integral equation
\bea
&& \hspace{-2cm} 
(M_n^2-4(m^2+\mu^2))\tilde{\psi}_n(\mu,\cos\theta;J_z,\la_1,\la_2)
+\sum_{J'_z,\la'_1,\la'_2}\int_{D}d\mu'\int_{-1}^{+1}d\cos\theta'
\frac{\mu^{'2}}{2}
\frac{m^2+\mu^{'2}(1-\cos^2\theta')}{(m^2+\mu^{'2})^{3/2}}\nn\\
&& \hspace{0.5cm}
\times <\mu,\cos\theta;J_z,\la_1,\la_2|\tilde{V}^{eff}|
\mu',\cos\theta';J'_z,\la'_1,\la'_2>
\tilde{\psi}_n(\mu',\cos\theta';J'_z,\la'_1,\la'_2)=0\nn\\
&&
\leea{r48}
The integration domain $D$ \eq{r6} is given now by 
$\mu\in [0;\frac{\La}{2}]$.
Neither $L_z$ nor $S_z$ are good quantum numbers; therefore
we set $L_z=J_z-S_z$.

The wave function is normalized
\bea
&& \sum_{J_z,\la_1,\la_2}\int d\mu~d\cos\theta
\tilde{\psi}_n^{*}(\mu,\cos\theta;J_z,\la_1,\la_2)
\tilde{\psi}_{n'}(\mu,\cos\theta;J_z,\la_1,\la_2)
=\de_{nn'}
\leea{r49}
where $n$ labels all quantum numbers.

The integral equation \eq{r48} is used further to calculate
the positronium mass spectrum numerically.

\section{Brodsky-Lepage prescription of the light-front dynamics:
effective electron-positron interaction}
\label{sec6.2}

In this section we write the effective electron-positron interaction,
using the light-front prescription
as formulated by Brodsky and Lepage \cite{BrLe}. The problem in 
the light-front field theory is, that parity and rotational invariance
are not manifest symmetries, leaving the possibility that approximations
or incorrect renormalization 
(in our case not gauge invariant regularization)
might lead to violations of these symmetries for physical
observables. The formulation for light-front field theory of 
Brodsky, Lepage
is for practical use, since it enables to pick up the rotation violating
part in the electron-positron interaction and to cancel its contribution
in the mass spectrum.

In what follows we write the second order solution
of the flow equations in a useful for our calculations form 
(subsection $6.2.1$)
and exploit it below to calculate the effective 
electron-positron interaction,
using Brodsky-Lepage prescription of light-front QED (subsection $6.2.2$).

\subsection{First and second order solutions
of the flow equations}
In this section we express the solutions of the flow equations
in the first and second orders of perturbation theory
through the similarity function. The similarity function
defines the 'rate' how fast the 'rest' sectors are eliminated
with the flow parameter $l$ (or effective UV-cutoff $\la$);
exact definition is given below.

We consider the flow equations, written in the form \eq{f10}
\bea
&& \frac{dH_{ij}}{dl}=[\eta,H_d+H_r]_{ij}-(E_i-E_j)[H_d,H_r]_{ij}
+\frac{du_{ij}}{dl}\frac{H_{ij}}{u_{ij}}\nn\\
&& \eta_{ij}=[H_d,H_r]_{ij}
+\frac{1}{E_i-E_j}\left(-\frac{du_{ij}}{dl}\frac{H_{ij}}{u_{ij}}\right)
\leea{f10a}
where the following conditions on the cutoff function
in 'diagonal' and 'rest' sectors, resp., are imposed
\bea
&& u_{dij}=1\nn\\
&& u_{rij}=u_{ij}
\leea{f11a}
In the perturbative frame we break the Hamiltonian as 
\bea
&& H=H_{0d}+\sum_n(H_{d}^{(n)}+H_{r}^{(n)})
\leea{f15a}
where $H^{(n)}\sim e^n$, $e$ is the bare coupling constant
(here we do not refer to the definite field theory).
To the order of $n$ in coupling constant
the flow equations in both sectors are given
\bea
&& \frac{dH_{dij}^{(n)}}{dl}=\sum_k[\eta^{(k)},H_r^{(n-k)}]_{dij}\nn\\
&& \eta_{dij}^{(n)}=0 \nn\\
&& \frac{dH_{rij}^{(n)}}{dl}=\sum_k[\eta^{(k)},H_d^{(n-k)}
+H_r^{(n-k)}]_{rij}
-(E_i-E_j)\sum_k[H_d^{(k)},H_r^{(n-k)}]_{rij}
+\frac{du_{ij}}{dl}\frac{H_{rij}^{(n)}}{u_{ij}}\nn\\
&& \eta_{rij}^{(n)}=\sum_k[H_d^{(k)},H_r^{(n-k)}]_{rij}
+\frac{1}{E_i-E_j}\left(-\frac{du_{ij}}{dl}\frac{H_{rij}^{(n)}}
{u_{ij}}\right)\nn\\
&& \eta^{(n)}=\eta^{(n)}_d+\eta^{(n)}_r
\leea{f16}
We solve these equations in the leading order for the Fock state
conserving sector.
\bea
&& \frac{dH_{dij}^{(2)}}{dl}=[\eta_r^{(1)},H_r^{(1)}]_{dij}\nn\\
&& \eta_{rij}^{(1)}=-\frac{1}{E_i-E_j}\frac{dH_{rij}^{(1)}}{dl}\nn\\
&& \frac{dH_{rij}^{(1)}}{dl}=\frac{du_{ij}}{dl}\frac{H_{rij}^{(1)}}{u_{ij}}
\leea{f17}
and $H_{rij}^{(1)}=H_{rji}^{(1)}$.
Explicitly one has
\bea
&& \frac{dH_{dij}^{(2)}}{dl}=-\sum_k\left(
\frac{1}{E_i-E_k}\frac{dH_{rik}^{(1)}}{dl}H_{rjk}^{(1)}
+\frac{1}{E_j-E_k}\frac{dH_{rjk}^{(1)}}{dl}H_{rik}^{(1)}
\right)_d\nn\\
&& H_{rij}^{(1)}(l)=H_{rij}^{(1)}(l=0)\frac{f_{ij}(l)}{f_{ij}(l=0)}
\leea{f18}
where we have introduced the function $f_{ij}$ defining the leading order
solution for the 'rest' part. Further we refer to it as similarity function.
Here
\bea
&& f_{ij}(l)=u_{ij}(l)={\rm e}^{-(E_i-E_j)^2l}
\leea{f19}
The similarity function $f_{\la}(\De)$ has the same behavior
(when $\la\rightarrow\infty$~ $f_{\la}(\De)=1$,
and when $\la\rightarrow 0$~ $f_{\la}(\De)=0$)
as the cutoff function $u_{\la}(\De)$.

Making use of the connection $l=1/\la^2$, we get
\bea
&& \frac{dH_{dij}^{(2)}}{d\la}=-\sum_k\left(
\frac{1}{E_i-E_k}\frac{dH_{rik}^{(1)}}{d\la}H_{rjk}^{(1)}
+\frac{1}{E_j-E_k}\frac{dH_{rjk}^{(1)}}{d\la}H_{rik}^{(1)}
\right)_d\nn\\
&& H_{rij}^{(1)}(\la)=H_{rij}^{(1)}(\La\rightarrow\infty)
\frac{f_{ij}(\la)}{f_{ij}(\La\rightarrow\infty)}
\leea{f20}
Neglecting the dependence of the energy $E_i$ on the cutoff, 
one has
\bea
&& H_{dij}^{(2)}(\la)=H_{dij}^{(2)}(\La\rightarrow\infty)
+\sum_k\left(H_{rik}^{(1)}(\La\rightarrow\infty)
H_{rjk}^{(1)}(\La\rightarrow\infty)\right)_d\nn\\
&& \times\left(
\frac{1}{E_i-E_k}\int_{\la}^{\infty}
\frac{df_{ik}(\la')}{d\la'}f_{jk}(\la')d\la'
+\frac{1}{E_j-E_k}\int_{\la}^{\infty}
\frac{df_{jk}(\la')}{d\la'}f_{ik}(\la')d\la'
\right)_d
\leea{f21a}
where $\La$ is the bare cutoff; the sum $\sum_k$ is over
all intermediate states; and the label 'd' denotes 
the 'diagonal' sector.

In the case of other unitary transformations,
\eq{f12},\eq{f13}, the similarity functions $f(\la)$ are given
\bea
&& \frac{dH_{ij}}{d\la}=u_{ij}[\eta,H_d+H_r]_{ij}
+r_{ij}\frac{du_{ij}}{d\la}\frac{H_{ij}}{u_{ij}}\nn\\
&& \eta_{ij}=\frac{r_{ij}}{E_i-E_j}
\left([\eta,H_d+H_r]_{ij}-\frac{du_{ij}}{d\la}
\frac{H_{ij}}{u_{ij}}\right)\nn\\
&& f_{ij}(\la)=u_{ij}(\la){\rm e}^{r_{ij}(\la)}\nn\\
&& u_{ij}+r_{ij}=1
\leea{f12a}
 and 
\bea
&& \frac{dH_{ij}}{d\la}=u_{ij}[\eta,H_d+H_r]_{ij}
+\frac{du_{ij}}{d\la}\frac{H_{ij}}{u_{ij}}\nn\\
&& \eta_{ij}=\frac{1}{E_i-E_j}
\left(r_{ij}[\eta,H_d+H_r]_{ij}
-\frac{du_{ij}}{d\la}\frac{H_{ij}}{u_{ij}}\right)\nn\\
&& f_{ij}(\la)=u_{ij}(\la)
\leea{f13a}
where $u(\la)$ is the cutoff function. One can choose
$u_{ij}=\theta(\la-|\De_{ij}|)$, where
$\De_{ij}=\sum_{k=1}^{n_2}E_{i,k}-\sum_{k=1}^{n_1}E_{j,k}$.

The equations \eq{f21a} are the same for all unitary
transformations, given above 
up to the choice of the similarity function $f(\la)$.
Specifying the function $f(\la)$ 
we get the interaction $H_d^{(2)}$, generated
by different unitary transformations.

\subsection{Effective electron-positron interaction} 
The exchange channel brings the dominant contribution 
to the mass spectrum.
We focus therefore on the electron-positron interaction
in the exchange channel.
We use \eq{f21a} to calculate the generated
electron-positron interaction in the second order in $e$.
In this case the 'diagonal' sector (denoted 'd')
is $|e\bar{e}>$ sector, and
the matrix element is given fig.($4$)
$<p_1,\la_1;p_2,\la_2|...|p'_1,\la'_1;p'_2,\la'_2>$;
$H_r^{(1)}(\La\rightarrow\infty)$ is the electron-photon
coupling term with the bare coupling constant $e$;
the initial value of generated interaction
is given $H_d^{(2)}(\La\rightarrow\infty)=0$.
The generated interaction is given
\bea
&& \hspace{-1cm} 
V_{\la}^{gen}=-e^2<\ga^{\mu}\ga^{\nu}>
\bigg[\frac{\theta_{\varepsilon}(q^+)}{q^+}D_{\mu\nu}(q)
\left(\frac{\Theta_{\la}(D_1,D_2)}{D_1}
+\frac{\Theta_{\la}(D_2,D_1)}{D_2}\right)\nn\\
&& +\frac{\theta_{\varepsilon}(-q^+)}{-q^+}D_{\mu\nu}(-q)
\left(\frac{\Theta_{\la}(-D_1,-D_2)}{-D_1}
+\frac{\Theta_{\la}(-D_2,-D_1)}{-D_2}\right)\bigg]\nn\\
&& \hspace{-1.5cm} 
\Theta_{\la}(D_1,D_2)=\int_{\la}^{\infty}
\frac{df_{\la'}(D_1)}{d\la'}f_{\la'}(D_2)d\la'
\leea{r8}
where we sum ($\sum_k$ in \eq{f21a})
the two terms corresponding to the two
time-ordered diagrams with $q^+ >0$ and $q^+ <0$; $\la$ is the 'running'
cutoff, that defines the continuous step of the unitary transformation;
$f_{\la}(\De)$ is the similarity function, arising 
from the unitary transformation; the function 
$\theta_{\varepsilon}(q^+)$
restricts the longitudinal momentum of intermediate photon.
The similarity function $f_{\la}(\De)$ and the cutoff function
$\theta_{\varepsilon}(q^+)$ are specified below.\footnote
{The similarity function $f_{\La}$ plays the role of transverse UV regulator
(see the chapter \ref{ch7}), 
and the function $\theta_{\varepsilon}$ regulates
the longitudinal IR divergences. On a tree level there is no UV divergences
in four fermion interaction, but the longitudinal IR divergences are present.  
}   
$D_{\mu\nu}(q)=\sum_{\la}\epsilon_{\mu}(\la,q)
\epsilon_{\nu}^{\star}(\la,q)$ is the polarization sum;
the null vector  
$\eta_{\mu}=(0,\eta_{+},0,0)$, $\eta^{\mu}\eta_{\mu}=0$ is given below; 
the energy denominators in the exchange channel 
are given (fig.($4$))
$D_1=p_1^{'-}-p_1^--q^-$ and $D_2=p_2^--p_2^{'-}-q^-$;
$q=p'_1-p_1$ is the exchanged photon momentum, with $q^-=\frac{q_{\perp}^2}{q^+}$.
The notation $<\ga^{\mu}\ga^{\nu}>$ is introduced for the current-current term.
In the exchange channel this matrix element is given fig.($4$)
\bea
&& <\ga^{\mu}\ga^{\nu}>|_{exch}
=\frac{\bar{u}(p_1,\la_1)}{\sqrt{p_1^{+}}}\ga^{\mu}
\frac{u(p_1^{'},\la_1^{'})}{\sqrt{p_1^{' +}}}
\frac{\bar{v}(p_2^{'},\la_2^{'})}{\sqrt{p_2^{' +}}}\ga^{\nu}
\frac{v(p_2,\la_2)}{\sqrt{p_2^{+}}}
P^{+ 2}
\leea{r12}
where $p_i,p_i^{'}$ are light-front three-momenta carried by the constituents,
$\la_i,\la_i^{'}$ are their light-front helicities,
$u(p_1,\la_1),v(p_2,\la_2)$ are their spinors (see below);
index $i=1,2$ refers to electron and positron, respectively;
$P=(P^+,P^{\perp})$ is light-front positronium momentum.

Below we use the light-front conventions formulated by 
Lepage and Brodsky \cite{BrLe} (see also \cite{BrPaPi}).
The polarization sum is given
\bea
&& D_{\mu\nu}(q)=\sum\limits_{\lambda}
       \epsilon_{\mu} (q,\lambda)
       \epsilon^{\star}_{\nu} (q,\lambda)
       = - g_{\mu\nu}
       +{\eta_{\mu} q_{\nu}+\eta_{\nu} q_{\mu}
       \over q^+}
\leea{}
where the metric tensor(s) 
\begin {equation}
       g^{\mu\nu}=\pmatrix{0&2&0&0 \cr 
       2&0&0&0 \cr 0&0&-1&0 \cr 0&0&0&-1 \cr}
       \qquad{\rm and}\qquad
       g_{\mu\nu}=\pmatrix{0&{1\over2}&0&0\cr 
       {1\over2}&0&0&0\cr  0&0&-1&0\cr 0&0&0&-1\cr}
\ .\end {equation}
and the null vector is $\eta^{\mu}=(0,2,\vec{0})$. 
The Dirac spinors are given
\begin{eqnarray}
       u(p, \lambda) = {1\over \sqrt{p^+} } 
       \left(p^+ + \beta m + \vec \alpha_{\!\perp} 
       \vec p_{\!\perp}\right) \times
       \cases{ \chi(\uparrow), &for $\lambda=+1$, \cr
       \chi(\downarrow), &for $\lambda=-1$, \cr} 
\\
       v(p, \lambda) = {1\over \sqrt{p^+} } 
       \left(p^+ - \beta m + \vec \alpha_{\!\perp} 
       \vec p_{\!\perp}\right) \times
       \cases{\chi(\downarrow), &for $\lambda=+1$, \cr
       \chi(\uparrow), &for $\lambda=-1$. \cr}
\end{eqnarray}
Here $\beta=\ga^0$, $\vec{\alpha}=\ga^0\vec{\ga}$; and
the two $\chi$-spinors are
\bea
       \chi(\uparrow) = {1\over\sqrt{2} } \, 
      \left(\begin{array} {r}
      1 \\  0 \\ 1 \\ 0 \end{array}\right)
       \qquad{\rm and}\qquad
       \chi(\downarrow) = {1\over\sqrt{2}} \, 
      \left(\begin{array} {r}
      0 \\  1 \\ 0 \\ -1 \end{array}\right)
\ .\leea{}
These conventions are used in Appendix \ref{appB} to calculate the
matrix elements of the effective interactions
\footnote{
In the chapters \ref{chapter3} and \ref{ch4} we have used the 
prescription of the light-front field theory 
as formulated by Zhang and Harindranath \cite{ZhHa}.
They have used the following conventions:
the polarization sum is
\bea
&& D_{\mu\nu}(q)=\frac{q^{\perp 2}}{q^{+ 2}}\eta_{\mu}\eta_{\nu}
+\frac{1}{q^+}(\eta_{\mu}q^{\perp}_{\nu}+\eta_{\nu}q^{\perp}_{\mu})
-g^{\perp}_{\mu\nu}
\leea{}
with the null vector $\eta_{\mu}=(0,1,\vec{0})$;  
the four-component spinors $u(p,\la)$, $v(p,\la)$ are given
through the two-component spinors $\chi_{\la}$ in
\eq{footnote3} and \eq{footnote2}, chapter \ref{chapter3}.

Using the above equations, we get
\bea
&& <\ga^{\mu}\ga^{\nu}>|_{exch} D_{\mu\nu}(q)=M_{2ii}^{ex}
\leea{}
where 
$M_{2ij}^{(ex)}=[\chi_{\la_1}^+
\Gamma^i(p'_1,p_1,p'_1-p_1)\chi_{\la'_1}]\,
[\chi_{\bar{\la}^{'}_2}^+\Gamma^j(-p_2,-p'_2,-(p'_1-p_1))
\chi_{\bar{\la}_2}]$
is the matrix element, obtained
in the chapter \ref{ch4} \eq{gi11} direct from the two-component 
field theory \eq{ch2}-\eq{ch6a}.
We reproduce then the form of generated interaction,
obtained in the chapter \ref{ch4} \eq{gi13a}.
}.

Using the symmetry
\bea
&& f_{\la}(-D)=f_{\la}(D)\nn\\
&& D_{\mu\nu}(-q)=D_{\mu\nu}(q)
\leea{r9}
we have the following generated interaction 
in the electron-positron sector 
\bea
&& \hspace{-2cm} V_{\la}^{gen}=-e^2<\ga^{\mu}\ga^{\nu}>
\frac{\theta_{\varepsilon}(q^+)+\theta_{\varepsilon}(-q^+)}{q^+}
D_{\mu\nu}(q)\left(
\frac{\Theta_{\la}(D_1,D_2)}{D_1}
+\frac{\Theta_{\la}(D_2,D_1)}{D_2}
\right)\nn\\
&&
\leea{r10}
We specify the cutoff function $\theta_{\varepsilon}$ in a form \cite{osu1}
\bea
&& \theta_{\varepsilon}=\theta(q^+-\varepsilon)F(q^+,q_{\perp})
\leea{r10a}
where $\varepsilon$ is 'small', say $\varepsilon\sim \frac{m^2}{\La}$
and $\La$ is the bare cutoff. 
For our purposes we do not need the explicit form 
of the function $F(q^+,q_{\perp})$, it defines the upper boundary
for $|q^+|$. The combination $(\theta_{\varepsilon}(q^+)+\theta_{\varepsilon}(-q^+))$
restricts $|q^+|$ to be above $\varepsilon$. 
In the integral in $dq^+$
this ensures the symmetric cutoff for longitudinal photon momentum
\bea
&& \int_{-\infty}^{+\infty}dq^+ \rightarrow \int_{-\infty}^{-\varepsilon}dq^+
+\int_{\varepsilon}^{+\infty}dq^+
\leea{r10b}
The resulting generated interaction is given at the value $\la=0$,
$V^{gen}=V^{gen}_{\la=0}$. To get the effective electron-positron interaction
in the light-front dynamics we sum the resulting generated
interaction and instantaneous photon exchange arising in the light-front QED
\bea
&& V^{eff}=V^{gen}+V^{inst}
\leea{} 
where each term is defined
\bea
&& V^{gen} = -e^2<\ga^{\mu}\ga^{\nu}>
\frac{1}{q^+}D_{\mu\nu}(q)
\left(\frac{\Theta(D_1,D_2)}{D_1}+\frac{\Theta(D_2,D_1)}{D_2}\right)\nn\\
&& V^{inst} = -e^2<\ga^{\mu}\ga^{\nu}>
\frac{1}{q^{+2}}\eta_{\mu}\eta_{\nu}\\
&&  
\Theta(D_1,D_2)=\Theta_{\la=0}(D_1,D_2)=
\int_{0}^{\infty}\frac{df_{\la'}(D_1)}{d\la'}f_{\la'}(D_2)d\la'
\leea{r11}  
For both interactions the prescription \eq{r10a} (\eq{r10b}) to treat
infrared divergences, small $q^+$, is imposed.
We combine all the terms together with the result
\bea
 V^{eff} &=& e^2<\ga^{\mu}\ga^{\nu}>g_{\mu\nu}\frac{1}{q^+}
\left(\frac{\Theta(D_1,D_2)}{D_1}+\frac{\Theta(D_2,D_1)}{D_2}\right)\nn\\
 &-& e^2<\ga^{\mu}\ga^{\nu}>\eta_{\mu}\eta_{\nu}\frac{1}{2q^{+2}}
(D_1-D_2)\left(\frac{\Theta(D_1,D_2)}{D_1}-\frac{\Theta(D_2,D_1)}{D_2}\right)
\leea{r11a}
This equation does not depend on the explicit form
of $\Theta$-factor.
The generated interaction has two types of infrared singularities:
$\frac{1}{q^{+2}}$ and $\frac{1}{q^+}$ types. We see that the $\frac{1}{q^{+2}}$
singularity of generated term is cancelled exactly by the instantaneous term
in the effective electron-positron interaction. Further we show  
that the cutoff condition \eq{r10a} ensures the cancellation of $\frac{1}{q^+}$ type
singularity in the physical observables, calculated from 
the effective electron-positron interaction.

We introduce
\bea
&& 2d=D_1-D_2~;~~D=\frac{D_1+D_2}{2}
\leea{r15}
that gives
\bea
&& D_1=D+d~;~~D_2=D-d
\leea{r16}
As $q^+\rightarrow 0$ we expand the effective interaction \eq{r11a} in the series 
in terms of $\frac{d}{D}<<1$ 
($d\sim (M_0^{'2}-M_0^2)/P^+$; $D\sim (-q^-)$)
\bea
\hspace{-2cm}
V^{eff} &=& e^2<\ga^{\mu}\ga^{\nu}>g_{\mu\nu}\frac{1}{q^+D}\nn\\
&& \times
\left(1-\frac{d}{D}[\Theta(D_1,D_2)-\Theta(D_2,D_1)]+(\frac{d^2}{D^2})
-(\frac{d^3}{D^3})[\Theta(D_1,D_2)-\Theta(D_2,D_1)]\right)\nn\\
\hspace{-2cm}
&+& e^2<\ga^{\mu}\ga^{\nu}>\eta_{\mu}\eta_{\nu}\frac{1}{q^{+2}}\nn\\
&& \times
\left(-\frac{d}{D}[\Theta(D_1,D_2)-\Theta(D_2,D_1)]+(\frac{d^2}{D^2})
-(\frac{d^3}{D^3})[\Theta(D_1,D_2)-\Theta(D_2,D_1)]\right)\nn\\
&+&O\left(m^2\frac{d^4}{D^4}\right)\nn\\
&=& V_0^{eff}+\sum_i\De V_{g_{\mu\nu}}^{(i)}
+\sum_i\De V_{\eta_{\mu}\eta_{\nu}}^{(i)}
\leea{r17}
where we have used the identities
\bea
&& \hspace{-1cm}
\frac{\Theta(D_1,D_2)}{D_1}+\frac{\Theta(D_2,D_1)}{D_2}=
\frac{1}{2}\left(\frac{1}{D_1}+\frac{1}{D_2}\right)
+\frac{1}{2}\left(\frac{1}{D_1}-\frac{1}{D_2}\right)
(\Theta(D_1,D_2)-\Theta(D_2,D_1))\nn\\
&& \hspace{-1cm}
\Theta(D_1,D_2)+\Theta(D_2,D_1)=1
\leea{r18}
index $(i)$ denotes the order with respect of $\frac{d}{D}$.
The leading order term is given
\bea
&& V_0^{eff}=e^2<\ga^{\mu}\ga^{\nu}>g_{\mu\nu}\frac{1}{q^+D}
\leea{r18a}
In the leading order of nonrelativistic approximation $|\vec{p}|/m<<1$
this term \eq{r18a} gives the $3$-dimensional Coulomb interaction 
\cite{GuWe}
\bea
&& V_0^{eff}\approx -\frac{16e^2m^2}{\vec{q}^2}
\leea{}
where $\vec{q}(q_z,\vec{q}_{\perp})=\vec{p'}-\vec{p}$ is the exchanged momentum.
Hence the effective electron-positron
interaction \eq{r11a} produces Bohr energy levels \cite{GuWe}.
The nonrelativistic expansion of the term
\eq{r18a} up to the second order $O\left(\frac{\vec{p}^2}{m^2}\right)$
gives the familiar Breit-Fermi spin-spin and tensor interactions 
\cite{BrPerotation},
that insures the correct spin-splittings for the positronium 
ground state \cite{JoPeGl}.

Corrections $\De V_{g_{\mu\nu}}^{(i)}$ and
$\De V_{\eta_{\mu}\eta_{\nu}}^{(i)}$ arise due to the unitary transformation
performed, i.e. that are the corrections due to the energy denominators
in the $``g_{\mu\nu}''$ term and the $``\eta_{\mu}\eta_{\nu}''$ term.
The first order corrections $O(d/D)$ were estimated 
in \cite{BrPerotation},\cite{GuWe3}, using the explicit form for similarity
function $f_{\la}(D)$. We are interested here in general properties 
(independent on the choice of $f_{\la}(D)$)
of the effective interaction, 
particularly in the origin of infrared divergences.

In \eq{r17}
the product of $''\eta_{\mu}\eta_{\nu}''$ term with the term 
$\sim (\frac{d}{D})$ gives the singularity of $\frac{1}{q^+}$ type
(when $(\Theta(D_1,D_2)-\Theta(D_2,D_1))\sim const$ with respect to $(d/D)$).
To find the corresponding infrared counterterm we integrate
the singular term over 'external legs' (that correspond in QED to physical
particles) with the wave packet function,
$\int dq^+d^2q_{\perp}\phi(p+q)$, where $p\sim p_1,p_2$. 
The integration of $\frac{1}{q^+}$ type term leads to logarithmic 
infrared divergences,
that are cancelled from small positive and negative longitudinal momenta
due to the symmetric cutoff condition \eq{r10b}.
Therefore, there are no infrared counterterms to be introduced
in the second order $O(e^2)$ in the effective interaction. Moreover,
this is true in the order $O(e^2)$ for all tree level diagrams in QED.

The same cancellation of the infrared divergent contribution from 
$''\eta_{\mu}\eta_{\nu}''$ term occurs, due to the symmetric cutoff,
in the spectrum of masses for positronium.

What about the finite corrections from $''\eta_{\mu}\eta_{\nu}''$ term?
The leading order finite correction from $''\eta_{\mu}\eta_{\nu}''$ term
\eq{r17} is $\sim \eta_{\mu}\eta_{\nu}\frac{1}{q^{+2}}(\frac{d}{D})^2$,
which is of the order $e^2q^0$.
The interaction of light-front field
theory $V^{PT}$ \eq{r23} and the effective interaction
$V^{eff}$ \eq{r11a},
generated by the unitary transformation, have both
a leading Coulomb behavior,
but they differ by spin-independent
$''\eta_{\mu}\eta_{\nu}''$ term
(excluding the divergent part) in the order $e^2q^0$,
which contributes to the mass in the order $\alpha^4$.
In the order of fine structure splitting $\alpha^4$
also terms of order $e^4q^{-1}$ and $e^6q^{-2}$ are important.
These terms arise from the next orders transformation of both
electron-photon vertex and also instantaneous term.
Since the Coulomb interaction, that is the only spin-independent
part of electron-positron interaction, arise from $''g_{\mu\nu}''$
term in the effective interaction $V^{eff}$,
we expect that the spin-independent
$''\eta_{\mu}\eta_{\nu}''$ term in the order $e^2q^0$
will be compensated in the mass spectrum by the corresponding
terms of the order $e^4$ and $e^6$.

Compare  effective electron-positron interaction \eq{r11a}, 
generated in the second order in $e$ by flow equations,
with the electron-positron interaction arising from
perturbative photon exchange.
The electron-positron interaction in the light-front 
perturbation theory is given by a sum of dynamical photon exchange
$V^{phot}$ and the instantaneous interaction,
$V^{PT}=V^{phot}+V^{inst}$. The instantaneous term is given
by \eq{r11}, and the $V^{phot}$ term, defined by the diagram with 
one photon exchange, is \cite{BrLe}  
\bea
&& V^{phot} = -e^2<\ga^{\mu}\ga^{\nu}>\left(
\frac{\theta_{\varepsilon}(q^+)}{q^+}D_{\mu\nu}(q)\frac{1}{\tilde{D}_{+}}
+\frac{\theta_{\varepsilon}(-q^+)}{-q^+}D_{\mu\nu}(-q)\frac{1}{\tilde{D}_{-}}
\right)
\leea{r19}
where the energy denominators are given 
$\tilde{D}_{\pm}=\sum_{inc}p^- -\sum_{interm}p^-$, with
the sums over the light-front energies, $p^-$,
of incident (inc) and intermediate (interm) particles;
and $\pm$ denotes two different time orderings of the photon.
For the process depicted on fig.($4$) we have 
$\tilde{D}_{+}=P^--p_1^{-}-p_2^{'-}-q^-$ and
$\tilde{D}_{-}=P^--p_1^{'-}-p_2^{-}+q^-$, where $q=p'_1-p_1$,
$P^-=M_n^2/P^+$~~~ (let $P^{\perp}=0$).
One can rewrite \eq{r19} in the form \cite{GuWe} 
(also see Appendix \ref{appB})
\bea
&& V^{phot} = -e^2<\ga^{\mu}\ga^{\nu}>
\frac{\theta_{\varepsilon}(q^+)+\theta_{\varepsilon}(-q^+)}{q^+}
D_{\mu\nu}(q)\frac{1}{\tilde{D}}
\leea{r19a}
where 
\bea
&& \theta_{\varepsilon}(q^+)\tilde{D}_{+}
=-\theta_{\varepsilon}(-q^+)\tilde{D}_{-}
=\tilde{D}
\leea{}
Combining instantaneous \eq{r11} and one photon exchange \eq{r19a} terms 
together, we have
\bea
&& V^{PT} = e^2<\ga^{\mu}\ga^{\nu}>g_{\mu\nu}
\frac{1}{q^+\tilde{D}}
 - e^2<\ga^{\mu}\ga^{\nu}>\eta_{\mu}\eta_{\nu}\frac{1}{q^{+2}}
\left(1-\frac{D}{\tilde{D}} \right)
\leea{r20}
where the prescription \eq{r10a} (\eq{r10b}) to treat infrared divergences
is imposed.

The $''\eta_{\mu}\eta_{\nu}''$ term carries the infrared divergent
part of the interaction $V^{PT}$. 
We approximate the positronium mass as average over the masses
of initial and final free (not bound) states
\bea
&& M_n^2=\frac{M_0^2+M_0^{'2}}{2}
\leea{r21}
where $p_1^-+p_2^-=M_0^2/P^+$,
$p_1^{'-}+p_2^{'-}=M_0^{'2}/P^+$,
$P^-=M_n^2/P^+$,
and $P(P^+,P_{\perp}=0)$ is the positronium momentum.
In this approximation the energy denominator of perturbative
photon exchange diagram is
\bea
&& \tilde{D}\rightarrow D=\frac{D_1+D_2}{2}
\leea{r22} 
It is obvious, when one notes (see also Appendix \ref{appB})
\bea
 D &=& \frac{D_1+D_2}{2}=\frac{p_1^{'-}-p_1^-+p_2^--p_2^{'-}}{2}-q^-
 =\frac{p_1^{'-}+p_2^{'-}+p_1^-+p_2^-}{2}-p_1^--p_2^{'-}-q^-\nn\\
  &=& -\left(
\frac{p_1^{'-}+p_2^{'-}+p_1^-+p_2^-}{2}-p_1^{'-}-p_2^{-}+q^-
\right)
\leea{r22a}
Given the \eq{r22}, the instantaneous interaction $V^{inst}$ \eq{r11},
which is the source of infrared divergences, is precisely canceled
by the part in $V^{phot}$ corresponding to emission and absorption of 
longitudinal photons. The resulting electron-positron interaction,
obtained in the light-front perturbation theory,
is given in this approximation \eq{r21}
\bea
&&  V^{PT} = e^2<\ga^{\mu}\ga^{\nu}>g_{\mu\nu}
\frac{1}{q^+ D}
\leea{r23}
By the special choice of similarity function 
this answer \eq{r23} can be generated
by the flow equations. When the similarity function is
\bea
&& f_{\la}(D)={\rm e}^{-\frac{DD'}{\la}}\nn\\
&& D' = sign D 
\leea{}
the $\Theta$-factor \eq{r11} is $\Theta_1=\frac{D_1}{D_1+D_2}$, and
$''\eta_{\mu}\eta_{\nu}''$ term is equal to zero in \eq{r11a}.
The effective electron-positron interaction \eq{r11a}
is given by the result of perturbation theory \eq{r23}
\bea
&& V^{eff}|_{f_{\la}(D)}=V^{PT}
\leea{}
where the condition \eq{r21} (\eq{r22}) is imposed.
This is the remarkable result, that shows that specifying the unitary
transformation in a proper way one can get rid of 
$''\eta_{\mu}\eta_{\nu}''$ term in the effective interaction \eq{r11a},
the term that causes the infrared divergences in longitudinal direction. 
In all other cases we need to introduce the symmetric cutoff condition
\eq{r10a} (\eq{r10b}) on $q^+$ to cancel the divergent contribution
in physical observables. 

What is the status of $''\eta_{\mu}\eta_{\nu}''$ term
in the effective electron-positron interaction $V^{eff}$ \eq{r11a}?
The $''\eta_{\mu}\eta_{\nu}''$ term describes instant emission
and absorption of 'longitudinal' photon and is specific
in the light-front gauge computations.
As was discussed above, this term is not desirable
in the electron-positron interaction and can be considered
as a consequence of the unitary transformation performed.
The physical reason for its appearance is the violation
of Lorenz and gauge symmetries by the derivation of the effective,
renormalized Hamiltonian. The first problem in the light-front field
theory is that whenever the generator of a symmetry is dynamical
(contains interactions) it is somewhere between very difficult
and impossible to monitor and maintain that symmetry 
at each step of a calculation - unless of course
one can solve the theory exactly. This concerns 
parity and rotational invariance, that are not manifest symmetries
on the light-front. The second problem is that whenever
we use the Hamiltonian technique, the naive regularization
by introducing the cutoffs breaks the gauge invariance
(and also Lorenz covariance),
and forces the bare Hamiltonian to contain a larger than normal
suite of counterterms to enable a finite limit
as the cutoffs are removed. One way is, that the counterterms 
are then adjusted to reproduce physical observables and to restore
the symmetries broken by cutoffs. The other way is
to find the gauge invariant procedure for regularization
of Hamiltonians. The attempt in the latter direction
was made by Brodsky, Hiller, and McCartor \cite{BrHiMcCa}.
Solving the Yukawa theory they have tried to preserve more symmetries
by using Pauli-Villars procedure for regularization.

In the next section we use the effective electron-positron 
interaction $V^{eff}$ \eq{r11a}, namely its $''g_{\mu\nu}''$ part, 
to calculate mass spectrum and wave functions for positronium.

\section{Mass spectrum and wave functions of positronium}
\label{sec6.3}

We start with the effective electron-positron interaction,
generated in the second order in coupling by flow equations, \eq{r11a}
\bea
  V^{eff} &=& e^2<\ga^{\mu}\ga^{\nu}>g_{\mu\nu}\frac{1}{q^+}\left(
\frac{\Theta(D_1,D_2)}{D_1}
+\frac{\Theta(D_2,D_1)}{D_2}
\right)\nn\\
 &-& e^2<\ga^{\mu}\ga^{\nu}>\eta_{\mu}\eta_{\nu}\frac{1}{2q^{+2}}
(D_1-D_2)\left(
\frac{\Theta(D_1,D_2)}{D_1}
-\frac{\Theta(D_2,D_1)}{D_2}
\right)
\leea{m1}
where we impose the prescription \eq{r10a} (\eq{r10b}) to treat
infrared divergences, small $q^+$.

We use Jacobi momenta, depicted on fig.($4$)
\bea
&& p_1(xP^+,x\vec{P}_{\perp}+\vec{k}_{\perp})\nn\\
&& p_2((1-x)P^+,(1-x)\vec{P}_{\perp}-\vec{k}_{\perp})
\leea{m2}
and corresponding for the momenta $p'_1,p'_2$;
here $x$ is the light-front fraction of electron momentum
and $P(P^+,\vec{P}_{\perp})$ is the total momentum of positronium.
For convenience we introduce
\bea
&& D_1=\frac{\De_1}{P^+}=-\frac{\tilde{\De}_1}{q^+};~~
D_2=\frac{\De_2}{P^+}=-\frac{\tilde{\De}_2}{q^+}
\leea{m3}
and from now on we use the rescaled value of the cutoff
$\la\rightarrow \la^2/P^+$.
The effective electron-positron interaction is written then
\bea
  V^{eff} = &-& e^2<\ga^{\mu}\ga^{\nu}>g_{\mu\nu}
\left(\frac{\Theta(\Delta_1,\Delta_2)}{\tilde{\De}_1}
+\frac{\Theta(\Delta_2,\Delta_1)}{\tilde{\De}_2}\right)\nn\\
 &-& e^2<\ga^{\mu}\ga^{\nu}>\eta_{\mu}\eta_{\nu}
\frac{1}{2q^{+2}}(\tilde{\De}_1-\tilde{\De}_2)
\left(\frac{\Theta(\Delta_1,\Delta_2)}{\tilde{\De}_1}
-\frac{\Theta(\Delta_2,\Delta_1)}{\tilde{\De}_2}\right)\nn\\
\Theta(\Delta_1,\Delta_2) &=& 
\int_{0}^{\infty}\frac{df_{\la'}(\De_1)}{d\la^{'2}}f_{\la'}(\De_2)d\la^{'2}
\leea{m4}

where the energy denominators in \eq{m4} read
\bea
\tilde{\De}_1 &=& \frac{(x\vec{k}'_{\bot}-x'\vec{k}_{\bot})^2
+m^2(x-x')^2}{xx'}\; ; \qquad
 \tilde{\De}_2=\tilde{\De}_1|_{x\rightarrow(1-x),x'\rightarrow(1-x')} \nn \\
&&\De_1=\frac{\tilde{\De}_1}{x-x'} \; ; \qquad
 \De_2=\frac{\tilde{\De}_2}{x-x'} 
\leea{m4a}
Note $\tilde{\De}_1,~\tilde{\De}_2$ are positive definite.

To perform the angular integration \eq{r43} analytically
we choose the similarity function as follows
\bea
&& f_{\la}(\De)=u_{\la}(\De)=\theta(\la^2-|\De|)
\leea{m5}
where for simplicity we use the sharp cutoff function $u_{\la}(\De)$.
Then the effective interaction reads
\bea
 V^{eff}=
   &-& e^2<\ga^{\mu}\ga^{\nu}>g_{\mu\nu}
\left(\frac{\theta(\tilde{\De}_1-\tilde{\De}_2)}{\tilde{\De}_1}
+\frac{\theta(\tilde{\De}_2-\tilde{\De}_1)}{\tilde{\De}_2}
\right)\nn\\
   &-& e^2<\ga^{\mu}\ga^{\nu}>\eta_{\mu}\eta_{\nu}
\frac{1}{2q^{+2}}|\tilde{\De}_1-\tilde{\De}_2|
\left(\frac{\theta(\tilde{\De}_1-\tilde{\De}_2)}{\tilde{\De}_1}
+\frac{\theta(\tilde{\De}_2-\tilde{\De}_1)}{\tilde{\De}_2}
\right)
\leea{m6}
where we have used $\theta(x)-\theta(-x)=sign(x)$
\footnote{
In this particular case, when the similarity function is given by 
\eq{m5}, the divergence arising is of the $\frac{1}{|q^+|}$ type, that
prevents from the first glance the cancellation of logarithmic
divergences in the integral 
\bea
&& \int_{|q^+|>\varepsilon} dq^+
\leea{abc} 
We have chosen the sharp cutoff  
for simplicity, but generally the smooth cutoffs are preferred
to avoid non-analyticities in the structure of counterterms
\cite{osu1}. We demand only, that the cutoff function
$u(x)$ with $x=\frac{|D|}{\la}\rightarrow\frac{|\De|}{\la^2}$ 
is $1$ for small arguments and
vanishes for large arguments. Explicit one can choose 
\bea
&& u(x)=1~~for~~ 0\leq x\leq \frac{1}{3}\nn\\
&& u(x)~monoton~falls~from~1~to~0~for~ \frac{1}{3}\leq x\leq \frac{2}{3}\nn\\
&& u(x)=0~~for~~ \frac{2}{3}\leq x \leq 1
\leea{a}
and to simplify the calculation of the integral \eq{m4}, we assume
$\frac{du(x)}{d\la^2}\sim\de(1-x)$. Then, when
the similarity function (the function that defines the first order
solution in the perturbation theory of the flow equation) 
is equal to the cutoff function
$f(x)=u(x)$, the effective electron-positron interaction reads
\bea
 V^{eff}=
   &-& e^2<\ga^{\mu}\ga^{\nu}>g_{\mu\nu}
\left(\frac{u(\tilde{\De}_1/\tilde{\De}_2)}{\tilde{\De}_1}
+\frac{u(\tilde{\De}_2/\tilde{\De}_1)}{\tilde{\De}_2}
\right)\nn\\
   &-& e^2<\ga^{\mu}\ga^{\nu}>\eta_{\mu}\eta_{\nu}
\frac{1}{2q^{+2}}(\tilde{\De}_1-\tilde{\De}_2)
\left(\frac{u(\tilde{\De}_1/\tilde{\De}_2)}{\tilde{\De}_1}
-\frac{u(\tilde{\De}_2/\tilde{\De}_1)}{\tilde{\De}_2}
\right)
\leea{b}
In this case the prescription \eq{abc} ensures the cancellation of
logarithmic divergences, and one does not need to introduce
counterterms. 
In the integral equation \eq{r4} we have the following restriction
of the integration domain
\bea
&& \int dx'\rightarrow\int_{|x'-x|>\varepsilon/P^+} d(x'-x)
\leea{c}
resulting, that the divergent part in $''\eta_{\mu}\eta_{\nu}''$ term
of the effective interaction \eq{b} does not contribute to mass spectrum.
}.

The general matrix elements for the effective interaction \eq{m6}
depending on the angles $\varphi,\varphi'$ 
(actually on the difference $(\varphi -\varphi')$)
$<x,k_{\perp},\varphi;\la_1,\la_2|V^{eff}|
x',k'_{\perp},\varphi';\la'_1,\la'_2>$
and also the matrix elements of the effective interaction
after the angular integration \eq{r43}
for the total momentum $J_z$
$<x,k_{\perp};J_z,\la_1,\la_2|\tilde{V}^{eff}|
x',k'_{\perp};J'_z,\la'_1,\la'_2>$
are given in Appendix \ref{appB}.

The positronium spectrum is calculated numerically,
using the integral equation \eq{r48} with the matrix elements
given in Appendix \ref{appB}.
We use for the numerical integration
the Gauss-Legendre algorithm (Gaussian quadratures). 
To improve the numerical convergence
the technique of Coulomb counterterms is included.
The problem has been solved for all components of the total
angular momentum, $J_z$. Since we calculate the values
of an invariant mass squared, a strong coupling constant
$\alpha=0,3$ has been chosen.
The latter means also, that the method 
of flow equations is applicable in non-perturbative regime.
We get the ionization threshold at $M^2\sim 4m^2$,
the Bohr spectrum, and what is more important,
the fine structure.
The agreement is quantitative (for the lowest eigenvalues), 
particularly for the physical value of the fine structure constant
$\alpha=\frac{1}{137}$.

The figure ($5$) is the summary of the spectra for different 
components of the total angular momentum, $J_z$.
As one can see, certain mass eigenvalues at 
$J_z=0$ are degenerate with certain eigenvalues at other $J_z$
to a very high degree of numerical precision. As an example,
consider the second lowest eigenvalue for $J_z=0$.
It is degenerate with the lowest eigenvalue for $J_z=\pm 1$,
and can thus be classified as a member of the triplet with $J=1$.
Correspondingly, the lowest eigenvalue for $J_z=0$ having no companion
can be classified as the singlet state with $J=0$.
Quite in general one can interpret $2J_{z,max}+1$ degenerate multiplets 
as members of a state with total angular momentum $J$. 
One can get the quantum number
of total angular momentum $J$ from the number of degenerate states
for a fixed eigenvalue $M_n^2$. 
Such form of spectrum is driven
by rotational invariance. It is a remarkable result, 
we restore the rotational symmetry in 
the light-front positronium calculations on the level 
of mass spectrum.

The integral equation is approximated by Gaussian quadratures, 
and the results are studied as a function of the number
of integration points $N$, as displayed in Figures ($6$) and ($7$).
On the fig.($6$) the mass squared eigenvalues $M_n^2$ for $J_z=0$
are shown as functions of the number of integration points
$N=N_1=N_2$. One sees, that the results stabilize
themselves quickly. 
To find out if the degeneracy obtained is merely a numerical
artifact, or a property of the positronium model,
consider fig.($7$). The mass (squared) discrepancy
between the $J_z=0$ and $J_z=1$ eigenvalues is plotted
versus the number of integration points N for three different
states. It is important, that the convergence
of the value $\De M^2(1^{3}S_1)$ with N occurs. But
the mass gap does not converge to zero, rather to some small value
$\sim 10^{-5}$. The mass gap for $2^{1}P_1$ state vanishes as $N$
increases, while for $2^{3}P_1$ state $\De M^2$ converges
again to a finite small number $\sim 10^{-4}$.
Kaluza and Pirner \cite{KaPi} and Trittmann and Pauli \cite{TrPa}
found that in light-front perturbation theory there is a discrepancy
between the case of $J_z=0$ and $J_z=1$.

\chapter{Renormalization in the light-front QED}
\label{ch7}

\section{Introduction.}
\label{7.1}

In this section we consider  
the renormalization group problem for light-front QED,
arising first in the second order in coupling. Renormalization
in the light-front field theory is much more complicated than in covariant
Feynman perturbation theory, and is nontrivial already in the leading order
for QED. Generally, there are three renormalization problems
that are signaled by the divergence of the free energy:\\
$(1)$~ $\vec{p}_{\perp}^2$;\\
$(2)$~ $p^+\rightarrow 0$;\\
$(3)$~ $p^+ =0$.\\

The divergences from $(1)$ are 'ultraviolet' (UV), producing counterterms
that are local in the transverse direction. While the counterterm
structure is richer than what is required in manifestly 
covariant perturbation theory, in perturbative regime 
one has that masses are relevant operators, and other gauge interactions
are marginal.

The divergences from $(2)$ are 'infrared' (IR), producing counterterms
that are nonlocal at least in the longitudinal direction and which
may also be nonlocal in the transverse direction.
These infrared divergences do not appear in Feynman perturbation theory
in covariant gauges. They must cancel perturbatively
for gauge invariant matrix elements in QED. Coupling coherence
fixes the infrared divergent part of the Hamiltonian as a power series
in the gauge coupling, allowing us to next consider
re-summations that preserve cancellation of all infrared
divergences.

Finally, $(3)$ is the vacuum problem, that we do not consider here.
The light-front vacuum is 'trivial', except for zero modes.
The light-front gauge $A^+=0$
has a singularity at longitudinal momentum $p^+=0$, i.e.
the zero-modes $p^+=0$ should be treated specially, that is called
the problem of zero-modes. One can connect $(3)$ and $(2)$,
but we ignore the problem of zero modes in QED (at least here).               
                   
In light-front renormalization group transverse and longitudinal 
directions scale separately, and transverse scaling runs the cutoff.
We assume transverse locality and identify relevant and marginal
operators using power counting, which is valid  in the perturbative
regime.
The longitudinal scale is not important for the perturbative
classification of operators. This means, that entire functions of
longitudinal momenta (their dimensionless ratios), including any
longitudinal momentum scale introduced by cutoff, appear
in the relevant and marginal operators. The simplest way to adjust 
the new operators is to fix broken by the cutoff covariance 
(Lorenz covariance) and gauge invariance in physical observables.
Generally, infinite number of counteterms must be added
to the Hamiltonian to restore these symmetries in observables,
and to obtain finite results. As a result, the renormalizability
and as a consequence predictive power of the theory seem
to be problematic.   

Coupling coherence, developed by Perry and Wilson \cite{PeWi}, says
how to deal in practice with the functions that appear 
in marginal and relevant light-front operators and how to reduce 
the number (infinite number) of couplings, that arise
by renormalization scaling when the symmetries of the theory 
(rotational and gauge invariance) are broken  by cutoffs.
We demand that these additional counterterms required to restore
the symmetries run coherently with the ``canonical renormalizable
couplings'' - that is we do not allow them to explicitly depend
on the cutoff but only implicitly through the 
``canonical renormalizable couplings''  
dependences on the cutoff.
In other words, coupling coherence provides the condition
under which a finite number of running variables determines
the renormalization group trajectory of the renormalized Hamiltonian.
Also it is conjectured that coupling coherence restores  (or reveal)
symmetries broken by the regulator or vacuum.

Coupling coherent Hamiltonian written in terms of dimensionless
couplings for $\la<<\La$ satisfies
\bea
&& \hspace{-1.5cm}
H_{\la}=U(\la,\La)H_{\La}(\La,e_{\La},m_{\La},w_{\La},c(e_{\La},m_{\La}))
U^+(\la,\La)\rightarrow
H_{\La}(\la,e_{\la},m_{\la},w(e_{\la},m_{\la}),c(e_{\la},m_{\la}))\nn\\
\leea{f1}
where $U(\la,\La)$ is the unitary transformation defined by flow equations.
The additional requirement to \eq{f1} is that all dependent couplings
(represented by '$c$' in the argument of the Hamiltonians)
vanish when the independent marginal couplings are set to zero.
In \eq{f1}, $e_{\La}$ and $m_{\La}$ are independent dimensionless
marginal and relevant couplings, respectively; $w_{\La}$
represents the infinite set of independent dimensionless 
irrelevant couplings; $c(e_{\la},m_{\la})$ represents the infinite
set of dependent dimensionless relevant, marginal and irrelevant couplings.
More about classification of couplings see \cite{PeWi}.

The initial bare Hamiltonian $H_{\La}$ does not satisfy \eq{f1} 
{\it a priori}, its form changes under the action of transformation
$U(\la,\La)$.$H_{\La}$ must be adjusted until its form does not change.
This ``adjustment'' is the process of renormalization.
Coupling coherence is a highly nontrivial constraint on the theory
and to date has only been solved perturbatively.
Further we present the solution of \eq{f1} for QED 
in the second order in coupling (we calculate the electron and photon
'mass' operators), that turns out to be simple
because coupling constant does not run until the third order.    

In the second order in coupling the program we proceed to find
renormalized electron and photon masses is the following:\\
$(1)$~ regularize the canonical light-front QED Hamiltonian.
In this case, we introduce the bare cutoff $\La$,
that regulates the divergences in transverse direction,
and introduce the second order mass counterterm $\de M_{\La}^{(2)}$.
We start with the bare cutoff mass
$m_{\La}^2=m^2+\de M_{\La}^{(2)}$, where 
$m$ is the mass term in free Hamiltonian $H_0$;\\
$(2)$~ perform the unitary transformation $U(\la,\La)$.
The second-order change in the Hamiltonian is given in 
\eq{renormalization} (section \ref{sec4.2}), which
gives in one-body sector the shift to the electron (photon)
mass
\bea 
&& \sim\int_{\La}^{\la}<[\eta^{(1)},H_{ee\ga} ]>_{one-body}d\la'
 \sim -(\de\Sigma_{\la}-\de\Sigma_{\La})
\leea{f2}
where $\eta^{(1)}$ is the first order generator of transformation,
defined by flow equations as $\eta^{(1)}=[H_0,H_{ee\ga}]$,
$H_{ee\ga}$ is the electron-photon vertex operator,
and $\de\Sigma_{\la}$ is proportional
to the value of the integral in \eq{f2}
at the point $\la$ and defines mass correction, namely self energy term.
Physically, in the electron sector the energy scales $\La$
down to $\la$ from photon emission have been 'integrated out'
and placed in effective interactions (the corresponding in the
photon sector);\\
$(3)$~ renormalize the Hamiltonian, i.e. adjust the Hamiltonian 
that its form does not change under the unitary transformation \eq{f1} 
using coupling coherence. As a result we get the renormalized
(in the given order) Hamiltonian $H_{\la}$.
In this case,
since the electron-photon coupling does not run until third order,
the constraint \eq{f1} is simple to solve: to second order 
the self-energy must exactly reproduce itself with $\La\rightarrow\la$
\bea
&& [m^2-(\de\Sigma_{\la}-\de\Sigma_{\La})+\de M_{\La}^{(2)}+O(e^4)]
  =[m^2+\de M_{\La}^{(2)}+O(e^4)]_{\La\rightarrow\la}\nn\\
&& =[m^2-\de\Sigma_{\la}+O(e^4)]
\leea{f3}
this fixes the counterterm
\bea
&& \de M_{\La}^{(2)}=-\de\Sigma_{\La}+O(e^4)
\leea{f4}
i.e. in the second order electron (photon) mass squared 
runs coherently with the cutoff according to \eq{f1} as
\bea
&& m_{\la}^2=m^2-\de\Sigma_{\la}
\leea{f5}
$(4)$~ using the renormalized
Hamiltonian $H_{\la}$ calculate physical observables, that due to the
constraint of coupling coherence are cutoff and renormalization scale
independent (are finite as $\La\rightarrow\infty$ and 
$\varepsilon\rightarrow 0$, where $\La$ is UV regulator
in transverse direction and $\varepsilon$ is IR regulator
in longitudinal direction (generally, it can be a set of regulators), 
and are independent of $\la$
or any other scale introduced by renormalization),
and manifest the symmetries of the theory broken  
by cutoffs.
We calculate the physical mass of electron (photon), that
includes the running mass $m_{\la}$ from $H_{\la}$
and in the second order the perturbation theory corrections.
In the electron sector these corrections arise from 
the low-energy photon emission, since $H_{\la}$ still has
the photon emission interaction below $\la$ 
of the form $f_{\la}\int d^2x^{\perp}dx^- H_{ee\ga}$.
(One has in the photon sector the corresponding interaction term, 
where the energy of 
$e\bar{e}$ pair is restricted by $\la$.)
It turns out, that the electron (photon) perturbative mass-squared shift
is exactly equal to the self-energy term obtained in $(3)$
\bea
&& \de m^2=\de\Sigma_{\la}+O(e^4)
\leea{f6}
therefore the physical mass-squared is given
\bea
&& m_{phys}^2=m_{\la}^2+\de m^2=m^2-\de\Sigma_{\la}+\de\Sigma_{\la}=m^2
\leea{f7}
where $m^2$ is the renormalized electron mass-squared
in the free Hamiltonian $H_0$. This result is to be expected,
that shows that the renormalization procedure 
using flow equations and coupling coherence is successful
(at least in the second order).

All details of calculation can be found in two sections that follow.
The renormalization scheme presented seems to work trivially simple 
in the second order. But there are difficulties, that one encounters
in practical calculations. The integral in \eq{f2} is infrared singular
(in both electron and photon sectors),
resulting in infrared divergent self-energy and hence infrared divergent 
running mass $m_{\la}$, with new types of divergences
(the electron mass has both linear and logarithmic infrared divergences).
This is in accordance with the discussion presented above.
The longitudinal cutoff violates boost invariance and the mass operator,
that is a function of a longitudinal momentum scale introduced
by the cutoff, is required to restore this symmetry in physical results.
Indeed, as shown by Perry \cite{Pe}, the divergent self-energy is exactly
canceled by perturbative mixing with small-x photons (see \eq{f6},\eq{f7}),
resulting in finite mass. 

We try to preserve boost invariance, that is manifest symmetry 
on the light-front, in the second order QED calculations 
on the level of renormalized Hamiltonian.
We take into account the diagrams arising from the normal ordering
of instantaneous interactions. 
The bare light-front QED Hamiltonian contains both 
the instantaneous interactions
and the electron-photon vertex, therefore instantaneous terms
must accompany
the latter also by scaling from the energies $\La$ down to $\la$ \eq{f1}. 
We get the running electron and photon masses free from infrared divergences,
moreover both electron and photon mass conterterms contain
the known from the covariant perturbation theory types of divergences.
Unfortunately, the electron wave function renormalization constant $Z_2$
contains mixing UV and IR logarithmic divergences and also pure IR 
logarithmic divergence;
the renormalization constant $Z_2$ in photon sector is IR finite.
It is argued in \cite{osu1}, that the mixing IR divergences are cancelled
in gauge invariant quantities. 
It is shown in \cite{ZhHa}, that the mixing divergences are also cancelled
completely in the old-fashioned Hamiltonian theory for the coupling constant
renormalization.
In $x^+$-ordered Hamiltonian theory,
the mixing divergences cannot and should not be removed as indicated
by the negativity of the second order correction of the wave function
renormalization constant, that must result in a physical theory \cite{ZhHa}.

Zhang and Harindranath \cite{ZhHa} performed the similar calculations 
of self-energy mass operators 
in light-front QCD, using perturbation theory frame.
They have tested different types of regulators to regularize
the transverse UV divergences; the regulator for IR divergences was
chosen in accordance with the prescription for boundary terms,
arising in the light-front Hamiltonian (see chapter \ref{chapter3}).
In the scheme discussed above the transverse regulator arise automatically
at the stage $(2)$, where the unitary transformation, dictated 
by flow equations, is performed. This is important, since 
operator classification and renormalization itself are 
driven by the scaling in transverse direction. The problem,
that still remains, is the violation of rotational and gauge symmetries
by this regulator, in other words, by the flow equations.
We leave this problem for future investigation.   
 
Further the detailed calculations of electron and photon
self energies and physical masses, using the program
outlined above, are presented.

\section{Flow equations for renormalization issues}
\label{5.2}

As was discussed above and in chapters $3$ and $4$, 
the commutator $[\eta^{(1)},H_{ee\ga}]$ 
contributes to the self-energy term, giving rise to the renormalization
of fermion and photon masses to the second order.
The flow equation for the electron (photon) light-cone energy $p^-$ is
\be
\frac{dp^-}{dl}=<[\eta^{(1)},H_{ee\ga}]>_{self~energy}
\; , \lee{ri1}
where the matrix element is calculated between the single
electron (photon) states $<p',s'|...|p,s>$. We drop the finite part
and define $\de p_{\la}^- = p^-(l_{\la})-<|H_0|>$. Integration over the
finite range gives
\be
\de p_{\la}^--\de p_{\La}^-=\int_{l_{\La}}^{l_{\la}}
<[\eta^{(1)},H_{ee\ga}]>_{self~energy}dl'
=-\frac{(\de\Sigma_{\la}(p)-\de\Sigma_{\La}(p))}{p^+}
\; , \lee{ri2}
that defines the cutoff dependent self energy $\de\Sigma_{\la}(p)$.
The mass correction and wave function renormalization constant
are given correspondingly, cf. \cite{ZhHa}
\bea
\de m_{\la}^2 &=& \left. p^+\de p^- \right|_{p^2=m^2}
 =-\de\Si_{\la}(m^2) \nn \\
Z_2 &=& \left. 1 + \frac{\partial \de p^-}{\partial p^-} \right|_{p^2=m^2}
\; . \leea{ri3}
The on-mass-shell condition is defined through the mass $m$ in the
free Hamiltonian $H_0$.

We show further, that to the second order $O(e^2)$ the electron and photon
masses and corresponding wave function renormalization constants
in the renormalized Hamiltonian vary in accordance with the result of $1$-loop
renormalization group equations. This can serve as evidence for the 
equivalence of the flow equations and Wilson's
renormalization. Therefore we have rewritten
the mass correction $\de m_{\la}^2$ through the self energy term,
arising in $1$-loop calculations of ordinary perturbative theory. The negative
overall sign stems from our definition of the flow parameter,
namely for $\De l>0$ we are lowering the cutoff 
\mbox{$dl=-\frac{2}{\la^3}d\la$}. 
 
We start with the bare cutoff mass \mbox{$m_{\La}^2=m^2+\de M_{\La}^{(2)}$},
where \mbox{$\de M_{\La}^{(2)}$} is the second order mass counterterm.
According to \eq{ri2},\eq{ri3} the electron (photon) mass runs
\be
m_{\la}^2 = m_{\La}^2 - [\de\Sigma_{\la}(m^2) - \de\Sigma_{\La}(m^2)]
\lee{ri4}
defining, due to renormalizability, the counterterm
\mbox{$\de M_{\La}^{(2)} =\de m_{\La}^2 = -\de\Sigma_{\La}(m^2)$}
and the dependence of the renormalized mass on the cutoff $\la$
\be
m_{\la}^2 = m^2 + \de m_{\la}^2 = m^2 - \de\Sigma_{\la}
\; . \lee{ri5}
We calculate explicitly the self-energy term.
The {\bf electron} energy correction contains several terms
\be
\de p_{\la}^-
= <p', s'|H - H_0|p, s>
= \left(\sum_{n=1}^3 \de p_{\la n}^-\right) \cdot \de^{(3)}(p-p') \de_{s s'}
\; . \lee{ri6}
The first term is induced by the flow equation in single electron
sector, namely comes from the commutator $[\eta{(1)}, H_{ee\ga}]$
\be
\de p_{1\la}^-=-\int_{l_{\la}}^{\infty}
<[\eta^{(1)},H_{ee\ga}]>_{self~energy}dl'=-\frac{\de\Sigma_{1\la}(p)}{p^+}
\; ; \lee{ri7}
it reads, cf. \eq{d9} in Appendix \ref{appC},
\bea
\de p_{1\la}^- &=& e^2 \int \frac{d^2k^{\bot} dk^+}{2(2\pi)^3}
\frac{\th (k^+)}{k^+} \th (p^+-k^+) \nn \\
&& \times \Ga^i(p - k, p, -k) \Ga^i(p, p - k, k) \,
\frac{1}{p^- - k^- - (p-k)^-} \times (-R)  
\; . \leea{ri8}
This term explicitly depends on the cutoff $\la$ ($l=1/\la^2$) through the
similarity function, that plays the role of a regulator in the loop integration
\be
R_{\la}=f_{p,k,\la}^2=\exp\left\{-2\left(\frac{\De_{p,k}}{\la}\right)^2\right\}
\; . \lee{ri9}
\Eq{ri8} corresponds to the first diagram in fig.($9$).

Two instantaneous diagrams, the second and third in fig.($9$),
contribute the cutoff independent (constant) terms.
They arise from normal-ordered instantaneous interactions
in the single electron sector and can be written as
\be
\de p_{n\la}=\de p_n(l=0)=c_n<\hat{O}\hat{O}^+>V_n^{inst}(l=0)
\ee
where $n=2,3$ corresponds to the second and third diagrams in \fig{eselfen},
$c_n$ is the symmetry factor, $<\hat{O}\hat{O}^+>$ stands for the boson
$(n=2)$ and  fermion $(n=3)$ contractions
(i.e. \mbox{$<\tilde{a}_k \tilde{a}_k^+> = \th(k^+)/k^+$} and
\mbox{$<\tilde{b}_p \tilde{b}_p^+> = \th(p^+)$}), 
and $V_n^{inst}(l=0)$ arises from normal-ordering 
of $H_{ee\ga\ga}$ for $n=2$ and of $H_{eeee}$ for $n=3$ (\eqs{ch16}{ch17}).

These two diagrams $\de p_n(l=0)$ define together with the first one 
$\de p_1(l=0)$ the initial condition for the total energy correction, \eq{ri6}.

Since the diagrams $n=2,3$ come from the normal-ordering canonical
Hamiltonian at $l=0$, they must accompany the first diagram
for any flow parameter $l$. In what follows we use for 
the instantaneous terms the same regulator $R$, \eq{ri9}
\bea
\de p_{2\la}^- &=& e^2\int \frac{d^2k^{\bot}dk^+}{2(2\pi)^3} \,
\frac{\th (k^+)}{k^+} \, \frac{\si^i\si^i}{[p^+-k^+]} \times (-R) \nn \\
\de p_{3\la}^- &=& e^2\int \frac{d^2k^{\bot}dk^+}{2(2\pi)^3} \, \th (k^+) \,
\frac{1}{2} \left( \frac{1}{[p^+ - k^+]^2} - \frac{1}{[p^+ + k^+]^2} \right) \times (-R)
\; . \leea{ri12}
We define the set of coordinates
\bea
x &=& \frac{k^+}{p^+} \nn \\
k &=& (xp^+, xp^{\bot} + \kappa^{\bot})
\; , \leea{ri13}
where \mbox{$p = (p^+, p^{\bot})$} is the external electron momentum.
Then the electron self energy diagrams, fig.($9$), \eq{d13}
in Appendix \ref{appC}, contribute
\bea
\hspace{-10mm} p^+\de p_{1\la}^- &=& -\frac{e^2}{8\pi^2}
 \int_0^1 dx \int d\kappa_{\bot}^2 \nn \\
&& \hspace{10mm} \times \left[
 \frac{p^2 - m^2}{\kappa_{\bot}^2 + f(x)} \left( \frac{2}{[x]} - 2 + x \right)
 -\frac{2m^2}{\kappa_{\bot}^2 + f(x)} + \left( \frac{2}{[x]^2} + \frac{1}{[1-x]} \right)
 \right]
 \times (-R) \nn \\
f(x) &=& xm^2 - x(1-x) p^2
\leea{ri14}
and
\bea
p^+ \de p_{2\la}^- &=& \frac{e^2}{8\pi^2} \int_0^{\infty}dx \int d\kappa_{\bot}^2
\left( \frac{1}{[x][1-x]} \right) \times (-R) \nn \\
&& \hspace{-7mm} \rightarrow \; \frac{e^2}{8\pi^2} \int_0^1dx \int d\kappa_{\bot}^2
\left( \frac{1}{[x]} \right) \times (-R) \nn \\
p^+ \de p_{3\la}^- &=& \frac{e^2}{8\pi^2} \int_0^{\infty}dx \int d\kappa_{\bot}^2
\left( \frac{1}{[1-x]^2} - \frac{1}{(1+x)^2} \right) \times (-R) \nn \\
&& \hspace{-7mm} \rightarrow \; \frac{e^2}{8\pi^2} \int_0^1dx \int d\kappa_{\bot}^2
\left( \frac{2}{[x]^2} \right) \times (-R)
\; ; \leea{ri15}
for details we refer to Appendix \ref{appC}. 
Note, that the transformation in the integrals over $x$
is performed before the regulator is taken into account \cite{ZhHa}.
(In the second integral the electron momentum is replaced by the gluon
one due to momentum conservation). The brackets '\mbox{\boldmath{$[\;]$}}'
denote the principle value prescription, defined further in \eq{ri21}.

The loop integral over $k$ \eqs{ri14}{ri15}
contains two types of divergences: UV in the transversal
coordinate $\kappa^{\bot}$ and IR in the longitudinal component $k^+$.
The physical value of mass must be IR-finite.
We show, that the three relevant diagrams together 
in fact give an IR-finite value for the renormalized mass; this enables to
determine counterterms independent of longitudinal momentum. 
In the wave function renormalization constant, however, the IR-singularity
is still present.
    
Define 
\be
\de_1=\frac{p^+}{P^+}
\; , \lee{ri16}
where $P=(P^+,P^{\bot})$ is the positronium momentum, $p$ the electron momentum. 
The transversal UV divergence is regularized through the unitary
transformation done, i.e. by the regulator $R$, \eq{ri9}
\be
R_{\la} =
\exp\left\{ -\left( \frac{\tilde{\De}_{p,k}}{\la^2\de_1} \right)^2 \right\}
\; \approx \;
\th(\la^2 \de_1 - |\tilde{\De}_{p,k}|)
\; , \lee{ri17}
where the cutoff is rescaled and defined in units of the
positronium momentum $P^+$, namely \mbox{$\la\rightarrow \sqrt{2}\la^2/P^+$},
and \mbox{$\De_{p,k}=p^--k^--(p-k)^-=\tilde{\De}_{p,k}/p^+$}. The rude approximation
for the exponential through a $\th$-function changes the numerical coefficient
within a few percent; nevertheless it is useful to estimate the integrals
in \eqs{ri14}{ri15} in this way analytically.
From \eq{ri17} we have for the sum of intermediate (electron and photon) state momenta 
(the external electron is on-mass-shell \mbox{$p^2=m^2$}) 
\be
\frac{\kappa^{\bot 2}}{[x]}+\frac{\kappa^{\bot 2}+m^2}{[1-x]} \; \leq \;
\la^2\de_1+m^2
\lee{ri18}
giving for the regulator
\bea
R_{\la} &=& \th(\kappa^{\bot 2}_{\la max}-\kappa^{\bot 2}) \,
\th(\kappa^{\bot 2}_{\la max})\nn\\
\kappa^{\bot 2}_{\la max}&=&x(1-x)\la^2\de_1-x^2m^2
\leea{ri19}
and \mbox{$\th(\kappa^{\bot 2}_{\la max})$} leads
to the additional condition for the longitudinal momentum
\bea
&& 0\leq x\leq x_{max} \nn \\
&& x_{max}=\frac{1}{1+m^2/(\la^2 \de_1)}
\leea{ri20}
implying that the singularity of the photon longitudinal momentum
for $x\rightarrow 1$ is regularized by the function $R_{\la}$.
This is the case due to the nonzero fermion mass present in \eq{ri18}
for the intermediate state with $(1-x)$ longitudinal momentum.
The IR-singularity when $x\rightarrow 0$ is still present; it
is treated by the principle value prescription \cite{ZhHa}
\be
\frac{1}{[k^+]}=\frac{1}{2}\left( \frac{1}{k^++i\varep P^+}
+\frac{1}{k^+-i\varep P^+}\right)
\; , \lee{ri21}
where $\varep=0_+$, and $P^+$ is the longitudinal part of the positronium
momentum (used here as typical momentum in the problem).
This defines the bracket '\mbox{\boldmath{$[\;]$}}'
in \eqs{ri14}{ri15}
\be
\frac{1}{[x]}=\frac{1}{2} \left( \frac{1}{x+i\frac{\varep}{\de_1}}+\\
\frac{1}{x-i\frac{\varep}{\de_1}} \right)
\; . \lee{ri22}
Making use of both regularizations for 
transversal and longitudinal components, we have for the first diagram,
\eq{ri14},
\bea
\de m_{1\la}^2 &\hspace{-1em}=\hspace{-1em}& p^+ \de p^- |_{p^2=m^2} \nn \\
\de m_{1\la}^2 &\hspace{-1em}=\hspace{-1em}&-\frac{e^2}{8\pi^2}
\bigg[ 3m^2 \ln \left( \frac{\la^2\de_1+m^2}{m^2} \right)
+ \frac{\la^2 \de_1}{\la^2\de_1+m^2} \left( \frac{3}{2}\la^2\de_1+m^2 \right)
\nn\\ 
&-& 2 \la^2 \de_1 \ln \left( \frac{\la^2\de_1}{\la^2\de_1+m^2} 
\frac{\de_1}{\varep} \right)
\bigg] \nn \\
\leea{ri23}
Note, that the third term has the mixing UV and IR divergences.
Combining the three relevant diagrams, fig.($9$), and integrating
with the common regulator, one obtains for the {\bf electron mass correction} 
\bea
\de m_{\la}^2 &=& p^+(\de p_1 + \de p_2 + \de p_3)|_{p^2 = m^2} =
-\de \Sigma_{\la}(m^2) \nn \\
\de m_{\la}^2 &=& -\frac{e^2}{8\pi^2}
\left\{ 3m^2 \ln \left( \frac{\la^2\de_1+m^2}{m^2} \right)
-\frac{\la^2 \de_1 m^2}{\la^2 \de_1 + m^2} \right\}
\; . \leea{ri24}
The mass correction is IR-finite (that gives rise to IR-finite
counterterms) and contains only a logarithmic UV-divergence. Namely,
when \mbox{$\la\de_1 \rightarrow\La \gg m$}
\be
\de m_{\La}^2=-\frac{3e^2}{8\pi^2}m^2 \ln \frac{\La^2}{m^2}
\; . \lee{ri25}
It is remarkable that we reproduce with the cutoff condition of \eq{ri18}
the standard result of covariant perturbative theory calculations  
including its global factor $3/8$. As was mentioned above, the difference
in sign, as compared with the $1$-loop renormalization group result, 
comes from scaling down from high to low energies 
in the method of flow equations.     

The similar regularization for the intermediate state momenta
in the self-energy integrals, called 'global cutoff scheme', 
was introduced by W.~M.~Zhang and A.~Harindranath \cite{ZhHa}.
In our approach the UV-regularization, that defines the concrete form of 
the regulator $R$, arises naturally from the method of flow equations, 
namely from the unitary transformation performed, where the generator
of the transformation is chosen as the commutator $\eta=[H_d,H_r]$.
Note also, that the regulator $R$, \eq{ri17}, in general
is independent of the electron momentum $p^+$ (rescaled cutoff
\mbox{$\la\de_1\longrightarrow\la$}), and therefore is boost invariant.

For the wave function renormalization constant, \eq{ri3}, one has
\be
\left. \frac{\partial\de p^-}{\partial p^-} \right|_{p^2=m^2} =
-\frac{e^2}{8\pi^2}\int_0^1\int d\kappa_{\bot}^2 \left[
\frac{2\frac{1}{x}-2+x}{\kappa_{\bot}^2+f(x)}
-\frac{x(1-x)2m^2}{(\kappa_{\bot}^2+f(x))^2}\right]_{p^2=m^2}
\times(-R)
\; , \lee{ri26}
that together with the regulator $R$, \eq{ri17}, results
\bea
Z_2 &\hspace{-0.5em}=\hspace{-0.5em}& 1 - \frac{e^2}{8\pi^2} \bigg[
\ln\frac{\la^2\de_1}{m^2} \cdot \left( \frac{3}{2}-2 \ln\frac{\de_1}{\varep} \right)
+ \ln\frac{\de_1}{\varep} \cdot \left( 2 - \ln\frac{\de_1}{\varep} \right)\nn\\
&+&
  F \left( \ln\frac{\la^2\de_1}{\la^2\de_1+m^2}; \frac{\la^2\de_1}{\la^2\de_1+m^2} \right)
\bigg] \nn \\
F &\hspace{-0.5em}=\hspace{-0.5em}& \ln\frac{\la^2\de_1}{\la^2 \de_1+m^2} \left(
\frac{1}{2} - \ln\frac{\la^2 \de_1}{\la^2 \de_1+m^2} \right)
+\frac{1}{2} \frac{\la^2 \de_1}{\la^2 \de_1 + m^2}
- 2 + 2 \int_0^{x_{max}}dx \frac{\ln x}{x-1}
\; . \leea{ri27}
As \mbox{$\la\de_1 \rightarrow \La \gg m$} the function $F$ tends to a constant
\be
F|_{\La \gg m} = C = -\frac{3}{2} +\frac{\pi^2}{3}
\; . \lee{ri28}
Therefore, by dropping the finite part, we obtain
\bea
Z_2 &=& 1 - \frac{e^2}{8\pi^2}
\left\{\ln\frac{\La^2}{m^2} \cdot \left( \frac{3}{2}
-2 \ln\frac{1}{\varep} \right)
+ \ln\frac{1}{\varep} \left( 2-ln\frac{1}{\varep} \right) \right\}
\; , \leea{ri29} 
where we have rescaled 
\mbox{$\frac{\varepsilon}{\de_1}\rightarrow\varepsilon$}. 
The electron wave function renormalization constant contains
logarithmic UV and IR divergences mixed, together with pure
logarithmic IR divergences. 
We mention, that the value of $Z_2$ is not sensitive to the form of regulator
applied; the same result for $Z_2$ was obtained with another choice of
regulator \cite{ZhHa}.

We proceed with renormalization to the second order in the {\bf photon} 
sector.
The diagrams that contribute to the photon self energy are shown in 
fig.($10$).
The commutator \mbox{$[\eta^{(1)},H_{ee\ga}]$}, 
corresponding to the first diagram,
gives rise to (\eq{d21} in Appendix \ref{appC})
\bea
\de q_{1\la}^ - \, \de^{ij} &=&
\frac{1}{[q^+]} e^2 \int \frac{d^2k^{\bot}dk^+}{2(2\pi)^3} \,
\th(k^+) \th(q^+-k^+) \\
&&\hspace{2em} \times \, Tr\left[\Ga^i(k,k-q,q) \Ga^j(k-q,k,-q)\right] \,
\frac{1}{q^- - k^- - (q-k)^-} \times(-R) \nn
\; , \leea{ri30}
where momenta are given in fig.($10$), and the regulator is
\be
R_{\la}=f_{q,k,\la}^2=\exp\left\{-2 \left(\frac{\De_{q,k}}{\la} \right)^2
\right\}
\; . \lee{ri31}
In full analogy with the electron self energy this also defines the regulator
for the second diagram with the instantaneous interaction, see 
fig.($10$),
\be
\de q_{2\la}^- \, \de^{ij} = \frac{1}{[q^+]} e^2 \int \frac{d^2k^{\bot}dk^+}{2(2\pi)^3}
\th(k^+) \, Tr(\si^i \si^j)
\left( \frac{1}{[q^+-k^+]} - \frac{1}{[q^++k^+]} \right) \times(-R)
\; . \lee{ri32}
We define the set of coordinates
\bea
&&\frac{(q-k)^+}{q^+}  =  x  \nn \\
&&k  =  ((1-x)q^+, (1-x)q^{\bot}+\kappa^{\bot}) \nn \\
&&(q - k)  =  (xq^+, xq^{\bot} - \kappa^{\bot})
\; , \leea{ri33}
where $q = (q^+, q^{\bot})$ is the external photon momentum. Then two diagrams
contribute (for details see Appendix \ref{appC}, \eq{d25}):
\bea
q^+ \, \de q_1^- &=&
-\frac{e^2}{8\pi^2} \int_0^1dx \int d\kappa_{\bot}^2 \nn \\
&& \times \left\{
 \frac{q^2}{\kappa_{\bot}^2+f(x)} \left( 2x^2 - 2x + 1 \right) +
 \frac{2m^2}{\kappa_{\bot}^2+f(x)}
 + \left( -2 + \frac{1}{[x][1-x]} \right)
\right\} \times (-R) \nn \\
f(x) &=& m^2 - x(1-x)q^2 \nn \\
q^+ \, \de q_2^- &=& \frac{e^2}{8\pi^2} \int_0^\infty dx \int d\kappa_{\bot}^2
\left( \frac{1}{[1-x]}-\frac{1}{1+x} \right) \times (-R) \nn \\
&&\hspace{-1.5em} \rightarrow \;
-\frac{e^2}{8\pi^2} \int_0^1dx \int d\kappa_{\bot}^2 \frac{2}{[x]} \times(-R)
\; . \leea{ri34}
Note, that the transformation in the second integral is done before
the regularization (by regulator the $R$) is performed \cite{ZhHa}.

Making use of the same approximation for the regulator as in the electron
sector, we obtain for the sum of intermediate (two electron) state momenta
\bea
&& \frac{\kappa_{\bot}^2+m^2}{x} + \frac{\kappa_{\bot}^2+m^2}{1-x}
 \leq \la^2\de_2 \nn \\
&& \de_2 = \frac{q^+}{P^+}
\; , \leea{ri35}
where the photon is put on mass-shell $q^2=0$ and the rescaled
cutoff \mbox{$\la \rightarrow \sqrt{2} \la^2/P^+$} has been used.
This condition means for the transversal integration
\bea
R_{\la} &=& \th(\kappa_{\la max}^{\bot 2}-\kappa^{\bot 2}) \,
\th(\kappa_{\la max}^{\bot 2}) \nn \\
\kappa_{\la max}^{\bot 2} &=& x(1-x)\la^2 \de_2-m^2
\; , \leea{ri36} 
and for the longitudinal integration
\bea
&& x_1 \leq x \leq x_2 \nn \\
&& x_1 = \frac{1-r}{2} \approx \frac{m^2}{\la^2\de_2} \nn \\
&& x_2 = \frac{1+r}{2} \approx 1 - \frac{m^2}{\la^2\de_2} \nn \\
&& r = \sqrt{1 - \frac{4m^2}{\la^2\de_2} }
\; , \leea{ri37}
where the approximate value is used when $m \ll \la$.
This shows that the condition of \eq{ri35} for two electrons with masses $m$ 
removes the light-front infrared singularities from \mbox{$x \rightarrow 0$}
and \mbox{$x\rightarrow 1$}. Thus, both UV and IR divergences are regularized
by the regulator R, \eq{ri36}.

The mass correction arising from the first diagram, \eq{ri34}, is
\be
\de m_{1\la}^2 = \frac{e^2}{8\pi^2} \, \frac{2}{3} \, \la^2 \de_2
\left( 1-\frac{4m^2}{\la^2\de_2} \right)^{3/2}
\; . \lee{ri38}
Combining together both diagrams with the same regulator, \eq{ri34}, we obtain
\be
\de m_{\la}^2 = \frac{e^2}{8\pi^2} \, \left( \frac{5}{3} \la^2 \de_2 \, r
- \frac{8}{3} m^2 \, r - 2m^2 \, \ln\frac{1+r}{1-r} \right)
\; , \lee{ri39}
where $r$ is defined in \eq{ri37}.
The result shows that the mass correction involves the quadratic
and logarithmic UV divergences, i.e. as $\la\de_2\rightarrow\La\gg m$
\be
\de m_{\La}^2 = \frac{e^2}{8\pi^2} \left( \frac{5}{3} \La^2 
-2m^2 \, \ln\frac{\La^2}{m^2} \right)
\; . \lee{ri40}
The wave function renormalization constant is defined through 
\be
\left. \frac{\partial\de q^-}{\partial q^-} \right|_{q^2=0} = 
- \frac{e^2}{8\pi^2} \int_0^1dx \int d\kappa_{\bot}^2
\left. \left\{ \frac{2x^2-2x+1}{\kappa_{\bot}^2+f(x)}
+ \frac{2m^2x(1-x)}{(\kappa_{\bot}^2+f(x))^2}
\right\} \right|_{q^2=0} \times (-R)
\; , \lee{ri41}
that, with the regulator $R$, \eq{ri36}, results
\be
Z_2 = 1 - \frac{e^2}{8\pi^2} \left( -\frac{2}{3}\ln\frac{1+r}{1-r}
+ \frac{10}{9} \, r + \frac{8}{9} \frac{m^2}{\la^2\de_2}\, r \right)
\; . \lee{ri42}
The photon wave function renormalization constant contains only
logarithmic UV divergence, indeed as \mbox{$\la\de_2\rightarrow\La\gg m$} 
\bea
&& Z_2=1-\frac{e^2}{8\pi^2}(-\frac{2}{3}\ln\frac{\La^2}{m^2})
\leea{ri43}
and is free of IR divergences (as is expected from the form of the
regulator $R$, \eq{ri35}).

\section{Mass renormalization}
\label{5.3}

Following light-cone rules the perturbative energy correction
of the electron with momentum $p$, coming from the emission and 
absorption of a photon with momentum $k$, is
\bea
&& \de \tilde{p}_{1\la}^-=\int\frac{d^2k^{\bot}dk^+}{2(2\pi)^3}\frac{\th(k^+)}{k^+}
\th(p^+-k^+)g_{p-k,p,\la}\Ga^i_{\la}(p-k,p,-k)
g_{p,p-k,\la}\Ga^i_{\la}(p,p-k,k)\nn\\
&&\hspace{5em} \times\frac{1}{p^--k^--(p-k)^-}
\; , \leea{mr1}
where $g_{ee\ga}$-coupling constant restricts the energy of the
photon.
Making use of the explicit form for the coupling, one has
\bea
\de \tilde{p}_{1\la}^- &=& e^2\int
 \frac{d^2k^{\bot}dk^+}{2(2\pi)^3}\frac{\th(k^+)}{k^+}
 \th(p^+-k^+) \\
&&\hspace{3em} \times \Ga^i_{\la}(p-k,p,-k) \,
 \Ga^i_{\la}(p,p-k,k)\frac{1}{p^--k^--(p-k)^-}
 \times (R) \nn
\; , \leea{mr1a}
where $R=f_{pk\la}^2$ plays the role of regulator. 
This expression coincide up to the overall sign with
the energy correction obtained in the previous section from the flow equations method.

Two instantaneous diagrams, arising from the normal-ordering Hamiltonian,
must be added to the first term with the same regulator $R$.
Then the full perturbative energy correction
$\de\tilde{p}_{\la}^-=\de\tilde{p}_{1\la}^-+\de\tilde{p}_{2\la}^-+
\de\tilde{p}_{3\la}^-$ is
\be
\de\tilde{p}_{\la}^-=-\de p_{\la}^-
\ee
where $\de p_{\la}^-$ is defined in \eq{ri6}. This means for the perturbative mass
correction
\be
\de m_{\la}^{PT2}=\de\Sigma_{\la}
\lee{mr1b}
and the self-energy term $\de \Sigma_{\la}$ is given in \eq{ri24}.

We combine the mass operator,
renormalized to the second order,      
\eq{ri5},
and the perturbation theory
correction, \eq{mr1b}, to obtain
the total physical mass to the order $O(e^2)$ 
\be
m_e^2=m_{\la}^2+\de m^2=(m^2+\de\Sigma_{\la})-\de\Sigma_{\la}
=m^2+O(e^4)
\; . \lee{mr3}
This means, that to the second order $O(e^2)$ the physical electron mass is,
up to a finite part, equal to the      
electron mass, that stands 
in the free (canonical) Hamiltonian $H_0$.

Along the same line one can proceed   
for the photon mass. One finds in the second order in coupling,
that the photon mass counterterm, obtained from the flow equations
\eq{ri38}-\eq{ri40}, is equal up to the overall sign
to the perturbative
photon mass correction. This means, that in the order $O(e^2)$
the physical photon mass
is equal to the photon mass term in the free Hamiltonian, i.e.
\be
m_{ph}^2=0+O(e^4)
\lee
where the photon mass in the canonical QED Hamiltonian
is equal to zero to preserve gauge invariance.

At the end we note, that the similarity function $f_{p_ip_f\la}$,
restricting the electron-photon vertex, plays 
the role of UV (and partially IR) regulator in the self energy integrals.
This means, that the regularization prescription of divergent 
integrals follows from the method of flow equations itself.
Moreover, the energy correction (i.e. mass correction
and wave function renormalization constant), obtained from the flow equations,
coincide up to the overall sign with the $1$-loop renormalization group
result. This is the remarkable result, indicating to the equivalence of
flow equations and Wilson's renormalization.

\chapter{Conclusions and outlook}

In this work we applied the method of flow equations to
QED on the light-front. 
We have outlined a strategy to derive an effective 
low-energy Hamiltonian on the light-front
by means of flow equations, considered in the light-front
dynamics.
Application of the
flow equations with the condition, that particle number conserving terms
are considered diagonal and those changing the particle number off-diagonal
led as in other cases to a useful effective Hamiltonian.

The main advantage of this procedure as compared 
with the similarity renormalization of Glazek and Wilson \cite{GlWi} 
is, that finally states of
different particle number are completely decoupled, since
the particle number violating contributions are eliminated down to $\la=0$.
Thus one is able to truncate the Fock space and the positronium problem 
reduces to a two particle problem, which was analyzed
analytically (since in leading order one obtains
the nonrelativistic Coulomb problem) in the chapter $4$,  
and numerically in the chapter $5$
for positronium bound states.

The effective Hamiltonian, obtained by the similarity transformation,
is band-diagonal in the energy space. The width of the band $\la$
introduces the artificial parameter in the procedure,
which is defined from the physical reasoning
($\la$ is low enough to neglect the contribution of high Fock states,
but is restricted from below to stay in perturbation theory region).
Flow equations as used here with the particle number 
conserving part of Hamiltonian to be diagonal, have no additional
parameter and converge well as $\la\rightarrow 0$ \cite{We}
to the effective Hamiltonian, which is block-diagonal in particle number
and are used therefore directly for the numerical calculations 
of the spectrum.

The procedure of elimination of nondiagonal blocks, that change
the number of quasiparticles, is performed not just in one step as
in the method of Tamm-Dancoff truncation but rather continuously
for the states with different energies in sequence.
This is the main advantage of the proposed method as compared
with Tamm-Dancoff truncation,
the possibility to perform simultaneously the ultraviolet renormalization
of the initial Hamiltonian.
In general, in the definite order of perturbation theory
all counterterms, associated with canonical operators of the theory
and also with possible new operators induced by unitary transformation,
can be obtained in the procedure \cite{GuWe}. Since different
sectors of the effective Hamiltonian are decoupled, 
one does not encounter the usual difficulties of Tamm-Dancoff truncation
and the methods related to it. Namely, the counterterms to be introduced
are 'sector-' and 'state-' independent \cite{HaOk}.
 
If one goes beyond
the tree approximation then one obtains terms with ultraviolet divergences
which have to be renormalized. 
In the second order in coupling the electron and photon
divergent mass corrections are generated by flow equations.
There are the same type of UV-divergences as obtained
in the covariant perturbation theory, the IR-divergences
in the longitudinal direction and mixed UV- and IR-divergences
in the mass terms. The two latter types of divergences
are specific for the light-front QED calculations.
They signal, that the boost invariance is broken,
since an explicit dependence on
the longitudinal IR-cutoff occur. To preserve boost invariance
in the renormalized Hamiltonian we take into account
the diagrams arising from the normal ordering 
of instantaneous interactions. Using then flow  equations
and coupling coherence we obtain the counterterms for
electron and photon masses, that are free from IR-divergences.
  
Simultaneously also terms describing interactions between more
than two particles are generated.
In this approach we were not faced with infrared problems,
except for the longitudinal IR-divergences, which arise due to
the light-front gauge formulation and
must be treated properly. These divergences, arising
on the level of effective QED Hamiltonian, show that
some symmetries of the initial theory,
that are not manifest on the light-front, are broken.
The first problem in the light-front field theory
is that whenever the generator of a symmetry is dynamical
(contains interactions) it is practically impossible
to monitor and maintain that symmetry at each step of a calculations-
unless of course one can solve the theory exactly. This concerns
parity and rotational invariance, that are not manifest
on the light-front. The second problem is that whenever
we use the Hamiltonian technique, the regularization by introducing
bare cutoffs breaks the gauge invariance
(and also Lorenz covariance), and forces the bare Hamiltonian
to contain a larger than normal suite of counterterms to 
enable a finite limit
as the cutoffs are removed.
One way is, that the counterterms are then adjusted to reproduce
physical observables and to restore the symmetries broken
by the cutoffs. The other way is to find the gauge 
invariant procedure for regularization of Hamiltonians.
The attempt in the latter direction was made by 
Brodsky, Hiller, McCartor \cite{BrHiMcCa}.
Solving the Yukawa theory they have tried to preserve more symmetries
by using Pauli-Villars procedure for regularization.

In our calculations, the effective electron-positron interaction,
obtained in the second order in coupling, contains
the IR-divergent term, describing instant emission and absorption
of 'longitudinal' photon. The physical reason for its appearance
is the violation of Lorenz and gauge symmetries by the derivation
of the effective, renormalized Hamiltonian. We use the symmetric
cutoff condition for IR-divergences in the effective interaction 
to cancel their contribution to the mass of positronium.
We find, that in this case the rotational symmetry is restored
on the level of positronium mass spectrum.    

In order to solve the flow equations analytically 
we were forced to apply in this work the perturbation theory expansion. 
One is able to improve this approach 
systematically by going to higher orders in the coupling.
It is a remarkable result, that the effective electron-positron Hamiltonian,
obtained in the second order in coupling, gives the correct 
Bohr spectrum and hyperfine splitting for positronium. 

We consider flow equations as a method which can also be used beyond
perturbation theory in a self-consistent way.
Examples in solid-state physics are the flow of the tunneling-frequency
in the spin-boson model \cite{KeMi1} and of the phonon energies
in the electron-phonon coupling \cite{LeWe}. Due to the flow 
the couplings decay even at resonance.

%
%

\begin{appendix}
\chapter{\label{appA}
Calculation of the commutator $[\eta^{(1)}(l), H_{ee\gamma}]$
 in the electron-positron sector}

Here we calculate the commutator $[\eta^{(1)}(l), H_{ee\gamma}]$
in the electron-positron sector.
The leading order generator $\eta^{(1)}$ is:
\bea
&& \hspace{-2cm}
\eta^{(1)}(l) = \sum_{\lambda s_1s_3}
\int d^3p_1 d^3p_3 d^3q
(\eta_{p_1p_3}^*(l)
\varepsilon_{\lambda}^i\tilde{a}_q +
\eta_{p_1p_3}(l) \varepsilon_{\lambda}^{i *}\tilde{a}_{-q}^+) \,
(\tilde{b}_{p_3}^+\tilde{b}_{p_1}+\tilde{b}_{p_3}^+\tilde{d}_{-p_1}^+ +
\tilde{d}_{-p_3}\tilde{b}_{p_1}+\tilde{d}_{-p_3}\tilde{d}_{-p_1}^+)\nonumber\\
&&\hspace{4em} \times \chi_{s_3}^+ \Gamma_l^i(p_1,p_3,-q) \chi_{s_1} \,
                      \delta_{q,-(p_1-p_3)}
\; , \leea{b1}
where
\be
\eta_{p_1p_3}(l)=-\Delta_{p_1p_3} \cdot g_{p_1p_3} =
\frac{1}{\Delta_{p_1p_3}} \cdot \frac{dg_{p_1p_3}}{dl}
\; , \lee{b2}
\mbox{$\Delta_{p_1p_3}=p_1^--p_3^--(p_1-p_3)^-$},
and the electron-photon coupling
\bea
&& \hspace{-2cm}
H_{ee\gamma} = \sum_{\lambda s_2s_4}
\int d^3p_2 d^3p_4 d^3q'
(g_{p_2p_4}^*(l)
\varepsilon _{\lambda'}^j\tilde{a}_{q'}+
g_{p_2p_4}(l)\varepsilon _{\lambda'}^{j *}\tilde{a}_{-q'}^+) \,
(\tilde{b}_{p_4}^+\tilde{b}_{p_2} +\tilde{b}_{p_4}^+\tilde{d}_{-p_2}^+ +
\tilde{d}_{-p_4}\tilde{b}_{p_2} +\tilde{d}_{-p_4}\tilde{d}_{-p_2}^+)\nonumber\\
&&\hspace{4em} \times\chi_{s_4}^+ \Gamma_l^i(p_2,p_4,-q') \chi_{s_2} \,
               \delta_{q',-(p_2-p_4)}
\; , \leea{b3}
where
\be
\Gamma_l^i(p_1,p_2,q) = 2\frac{q^i}{q^+} -
\frac{\sigma\cdot p_2^{\bot} - im_{p_1p_2}(l)}{p_2^+}\sigma^i -
\sigma^i\frac{\sigma\cdot p_1^{\bot} + im_{p_1p_2}(l)}{p_1^+}
\lee{b4}
and the tilde-fields are defined in \eq{ch19}.
The mass term $m_{p_1p_2}(l)$ starts to depend on $l$
in the second order in coupling constant. We omit 
the dependence of $m_{p_1p_2}$ and hence the dependence
of vertex matrix element $\Gamma^i(p_1,p_2,q)$ on $l$
in calculation of commutator $[\eta^{(1)}(l),H_{ee\gamma}]$
to the leading (second) order.
Further we use the identities for the polarisation vectors and spinors
\be
\sum_{\lambda} \varepsilon_{\lambda}^{i *} \varepsilon_{\lambda}^j
= \de^{ij} \; ,\qquad
\chi_s^+ \chi_{s'} = \de_{ss'}
\; . \lee{b5}

Using the commutation relations, \eq{ch10}, and identities \eq{b5} we have 
\bea
[\eta^{(1)}(l),H_{ee\gamma}] &=& \frac{1}{2} \,
 \left( - \eta_{p_1p_3} g_{p_2p_4}^* \frac{\theta(p_1^+-p_3^+)}{p_1^+ - p_3^+}
        + \eta_{p_1p_3}^* g_{p_2p_4} \frac{\theta(p_3^+-p_1^+)}{p_3^+-p_1^+} \right) \\
&& \times {\boldmath :} \,
 (- \tilde{b}_{p_3}^+ \tilde{d}_{-p_2}^+ \tilde{d}_{-p_4} \tilde{b}_{p_1}
  - \tilde{b}_{p_4}^+ \tilde{d}_{-p_1}^+ \tilde{d}_{-p_3} \tilde{b}_{p_2}
  + \tilde{b}_{p_3}^+ \tilde{d}_{-p_1}^+ \tilde{d}_{-p_4} \tilde{b}_{p_2}
  + \tilde{b}_{p_4}^+ \tilde{d}_{-p_2}^+ \tilde{d}_{-p_3} \tilde{b}_{p_1}) \,
          {\boldmath :} \nn \\
&& \times (\chi_{s_3}^+ \Gamma_l^i(p_1,p_3,p_1-p_3) \chi_{s_1}) \,
          (\chi_{s_4}^+ \Gamma_l^i(p_2,p_4,p_2-p_4) \chi_{s_2}) \:
          \delta_{p_1+p_2,p_3+p_4} \nn
\; , \leea{b6}
where the first two terms of the field operators contribute to
the exchange channel, and the next two to the annihilation channel.
We take into account both $s$- and $t$-channel terms to calculate
the bound states. The $:\;:$ stand for the normal ordering of the
fermion operators and $(\frac{1}{2})$ is the symmetry factor. The
sum over helicities $s_i$ and the 3-dimensional integration 
over momenta $p_i$, $i=1,..4$, according to \eq{ch20} is implied. 
We rewrite for both channels
\be
\hspace{0cm}
[\eta,H_{ee\gamma}]=
\left\{ \begin{array}{l}
  M_{2ij}^{(ex)}(\frac{1}{2})
       \left\{ \frac{\theta(p_1^+-p_3^+)}{(p_1^+-p_3^+)} \right.
       (\eta_{p_1,p_3}g_{-p_4,-p_2}^*-\eta_{-p_4,-p_2}^*g_{p_1,p_3}) \\
\hspace{2.0cm}
     + \left. \frac{\theta(-(p_1^+-p_3^+))}{-(p_1^+-p_3^+)}
       (\eta_{-p_4,-p_2}g_{p_1,p_3}^*-\eta_{p_1,p_3}^*g_{-p_4,-p_2})\ \right\} \\
\hspace{1.8cm}
     \times \delta^{ij}\delta_{p_1+p_2,p_3+p_4} \,
     b_{p_3s_3}^+d_{p_4\bar{s}_4}^+d_{p_2\bar{s}_2}b_{p_1s_1} \\
\\
 -M_{2ij}^{(an)}(\frac{1}{2})
       \left\{ \frac{\theta(p_1^++p_2^+)}{(p_1^++p_2^+)} \right.
       (\eta_{p_1,-p_2}g_{-p_4,p_3}^*-\eta_{-p_4,p_3}^*g_{p_1,-p_2}) \\
\hspace{2.2cm}
       + \left. \frac{\theta(-(p_1^++p_2^+))}{-(p_1^++p_2^+)}
       (\eta_{-p_4,p_3}g_{p_1,-p_2}^*-\eta_{p_1,-p_2}^*g_{-p_4,p_3}) \right\} \\
\hspace{2.0cm}
       \times \delta^{ij}\delta_{p_1+p_2,p_3+p_4} \,
       b_{p_3s_3}^+d_{p_4\bar{s}_4}^+d_{p_2\bar{s}_2}b_{p_1s_1}
\end{array} \right. 
\lee{b7}
where 
\bea
M_{2ij}^{(ex)}
&=& (\chi_{s_3}^+ \Gamma^i(p_1,p_3,p_1-p_3) \chi_{s_1}) \,
    (\chi_{\bar{s}_2}^+ \Gamma^j(-p_4,-p_2,-(p_1-p_3)) \chi_{\bar{s}_4}) \nn \\
\\
M_{2ij}^{(an)}
&=& (\chi_{s_3}^+ \Gamma^i(-p_4,p_3,-(p_1+p_2)) \chi_{\bar{s}_4}) \,
    (\chi_{\bar{s}_2}^+ \Gamma^j(p_1,-p_2,p_1+p_2) \chi_{s_1}) \nn
\; . \leea{b8}

The first term in the exchange channel with $p_1^+ > p_3^+$ corresponds to the 
light-front time ordering $x_1^+ < x_3^+$ 
with the intermediate state $P_k^-=p_3^- + (p_1-p_3)^- + p_2^-$,
the second term $p_1^+<p_3^+$ and $x_1^+>x_3^+$
has the intermediate state $P_k^- = p_1^- - (p_1 - p_3)^- + p_4^-$.
Both terms can be viewed as the retarded photon exchange.
The same does hold for the annihilation channel.

Consider only real couplings and take into account the symmetry
\be
\eta_{-p_4,-p_2}=-\eta_{p_4,p_2} \; ,\qquad g_{-p_4,-p_2}=g_{p_4,p_2}
\; . \lee{b9}
Then \mbox{$\left. \left< \! p_3s_3,p_4\bar{s}_4 \right|
[\eta^{(1)},H_{ee\gamma}] \left| p_1s_1,p_2\bar{s}_2 \right> \right.$},
the matrix element of the commutator between the free states of positronium
in the exchange and annihilation channel, reads
\be
<[\eta^{(1)},H_{ee\gamma}]> = 
\left \{\begin{array}{l}
 M_{2ii}^{ex} \, \frac{1}{(p_1^+-p_3^+)} \,
  (\eta_{p_1,p_3}g_{p_4,p_2} + \eta_{p_4,p_2}g_{p_1,p_3})
\\
-M_{2ii}^{an} \, \frac{1}{(p_1^++p_2^+)} \,
  (\eta_{p_1,-p_2}g_{p_4,-p_3} + \eta_{p_4,-p_3}g_{p_1,-p_2})
\end{array} \right. 
\; . \lee{b10}
where the conservation of $'+'$ and $'\perp'$ components
of the total momentum is implied, i.e.
$p_1^++p_2^+=p_3^++p_4^+$ and 
$p_1^{\perp}+p_2^{\perp}=p_3^{\perp}+p_4^{\perp}$.
We rewrite this expression through the corresponding $f$-functions
\bea
\eta_{p_1,p_3} g_{p_4,p_2} + \eta_{p_4,p_2} g_{p_1,p_3}
&=& e^2 \left[
  \frac{1}{\Delta_{p_1,p_3}} \frac{df_{p_1,p_3}(l)}{dl} f_{p_4,p_2}(l)
+ \frac{1}{\Delta_{p_4,p_2}} \frac{df_{p_4,p_2}(l)}{dl}f_{p_1,p_3}(l) \right] \nn \\
\\
\eta_{p_1,-p_2} g_{p_4,-p_3} +\eta_{p_4,-p_3} g_{p_1,-p_2}
&=& e^2 \left[
  \frac{1}{\Delta_{p_1,-p_2}}\frac{df_{p_1,-p_2}(l)}{dl}f_{p_4,-p_3}(l)
+ \frac{1}{\Delta_{p_4,-p_3}}\frac{df_{p_4,-p_3}(l)}{dl}f_{p_1,-p_2}(l) \right] \nn
\leea{b11}
with $\Delta_{p_1,p_2}=p_1^--p_2^--(p_1-p_2)^-$. 
This form in terms of the $f$-function is universal
for all unitary transformations. 

We calculate the matrix elements \mbox{$M_{2ii}$}, eq.~(188), for both channels.
Here we follow the notations introduced in \cite{JoPeGl}.

We make use of the identities
\be
\chi_s^+\sigma^i\sigma^j \chi_s = \delta^{ij}+is\varepsilon^{ij} \; ,\qquad
\chi_s^+\sigma^j\sigma^i \chi_s = \delta^{ij}+i\bar{s}\varepsilon^{ij}
\lee{b12}
with $\bar{s} = -s$ and $\chi_s^+ \chi_{s'} = \de_{ss'}$; also of
\be
\chi_{\bar{s}}^+ \sigma^i\chi_s = -\sqrt{2}s \varepsilon_s^i \; ,\qquad
\chi_s^+\sigma^i \chi_{\bar{s}} = -\sqrt{2}s \varepsilon_s^{i*}
\lee{b13}
with \mbox{$\varepsilon_s^* = -\varepsilon_{\bar{s}}$} and
\mbox{$\varepsilon_s^i \varepsilon_{s'}^i = -\delta_{s\bar{s'}}$}.

We use the standard light-front frame, fig.($3$),
\bea
&& p_1 = (xP^+,xP^{\bot}+\kappa_\bot) \; , \hspace{3em}
   p_2 = ((1-x)P^+,(1-x)P^{\bot}-\kappa_\bot) \; , \nn \\
&& p_3 = (x'P^+,x'P^{\bot}+\kappa'_\bot) \; , \hspace{1.5em}
   p_4 = ((1-x')P^+,(1-x')P^{\bot}-\kappa'_\bot)
\; , \leea{b14}
where \mbox{$P=(P^+,P^{\bot})$} is the positronium momentum.

Then, to calculate the matrix element $M_{2ii}$ in the {\bf exchange channel}, we find
\bea
P^+[\chi_{s_3}^+\Gamma^i(p_1,p_3,p_1-p_3)\chi_{s_1}]
&=& \chi_{s_3}^+ \left[ 2\frac{(\kappa_\bot-\kappa'_\bot)^i}{(x-x')}
 - \frac{\sigma \cdot \kappa'_\bot}{x'} \sigma^i+\sigma^i
\frac{\sigma \cdot \kappa_\bot}{x} +
          im\frac{x-x'}{xx'}\sigma^i
   \right] \chi_{s_1} \nn \\
&=& T_2^i \delta_{s_1s_3} +
 im\frac{x-x'}{xx'} (-\sqrt{2}) s_1 \varepsilon_{s_1}^i \delta_{s_1\bar{s}_3}
\; , \leea{b15}
and
\bea
&&\hspace{-1cm}
P^+[\chi_{\bar{s}_2}^+ \Gamma^i(-p_4,-p_2,-(p_4-p_2)) \chi_{\bar{s}_4}] \nn \\
&&\hspace{2cm} = \chi_{\bar{s}_2}^+
 \left[ 2 \frac{(\kappa_\bot-\kappa'_\bot)^i}{x-x'}
 + \left( \frac{\sigma \cdot \kappa_\bot}{1-x} \sigma^i
           + \sigma^i \frac{\sigma \cdot \kappa'_\bot}{1-x'} \right)
 - im \frac{x-x'}{(1-x)(1-x')} \sigma^i
 \right] \chi_{\bar{s}_4} \nn \\
&&\hspace{2cm} = - \left[ T_1^i\delta_{s_2s_4}
 + im \frac{x-x'}{(1-x)(1-x')} (-\sqrt{2}) s_2 \varepsilon_{s_2}^i
 \delta_{s_2\bar{s}_4} \right]
\; , \leea{b16}
where we have introduced
\bea
T_1^i & \equiv& -\left[ 2 
\frac{(\kappa_\bot-\kappa'_\bot)^i}{x-x'}+\frac{\kappa_\bot^i(s_2)}{(1-x)} +
 \frac{{\kappa'}_\bot^i(\bar{s}_2)}{(1-x')} \right] \nn \\
\\
T_2^i & \equiv& 2 \frac{(\kappa_\bot-\kappa'_\bot)^i}{x-x'}-
\frac{\kappa_\bot^i(s_1)}{x} -
 \frac{{\kappa'}_\bot^i(\bar{s}_1)}{x'} \nn
\leea{b17}
and
\be
\kappa_\bot^i(s)\equiv \kappa_\bot^i+is \, \varepsilon_{ij} \, \kappa_\bot^j
\; . \lee{b18}
Finaly we result 
\bea
P^{+2} \, M_{2ii}^{(ex)} &\hspace{-2mm}=\hspace{-2mm}& - \left\{
   \delta_{s_1s_3} \delta_{s_2s_4} T_1^{\bot} \cdot T_2^{\bot}
 - \delta_{s_1\bar{s}_2} \delta_{s_1\bar{s}_3} \delta_{s_2\bar{s}_4}
 2m^2 \frac{(x-x')^2}{xx'(1-x)(1-x')} \right. \\
&&\hspace{2em} \left.
 + im \sqrt{2}(x'-x) \left[ \delta_{s_1\bar{s}_3} \delta_{s_2s_4}
 \frac{s_1}{xx'}T_1^{\bot} \cdot \varepsilon_{s_1}^{\bot}
 + \delta_{s_1s_3} \delta_{s_2\bar{s}_4}
 \frac{s_2}{(1-x)(1-x')} T_2^{\bot} \cdot \varepsilon_{s_2}^{\bot} \right]
 \right\} \nn
\; . \leea{b19}
Whereas in the {\bf annihilation channel} we calculate
\bea
P^+[\chi_{s_3}^+ \Gamma^i(-p_4,p_3,-(p_1+p_2)) \chi_{\bar{s}_4}]
&=& \chi_{s_3}^+ \left[
 - \frac{\sigma \cdot \kappa'_\bot}{x'} \sigma^i + 
\sigma^i \frac{\sigma \cdot \kappa'_\bot}{1-x'}
 + im\frac{1}{x'(1-x')}\sigma^i \right] \chi_{\bar{s}_4} \nn \\
&=& T_3^i\delta_{s_3\bar{s}_4}
 + im\frac{1}{x'(1-x')} (-\sqrt{2}) s_4 \varepsilon_{s_4}^{i*}
 \delta_{s_3s_4}
\leea{b20}
and
\bea
\hspace{-1cm}
P^+[\chi_{\bar{s}_2}^+ \Gamma^i(p_1,-p_2,p_1+p_2) \chi_{s_1}]
&=& \chi_{\bar{s}_2}^+ \left[
 \frac{\sigma \cdot \kappa_\bot}{1-x} \sigma^i - 
\sigma^i \frac{\sigma \cdot \kappa_\bot}{x}
 - im\frac{1}{x(1-x)} \sigma^i \right] \chi_{s_1} \nn \\
&=& T_4^i\delta_{s_1\bar{s}_2}
 -im\frac{1}{x(1-x)} (-\sqrt{2}) s_1 \varepsilon_{s_1}^i
 \delta_{s_1s_2}
\; , \leea{b21}
where we have introduced
\bea
T_3^i & \equiv & -\frac{{\kappa'}_\bot^i(\bar{s}_3)}{x'}
 + \frac{{\kappa'}_\bot^i(s_3)}{1-x'} \nn \\
\\
T_4^i & \equiv & \frac{\kappa_\bot^i(\bar{s}_1)}{1-x}
 - \frac{\kappa_\bot^i(s_1)}{x} \nn
\; . \leea{b22}
We finally have
\bea
P^{+2} \, M_{2ii}^{(an)} &=& \delta_{s_1\bar{s}_2} \delta_{s_3\bar{s}_4}
 T_3^{\bot}\cdot T_4^{\bot}
 + \delta_{s_1s_2} \delta_{s_3s_4}\delta_{s_1s_3}
 2m^2 \frac{1}{xx'(1-x)(1-x')} \\
&& + im \sqrt{2} \left[\delta_{s_3\bar{s}_4}\delta_{s_1s_2}
 \frac{s_1}{x(1-x)} T_3^{\bot} \cdot \varepsilon_{s_1}^{\bot}
 - \delta_{s_3s_4} \delta_{s_1\bar{s}_2}
 \frac{s_3}{x'(1-x')} T_4^{\bot}\cdot \varepsilon_{s_4}^{\bot *} \right] \nn
\; . \leea{b23}

\chapter{Matrix elements of the effective interaction.Exchange channel}
\label{appB}

In this Appendix we follow the scheme of the work \cite{TrPa}
to calculate the matrix elements of the effective interaction
in the exchange channel.
Here, we list the general, angle-dependent matrix elements
defining the effective interaction in the exchange channel 
\eq{m1} (part I),
and the corresponding matrix elements
of the effective interaction for arbitrary $J_z$, after integrating out
the angles \eq{r43} (part II). The whole is given for the similarity
function $f_{\la}(\De)=u_{\la}(\De)=\theta(\la^2-|\De|)$ \eq{f13}
with the sharp cutoff.
The effective interaction generated by the unitary transformation
in the exchange channel reads \eq{m4},\eq{m6}
\bea
V_{LC}^{eff}
&=& -e^2<\ga^{\mu}\ga^{\nu}>g_{\mu\nu}
\left(\frac{\theta(a_1-a_2)}{\tilde{\De}_1}
+\frac{\theta(a_2-a_1)}{\tilde{\De}_2}
\right)\nn\\
&& -e^2<\ga^{\mu}\ga^{\nu}>\eta_{\mu}\eta_{\nu}\frac{1}{2q^{+2}}
\left(\frac{(a_1-a_2)\theta(a_1-a_2)}{\tilde{\De}_1}
+\frac{(a_2-a_1)\theta(a_2-a_1)}{\tilde{\De}_2}
\right)\nn\\
&=& -e^2<\ga^{\mu}\ga^{\nu}>g_{\mu\nu}
\left(\frac{\theta(a_1-a_2)}{\tilde{\De}_1}
+\frac{\theta(a_2-a_1)}{\tilde{\De}_2}
\right)\nn\\
&& -e^2<\ga^{\mu}\ga^{\nu}>\eta_{\mu}\eta_{\nu}\frac{1}{2q^{+2}}
|a_1-a_2|
\left(\frac{\theta(a_1-a_2)}{\tilde{\De}_1}
+\frac{\theta(a_2-a_1)}{\tilde{\De}_2}
\right)
\leea{a1}
where fig.($4$)
\bea
&& <\ga^{\mu}\ga^{\nu}>|_{exch}
=\frac{\bar{u}(p_1,\la_1)}{\sqrt{p_1^{+}}}\ga^{\mu}
\frac{u(p_1^{'},\la_1^{'})}{\sqrt{p_1^{' +}}}
\frac{\bar{v}(p_2^{'},\la_2^{'})}{\sqrt{p_2^{' +}}}\ga^{\nu}
\frac{v(p_2,\la_2)}{\sqrt{p_2^{+}}}
P^{+ 2}
\leea{a1a}
$q=p'_1-p_1$ is the momentum transfer. One has in \eq{a1}
\bea
&& \tilde{\De}_1=a_1-2k_{\perp}k_{\perp}^{'}\cos(\varphi-\varphi^{'})\nn\\
&& \tilde{\De}_2=a_2-2k_{\perp}k_{\perp}^{'}\cos(\varphi-\varphi^{'})\nn\\
&& \tilde{\De}=a-2k_{\perp}k_{\perp}^{'}\cos(\varphi-\varphi^{'})\nn\\
&& \vec{k}_{\perp}=k_{\perp}(\cos\varphi,\sin\varphi)
\leea{}
and
\bea
a_1 &=& \frac{x'}{x}k_{\perp}^2+\frac{x}{x'}k_{\perp}^{'2}
+m^2\frac{(x-x')^2}{xx'}\nn\\
&=& k_{\perp}^2+k_{\perp}^{'2}
+(x-x')\left(k_{\perp}^2(-\frac{1}{x})-k_{\perp}^{'2}(-\frac{1}{x'})\right)
+m^2\frac{(x-x')^2}{xx'}\nn\\
a_2 &=& \frac{1-x'}{1-x}k_{\perp}^2+\frac{1-x}{1-x'}k_{\perp}^{'2}
+m^2\frac{(x-x')^2}{(1-x)(1-x')}\nn\\
&=& k_{\perp}^2+k_{\perp}^{'2}
+(x-x')\left(k_{\perp}^2\frac{1}{1-x}-k_{\perp}^{'2}\frac{1}{1-x'}\right)
+m^2\frac{(x-x')^2}{(1-x)(1-x')}\nn\\
a &=& k_{\perp}^2+k_{\perp}^{'2}
+\frac{(x-x')}{2}\left(k_{\perp}^2(\frac{1}{1-x}-\frac{1}{x})
-k_{\perp}^{'2}(\frac{1}{1-x'}-\frac{1}{x'})\right)\nn\\
&+& m^2\frac{(x-x')^2}{2}\left(\frac{1}{xx'}
+\frac{1}{(1-x)(1-x')}\right)\nn\\
a &=& \frac{1}{2}(a_1+a_2)
\leea{a3}
The energy denominator ($\tilde{\De}$ and $a$ corresponding)
in the case of perturbative theory is given for completeness.

It is useful to display the matrix elements of the effective interaction
in the form of tables. The matrix elements depend on the one hand
on the momenta of the electron and positron, respectively, and on the other
hand on their helicities before and after the interaction.
The dependence on the helicities occur during the calculation
of these functions
$E(x,\vec{k}_{\perp};\la_1,\la_2|x',\vec{k}'_{\perp};\la'_1,\la'_2)$
in part I and 
$G(x,k_{\perp};\la_1,\la_2|x',k'_{\perp};\la'_1,\la'_2)$ in part II
as different Kronecker deltas \cite{BrLe}.
These functions are displayed in the form of helicity tables.
We use the following notation for the elements of the tables
\bea
&& F_i(1,2)~\rightarrow ~E_i(x,\vec{k}_{\perp};x',\vec{k}'_{\perp});~
G_i(x,k_{\perp};x',k'_{\perp})
\leea{a4}
Also we have used in both cases for the permutation of particle and
anti-particle
\bea
&&F_3^{*}(x,\vec{k}_{\perp};x',\vec{k}'_{\perp})
=F_3(1-x,-\vec{k}_{\perp};1-x',-\vec{k}'_{\perp})
\leea{a5}
one has the corresponding for the elements of arbitrary $J_z$;
in the case when the function additionally depends
on the component of the total angular momentum $J_z=n$
we have introduced
\bea
&& \tilde{F}_i(n)=F_i(-n)
\leea{a6}

\section{The general helicity table.}

To calculate the matrix elements of the effective interaction
in the exchange channel we use the matrix elements of the Dirac spinors
listed in Table B.$1$ \cite{BrLe}. 
Also the following holds
$\bar{v}_{\la'}(p)\ga^{\alpha}v_{\la}(q)
=\bar{u}_{\la}(q)\ga^{\alpha}u_{\la'}(p)$.

\begin{table}[htb]
\centerline{
\begin{tabular}{|c||c|}
\hline
\parbox{1.5cm}{ \[\cal M\] } & 
\parbox{12cm}{
\[
\frac{1}{\sqrt{k^+k^{'+}}}
\bar{u}(k',\lambda') {\cal M} u(k,\lambda)
\]
}
\\\hline\hline
$\gamma^+$ & \parbox{12cm}{
\[
\hspace{3cm}
2\delta^{\lambda}_{\lambda'}
\]}
\\\hline
$\gamma^-$ &
\parbox{12cm}{
\[
\hspace{-1cm}
\frac{2}{k^+k^{'+}}\left[\left(m^2+
k_{\perp}k'_{\perp}e^{+i\lambda(\varphi-\varphi')}\right)
\delta^{\lambda}_{\lambda'}
-m\lambda\left(k'_{\perp}e^{+i\lambda \varphi'}-
k_{\perp} e^{+i\lambda\varphi}\right)
\delta^{\lambda}_{-\lambda'}\right]
\]}
\\\hline
$\gamma_{\perp}^1$ & 
\parbox{12cm}{
\[
\left(\frac{k'_{\perp}}{k^{'+}}e^{-i\lambda\varphi'}+
\frac{k_{\perp}}{k^+}e^{+i\lambda\varphi}
\right)\delta^{\lambda}_{\lambda'}
+m\lambda\left(\frac{1}{k^{'+}}-\frac{1}{k^+}
\right)\delta^{\lambda}_{-\lambda'}
\]}
\\\hline
$\gamma_{\perp}^2$ &
\parbox{12cm}{
\[ 
i\lambda\left(\frac{k'_{\perp}}{k^{'+}}e^{-i\lambda\varphi'}-
\frac{k_{\perp}}{k^+}e^{+i\lambda\varphi}
\right)\delta^{\lambda}_{\lambda'}
+im\left(\frac{1}{k^{'+}}-\frac{1}{k^+}
\right)\delta^{\lambda}_{-\lambda'}
\]}\\
\hline
\end{tabular}
}
\caption[Matrix elements of the Dirac spinors]
{Matrix elements of the Dirac spinors.}
\end{table}

We introduce for the matrix elements entering in the effective
interaction \eq{a1}
\bea
2E^{(1)}(x,\vec{k}_{\perp};\la_1,\la_2|
x',\vec{k}_{\perp}^{'};\la_1^{'},\la_2^{'})
&=& <\ga^{\mu}\ga^{\nu}>g_{\mu\nu}=\nn\\
&=& \frac{1}{2}<\ga^+\ga^->+\frac{1}{2}<\ga^-\ga^+>
-<\ga_1^2>-<\ga_2^2>\nn\\
2E^{(2)}(x,\vec{k}_{\perp};\la_1,\la_2|
x',\vec{k}_{\perp}^{'};\la_1^{'},\la_2^{'}) 
&=& <\ga^{\mu}\ga^{\nu}>\eta_{\mu}\eta_{\nu}\frac{1}{q^{+2}}
=<\ga^+\ga^+>\frac{1}{q^{+2}}
\leea{a7}
where
\bea
&& <\ga^{\mu}\ga^{\nu}>=\frac{\bar{u}(x,\vec{k}_{\perp};\la_1)~\ga^{\mu}
~u(x',\vec{k}_{\perp}^{'};\la_1^{'})~
~\bar{v}(1-x',-\vec{k}_{\perp}^{'};\la_2^{'})~\ga^{\nu}
~v(1-x,-\vec{k}_{\perp};\la_2)}{\sqrt{xx'(1-x)(1-x')}}
\leea{a8}
These functions are displayed in the Table B.$2$.

\begin{table}[htb]
\centerline{
\begin{tabular}{|c||c|c|c|c|}\hline
\rule[-3mm]{0mm}{8mm}{\bf final : initial} & 
$(\lambda_1',\lambda_2')=\uparrow\uparrow$ 
& $(\lambda_1',\lambda_2')=\uparrow\downarrow$ 
& $(\lambda_1',\lambda_2')=\downarrow\uparrow$ &
$(\lambda_1',\lambda_2')=\downarrow\downarrow$ \\ \hline\hline
\rule[-3mm]{0mm}{8mm}$(\lambda_1,\lambda_2)=\uparrow\uparrow$ & $E_1(1,2)$  
& $E_3^*(1,2)$ & $E_3(1,2)$ & $0$ \\ \hline
\rule[-3mm]{0mm}{8mm}$(\lambda_1,\lambda_2)=\uparrow\downarrow$ & 
$E_3^*(2,1)$ & $E_2(1,2)$ & $E_4(1,2)$ 
& $-E_3(2,1)$ \\ \hline
\rule[-3mm]{0mm}{8mm}$(\lambda_1,\lambda_2)=\downarrow\uparrow$& $E_3(2,1)$ 
& $E_4(1,2)$ & $E_2(1,2)$  & 
$-E_3^*(2,1)$\\ \hline
\rule[-3mm]{0mm}{8mm}$(\lambda_1,\lambda_2)=\downarrow\downarrow$ & $0$ 
& $-E_3(1,2)$ & $-E_3^*(1,2)$ & 
$E_1(1,2)$\\ \hline
\end{tabular}
}
\protect\caption{General helicity table defining the effective interaction
in the exchange channel.}
\label{GeneralHelicityTable}
\end{table}

The matrix elements 
$E_i^{(n)}(1,2)=E_i^{(n)}(x,\vec{k}_{\perp};x',\vec{k}'_{\perp})$
$(n=1,2)$ are the following
\bea
E_1^{(1)}(x,\vec{k}_{\perp};x',\vec{k}_{\perp}^{'})
&=& m^2\left(\frac{1}{xx'}+\frac{1}{(1-x)(1-x')}\right)
+\frac{k_{\perp}k_{\perp}^{'}}{xx'(1-x)(1-x')}
{\rm e}^{-i(\varphi-\varphi^{'})}\nn\\
E_2^{(1)}(x,\vec{k}_{\perp};x',\vec{k}_{\perp}^{'})
&=& m^2\left(\frac{1}{xx'}+\frac{1}{(1-x)(1-x')}\right)
+k_{\perp}^2\frac{1}{x(1-x)}+k_{\perp}^{'2}\frac{1}{x'(1-x')}\nn\\
&+& k_{\perp}k_{\perp}^{'}
\left(\frac{{\rm e}^{i(\varphi-\varphi^{'})}}{xx'}
+\frac{{\rm e}^{-i(\varphi-\varphi^{'})}}{(1-x)(1-x')}\right)\nn\\
E_3^{(1)}(x,\vec{k}_{\perp};x',\vec{k}_{\perp}^{'})
&=& -\frac{m}{xx'}
\left(k_{\perp}^{'}{\rm e}^{i\varphi^{'}}
-k_{\perp}\frac{1-x'}{1-x}{\rm e}^{i\varphi}\right)\nn\\
E_4^{(1)}(x,\vec{k}_{\perp};x',\vec{k}_{\perp}^{'})
&=& -m^2\frac{(x-x')^2}{xx'(1-x)(1-x')}
\leea{a9}
and
\bea
&& E_1^{(2)}(x,\vec{k}_{\perp};x',\vec{k}_{\perp}^{'})
=E_2^{(2)}(x,\vec{k}_{\perp};x',\vec{k}_{\perp}^{'})
=\frac{2}{(x-x')^2}\nn\\
&& E_3^{(2)}(x,\vec{k}_{\perp};x',\vec{k}_{\perp}^{'})
=E_4^{(2)}(x,\vec{k}_{\perp};x',\vec{k}_{\perp}^{'})=0
\leea{a10}

\section{The helicity table 
for arbitrary $J_z$.}

Following the description given in the main text \eq{r43}
we integrate out the angles in the effective interaction 
in the exchange channel \eqs{m6}{a1}.
For the matrix elements of the effective interaction
for an arbitrary $J_z=n$ with $n\in {\bf Z}$~~~
$G(x,k_{\perp};\la_1,\la_2|x',k'_{\perp};\la'_1,\la'_2)=
<x,k_{\perp};J_z,\la_1,\la_2|\tilde{V}_{LC}^{eff}|
x',k'_{\perp};J'_z,\la'_1,\la'_2>|_{exch}$ in the exchange channel
one obtains the helicity Table B.$3$.

\begin{table}[htb]
\centerline{
\begin{tabular}{|c||c|c|c|c|}\hline
\rule[-3mm]{0mm}{8mm}{\bf final : initial} 
&$(\lambda'_1,\lambda'_2)=\uparrow\uparrow$
&$(\lambda'_1,\lambda'_2)=\uparrow\downarrow$
&$(\lambda'_1,\lambda'_2)=\downarrow\uparrow$
&$(\lambda'_1,\lambda'_2)=\downarrow\downarrow$\\\hline\hline
\rule[-3mm]{0mm}{8mm}$(\lambda_1,\lambda_2)=\uparrow\uparrow$
&$G_1(1,2)$&$G_3^*(1,2)$&$G_3(1,2)$&$0$\\\hline
\rule[-3mm]{0mm}{8mm}$(\lambda_1,\lambda_2)=\uparrow\downarrow$
&$G_3^*(2,1)$&$G_2(1,2)$&$G_4(1,2)$&$-\tilde{G}_3(2,1)$\\\hline
\rule[-3mm]{0mm}{8mm}$(\lambda_1,\lambda_2)=\downarrow\uparrow$
&$G_3(2,1)$&$G_4(1,2)$&$\tilde{G}_2(1,2)$&$-\tilde{G}_3^*(2,1)$\\\hline
\rule[-3mm]{0mm}{8mm}$(\lambda_1,\lambda_2)=\downarrow\downarrow$&$0$&$-
\tilde{G}_3(1,2)$&$-\tilde{G}_3^*(1,2)$&
$\tilde{G}_1(1,2)$\\
\hline
\end{tabular}
}
\protect\caption[Helicity table of the effective interaction
in the exchange channel for arbitrary $J_z = \pm n$, $x>x'$.]
{\protect\label{HelicityTableJz}Helicity table of the effective interaction
for $J_z = \pm n$, $x>x'$.}
\end{table}
\vspace{0.5cm}

Here, the functions $G_i(1,2)=G_i(x,k_{\perp};x',k'_{\perp})$
are given
\bea
G_1(x,k_{\perp};x',k_{\perp}^{'})&=& 
m^2\left(\frac{1}{xx'}+\frac{1}{(1-x)(1-x')}\right)Int_{a_1a_2}(|1-n|)\nn\\
&+& \frac{k_{\perp}k_{\perp}^{'}}{xx'(1-x)(1-x')}Int_{a_1a_2}(|n|)
+\frac{|a_1-a_2|}{(x-x')^2}Int_{a_1a_2}(|1-n|)\nn\\
G_2(x,k_{\perp};x',k_{\perp}^{'})&=&
(m^2\left(\frac{1}{xx'}+\frac{1}{(1-x)(1-x')}\right)
+k_{\perp}^2\frac{1}{x(1-x)}+k_{\perp}^{'2}\frac{1}{x'(1-x')})
Int_{a_1a_2}(|n|)\nn\\
&+& k_{\perp}k_{\perp}^{'}\left(\frac{1}{xx'}Int_{a_1a_2}(|1-n|)
+\frac{1}{(1-x)(1-x')}Int_{a_1a_2}(|1+n|)\right)\nn\\
&+& \frac{|a_1-a_2|}{(x-x')^2}Int_{a_1a_2}(|n|)\nn\\
G_3(x,k_{\perp};x',k_{\perp}^{'})&=&
-\frac{m}{xx'}
\left(k_{\perp}^{'}Int_{a_1a_2}(|1+n|)
-k_{\perp}\frac{1-x'}{1-x}Int_{a_1a_2}(|n|)\right)\nn\\
G_4(x,k_{\perp};x',k_{\perp}^{'})&=&
-m^2\frac{(x-x')^2}{xx'(1-x)(1-x')}Int_{a_1a_2}(|n|)
\leea{a11}
where we have introduced the functions
\bea
&& Int_{a_1a_2}(n)=\theta(a_1-a_2)Int_{a_1}(n)
+\theta(a_2-a_1)Int_{a_2}(n)\nn\\
&& Int_{a_i}(n)=\frac{\alpha}{\pi}(-A(a_i))^{-n+1}
\left(\frac{B(a_i)}{k_{\perp}k_{\perp}^{'}}\right)^{n}\nn\\
&& A(a_i)=\frac{1}{\sqrt{a_i^2-4k_{\perp}^2k_{\perp}^{'2}}}\nn\\
&& B(a_i)=\frac{1}{2}(1-a_iA(a_i))
\leea{a12}
and the functions $a_i$, $i=1,2$ are given in \eq{a3}.

The following integrals were used by the calculation
of the matrix elements \cite{TrPa}
\bea
&& \frac{1}{2\pi}\int_0^{2\pi}d\varphi\int_0^{2\pi}d\varphi'
\frac{\cos(n(\varphi-\varphi'))}
{a_i-2k_{\perp}k'_{\perp}\cos(\varphi-\varphi')}
=2\pi(-A(a_i))^{-n+1}\left(\frac{B(a_i)}{k_{\perp}k'_{\perp}}\right)^n\nn\\
&& \frac{1}{2\pi}\int_0^{2\pi}d\varphi\int_0^{2\pi}d\varphi'
\frac{\sin(n(\varphi-\varphi'))}
{a_i-2k_{\perp}k'_{\perp}\cos(\varphi-\varphi')}=0
\leea{a13}

The condition on the parameter space $(x,k_{\perp})$
due to the $\theta$-function, namely $\theta(a_1-a_2)$
(i.e. when $a_1>a_2$) reads
\bea
&& (x-x')\left(x(1-x)(k_{\perp}^{'2}+m^2)
-x'(1-x')(k_{\perp}^2+m^2)\right)>0
\leea{a17}
Making use of the coordinate change \eq{r46}
this is equivalent to
\bea
&& \left(\frac{\mu\cos\theta}{\sqrt{\mu^2+m^2}}
-\frac{\mu'\cos\theta'}{\sqrt{\mu^{'2}+m^2}}\right)
(\mu-\mu')<0
\leea{a18}

\chapter{\label{appC}
Fermion and photon self energy terms}
 
We calculate here the fermion and photon self energy terms,
arising from the second order commutator $[\eta^{(1)},H_{ee\ga}]$.

\paragraph{I.}
We first derive the {\bf electron self energy} terms.
Making use of the expressions for the generator of the unitary
transformation $\eta^{(1)}$ defined in \eq{gi1} and of $H_{ee\ga}$, \eq{ch14},
we obtain the following expression for the commutator in the
electron self energy sector
\bea
&&\hspace{-5em} \frac{1}{2} (\et_{p_1p_2}g_{p_2p_1}-\et_{p_2p_1}g_{p_1p_2}) \,
\biggl[ \th(p_1^+) \frac{\th(p_2^+-p_1^+)}{p_2^+ - p_1^+} \th(p^+_2) \,
               b^+_{p_2}b_{p_2}\chi^+_{s_2}\chi_{s_2}\nn\\
&&\hspace{6em}
      - \th(p_2^+) \frac{\th(p_1^+-p_2^+)}{p_1^+ - p_2^+} \th(p^+_1) \,
               b^+_{p_1}b_{p_1}\chi^+_{s_1}\chi_{s_1} \biggr]
M_{2ij}(p_1,p_2) \de^{ij}
\; , \leea{d1}
where
\be
M_{2ij}(p_1,p_2)=\Ga^i(p_1,p_2,p_1-p_2)\Ga^j(p_2,p_1,p_2-p_1)
\lee {d2}
and the momentum integration over $p_1,p_2$ is implied;
$1/2$ stands as the symmetry factor.
The matrix element of the commutator between the free fermion states is
\bea
&& <p_1,s_1|[\et^{(1)},H_{ee\ga}]|p_1,s_1>_{self energy} \nn \\
&&\hspace{5em} = -\int_{p_2}(\et_{p_1p_2} g_{p_2p_1} - \et_{p_2p_1}g_{p_1p_2}) \,
\th(p_2^+) \frac{\th(p_1^+-p_2^+)}{p_1^+-p_2^+} \,M_{2ii}(p_1,p_2)
\; , \leea{d3}
where the integration $\int_p$ is defined in \eq{ch20}. We use the expression
for the generator $\eta$ through the coupling, namely 
\be
\et_{p_1p_2} g_{p_2p_1} - \et_{p_2p_1} g_{p_1p_2} = \frac{1}{\De_{p_1p_2}}
\left( g_{p_1p_2} \frac{dg_{p_2p_1}}{dl} +
       g_{p_2p_1} \frac{dg_{p_1p_2}}{dl} \right)
\; . \lee{d4}
Change of the variables according to
\bea
p_1 &=& p \nn \\
p_2 &=& p_k \nn \\
p_1 - p_2 &=& k
\leea{d5}
brings the integral in \eq{d3} to the standard form of loop integration 
\be
-\int_k(\et_{p, p - k} g_{p - k, p} - \et_{p - k, p} g_{p, p - k}) \,
\th(p^+-k^+) \frac{\th(k^+)}{k^+} \, M_{2ii}(p,p-k)
\; . \lee{d6}
According to \eq{ri2}, the integral $\int_{l_{\la}}^{l_{\La}}$
of the commutator $[\eta^{(1)},H_{ee\ga}]$ defines the difference between
the energies (or energy corrections) \mbox{$\de p_{1\la}^--\de p_{1\La}^-$}.
Making use of
\be 
\int^{l_{\la}}_{l_{\La}} dl' (\et_{p_1p_2} g_{p_2p_1}-\et_{p_2p_1} g_{p_1p_2})
= \frac{1}{p_1^- - p_2^- - (p_1-p_2)^-} \,
(g_{p_1, p_2, \La} g_{p_2, p_1, \la} - g_{p_1, p_2, \la} g_{p_2, p_1, \La})
\lee{d7}
we have the following explicit expression: 
\bea 
&&\hspace{-1em} \de p_{1\la}^--\de p_{1\La}^- = e^2\int \frac{d^2k^{\bot}dk^+}{2(2\pi)^3} \,
\frac{\th (k^+)}{k^+}
\th (p^+-k^+) \, \frac{(-1)}{p^--k^--(p-k)^-} \\
&&\hspace{4em} \times \Ga^i(p-k,p,-k) \Ga^i(p,p-k,k) \,
\left[ \exp\left\{-2 \left( \frac{\De_{p,p-k}}{\la} \right)^2
\right\}
- \exp\left\{-2 \left( \frac{\De_{p,p-k}}{\La} \right)^2 \right\} \right] \nn
\; , \leea{d8}
where the solution for the $ee\ga$-coupling constant was used.
Therefore the electron energy correction corresponding to the first diagram,
fig.($8$), is 
\bea
\de p_{1\la}^- &=& e^2\int \frac{d^2k^{\bot}dk^+}{2(2\pi)^3} \,
 \frac{\th (k^+)}{k^+} \th (p^+-k^+) \\
&&\hspace{2em} \times \Ga^i(p-k,p,-k) \Ga^i(p,p-k,k) \,
 \frac{1}{p^--k^--(p-k)^-} \, \times (-R) \nn
\; , \leea{d9}
where we have introduced the regulator $R$, defining the cutoff condition
(see main text),
\be
R = \exp\left\{-2 \left( \frac{\De_{p,k}}{\la} \right)^2 \right\}
\lee{d10}
(note that $\De_{p,k}=\De_{p,p-k}$).
To perform the integration over $k=(k^+,k^{\bot})$ explicitly, choose
the parametrization
\bea
\frac{k^+}{p^+} &=& x \nn \\
k &=& (xp^+,xp^{\bot}+\kappa^{\bot})
\; , \leea{d11}
where $p=(p^+,p^{\bot})$ is the external electron momentum.
Then the terms occuring in $\de p_{1\la}^-$ are rewritten in the form
\bea
& \Ga^i(p-k,p,-k)\Ga^i(p,p-k,k)=
\frac{1}{(p^+)^2(1-x)^2} \left( \left( 4\frac{1}{x^2}-4\frac{1}{x}+2 \right)
\kappa_{\bot}^2+2m^2x^2 \right) & \nn \\
& \De_{p, p - k} = p^- - k^- - (p - k)^- = \frac{1}{p^+x(1-x)}
(x(1-x)p^2 - \kappa_{\bot}^2-xm^2) = \frac{\tilde{\De}_{p,p-k}}{p^+} &
\; . \leea{d12}
Therefore the integral for the electron energy correction
corresponding to the first diagram of fig.($8$) takes the form
\bea
p^+\de p_{1\la}^- &=&-\frac{e^2}{8\pi^2}\int_0^1dx\int d\kappa_{\bot}^2
\frac{(\frac{2}{x^2}-\frac{2}{x}+1)\kappa_{\bot}^2+m^2x^2}
{(1-x)(\kappa_{\bot}^2+f(x))}\times(-R) \\
&=& -\frac{e^2}{8\pi^2}\int_0^1dx\int d\kappa_{\bot}^2 \nn \\
&&\hspace{2em} \times \left[ \frac{p^2-m^2}{\kappa_{\bot}^2+f(x)}
 \left( \frac{2}{[x]}-2+x \right) - \frac{2m^2}{\kappa_{\bot}^2+f(x)}
 + \left( \frac{2}{[x]^2}+\frac{1}{[1-x]} \right)
\right] \times (-R) \nn
\; , \leea{d13}
where 
\be
f(x)=xm^2 - x(1-x)p^2
\; . \lee{d14}
In the last integral the principal value prescription  for $\frac{1}{[x]}$
as $x\rightarrow 0$ was introduced (see main text), to regularize
the IR divergencies present in the longitudinal direction. 

We thus have derived the expression for the energy correction
which has been used in the main text.

\paragraph{II.}
We repeat the same procedure for the {\bf photon self energy}.
The second order commutator $[\eta^{(1)},H_{ee\ga}]$ gives
the following expression in the photon self energy sector
\bea
&&\hspace{-2em} \frac{1}{2} (\et_{p_1p_2} g_{p_2p_1} - 
\et_{p_2p_1} g_{p_1p_2}) \cdot
\biggl[ \th(p_1^+) \th(-p_2^+) \frac{\th(p_1^+-p_2^+)}{(p_1^+-p_2^+)} \,
            a_{-q}^+ a_{-q} \varep_{\la}^{i*} \varep_{\la}^j \\
&&\hspace{9em}
      - \th(-p_1^+) \th(p_2^+) \frac{\th(p_2^+-p_1^+)}{(p_2^+-p_1^+)} \,
            a_{q}^+a_{q} \varep_{\la}^{i} \varep_{\la}^{j*} \biggr] \cdot
Tr M_{2ij}(p_1,p_2) \, \de_{q,-(p_1 - p_2)} \nn
\; , \leea{d15}
where $M_{2ij}(p_1,p_2)$ is defined in \eq{d2} and the trace acts in spin space;
the integration over the momenta $q$, $p_1$ and $p_2$ is implied.
The matrix element between the free photon states reads
\bea
&&\hspace{-2em} <q,\la|[\et^{(1)},H_{ee\ga}]|q,\la>_{self energy}\de_{ij} \\
&&\hspace{3em}
= - \frac{1}{q^+} \int_{p_1,p_2} (\et_{p_1p_2}g_{p_2p_1}-
\et{p_2p_1}g_{p_1p_2}) \,
\th(-p_1^+) \th(p_2^+) \, Tr M_{2ij}(p_1,p_2) \, \de_{q,-(p_1-p_2)} \nn
\; , \leea{d16}
that can be rewritten after the change of coordinates according to
\bea
p_1 &=& -k \nn \\
p_2 &=& -(k-q) \nn \\
p_2 - p_1 &=& q
\leea{d17}
in the following way
\be
\frac{1}{q^+}\int_k(\et_{k,k-q}g_{k-q,k}-\et_{k-q,k}g_{k,k-q}) \,
\th(k^+)\th(q^+-k^+) \, Tr M_{2ij}(k, k - q)
\; , \lee{d18}
where the symmetry
\bea
\et_{-p_1,-p_2} &=& -\et_{p_1,p_2} \nn\\
g_{-p_1,-p_2} &=& g_{p_1,p_2}
\leea{d19}
has been used. The integration of the commutator over $l$ in the flow equation 
gives rise to 
\bea
&&\hspace{-1em} (\de q_{1\la}^- - \de q_{1\La}^-) \de^{ij}
= \frac{1}{q^+} e^2 \int \frac{d^2k^{\bot}dk^+}{2(2\pi)^3}
\th(k^+) \th(q^+ - k^+) \, \frac{(-1)}{q^- -k^- - (q-k)^-} \\
&&\hspace{2em} \times \, Tr \left( \Ga^i(k,k-q,q) \Ga^j(k-q,k,-q) \right) \,
\left[ \exp\left\{ -2 \left( \frac{\De_{q,q-k}}{\la} \right)^2 \right\}
     - \exp\left\{ -2 \left( \frac{\De_{q,q-k}}{\La} \right)^2 \right\} \right] \nn
. \leea{d20}
This means for the photon energy correction 
\bea
\de q_{1\la}^-\de^{ij} &=&
\frac{1}{q^+}e^2 \int \frac{d^2k^{\bot}dk^+}{2(2\pi)^3}
\th(k^+) \th(q^+ - k^+) \\ 
&& \times \, Tr \left( \Ga^i(k,k-q,q) \Ga^j(k-q,k,-q) \right)
\frac{1}{q^- -k^- - (q-k)^-} \, \times(-R) \nn
\; , \leea{d21}
where the regulator $R$
\be
R=\exp\left\{ -2 \left( \frac{\De_{q,k}}{\la} \right)^2 \right\}
\lee{d22}
has been introduced. Define the new set of coordinates
\bea
\frac{(q-k)^+}{q^+} &=& x \nn \\
k &=& ((1-x)q^+,(1-x)q^{\bot}+\kappa^{\bot}) \nn \\
q - k &=& (xq^+,xq^{\bot} - \kappa^{\bot})
\; , \leea{d23}
where $q=(q^+,q^{\bot})$ is the photon momentum.
Then the terms present in $\de q_{1\la}^-$ are
\bea
& \Ga^i(k,k-q,q)\Ga^i(k-q,k,-q)
=\frac{2}{(q^+)^2x(1-x)^2}
\left( \left( 2x-2+\frac{1}{x} \right) \kappa^{\bot 2} + \frac{m^2}{x} \right)
 & \nn \\
& \De_{k-q,k} = q^- - k^- - (q - k)^- =
-\frac{\kappa^{\bot 2} + m^2}{q^+x(1-x)}+ \frac{q^2}{q^+} = 
\frac{\tilde{\De}_{k-q,k}}{q^+} &
\; . \leea{d24}
The integral for the photon energy correction corresponding
to the first diagram of fig.($9$) takes the form 
\bea
q^+\de q_{1\la}^- &\hspace{-3mm}=\hspace{-3mm}&
 -\frac{e^2}{8\pi^2}\int_0^1 \!\! dx 
 \int d\kappa_{\bot}^2
 \frac{(2x-2+\frac{1}{x})\kappa_{\bot}^2+\frac{m^2}{x}}
 {(1-x)(\kappa_{\bot}^2+f(x))} \!\!\times\! (-R) \\
&\hspace{-3mm}=\hspace{-3mm}&-\frac{e^2}{8\pi^2}\int_0^1 \!\! dx
 \int d\kappa_{\bot}^2
 \left\{ \frac{q^2}{\kappa_{\bot}^2+f(x)}
  \left( 2x^2-2x+1 \right) +\frac{2m^2}{[1-x]}
  + \left( -2+\frac{1}{[x][1-x]} \right) \right\} \!\!\times\! (-R) \nn
\leea{d25}
with
\be
f(x)=m^2-q^2x(1-x)
\; , \lee{d26}
and the principal value prescription, denoted by '\mbox{\boldmath{$[\;]$}}',
introduced to regularize the IR divergencies.

This is the form of the photon correction used in the main text.

\end{appendix}

  
\addcontentsline{toc}{chapter}{List of Figures}

Figure 1: Flow equations perform the block-diagonalization of the bare 
Hamiltonian of the canonical theory $H_B(\La)$ into a Hamiltonian consisting
of blocks with equal number of particles. For a finite value of $\la$ the
matrix elements of the 'particle number changing' sectors are squeezed 
into an 
energy band with roughly $|E_i-E_j|<\la$ (left hand side picture) and are 
eliminated completely as $\la \rightarrow 0$ (right hand side picture).

Table 1: The effective light front QED Hamiltonian matrix up to
second order in $e$ in the Fock space representation. The matrix elements
of the 'diagonal' (Fock state conserving) sectors are unrestricted in the
energy differences; the 'rest' (Fock state changing) sectors are squeezed
roughly in an energy band of width $\la$. Black dots correspond to zero
matrix elements in order $O(e^2)$. Instantaneous and disconnected
diagrams are not included.

Figure 2: The matrix elements of the effective Hamitonian,
obtained by the flow equations in the second order in coupling,
with corresponding diagrams. 
The diagrams $2 - 5$
belong to the 'diagonal' sector; the $1, 6, 7$ correspond
to the 'rest' sector (the $6, 7$ diagrams are drawn schematically,
namely the corresponding momentum change must be performed
to get the real 'rest' diagrams, depicted in Table 1.)
The photon momenta are $x^+$-ordered, from left to right. The similarity
function is chosen 
\mbox{$f_{p_ip_f,\la} \!=\! \exp(-\De_{p_ip_f\la}^2 \!/\!\la^2)$},
where \mbox{$\De_{p_ip_f\la} \!=\! \Si p^-_i \!-\! \Si p^-_f$}
(the index \mbox{$`i`$} denotes initial and \mbox{$`f`$} final states)
and 
\mbox{$\De_{p_1p_2\la} \!=\! p_1^- \!-\! p_2^- \!-\! (p_1 \!-\! p_2)^-$},
\mbox{$p^-\!=\! (p_{\bot}^2 \!+\! m_{\la}^2)\!/\!p^+$}.

Figure 3 : 
The effective electron-positron interaction in the exchange channel;
the diagrams correspond to the generated and the instantaneous interactions.

Figure 4: Another choice of the coordinates in the effective
electron-positron interaction, that are used in the light-front
integral equation.

Figure 5: Positronium spectrum for $-3\leq J_z\leq 3$,
$\alpha=0.3$ and $\La=1$. The annihilation channel is not
included. For an easier identification of the spin-parity
multiplets, the corresponding non-relativistic notation
$L_{J}^{J_z}$ is inserted. Masses are given in units
of the electron mass.

Figure 6: Stability of positronium spectrum for $J_z=0$,
without annihilation interaction. Eigenvalues $M_i^2$
for $\alpha=0.3$ and $\La=1$ are plotted versus $N$,
the number of Gaussian points.
Masses are in units of the electron mass.

Figure 7: Deviation of corresponding eigenvalues
for $J_z=0$ and $J_z=1$ ($\alpha=0.3$ and $\La=1$)
with growing number of integration points $N$.
The graphs shown $\De M^2=M_n^2(J_z=0)-M_n^2(J_z=1)$
for the states $1\!^{3}S_1$ (triagles),
$2\!^{3}P_1$ (squares), and $2\!^{1}P_1$ (circles).

Figure 8: Electron self energy term includes the diagram arising from
the commutator term $[\eta^{(1)},H_{ee\ga}]$ in the electron sector, 
and also two diagrams, the second and the third on the figure $5$, 
coming from the normal ordering of 
instantaneous interactions $H_{ee\ga\ga}$ and $H_{eeee}$, resp.

Figure 9: Photon self energy term includes the diagram coming from 
the commutator $[\eta^{(1)},H_{ee\ga}]$ in the photon sector,
and also the diagram from the normal ordering of 
the instantaneous interaction $H_{ee\ga\ga}$.

\newpage
\addcontentsline{toc}{section}{Figure 1: Block-diagonalization
of Hamiltonians}

\begin{figure}
$$
\fips{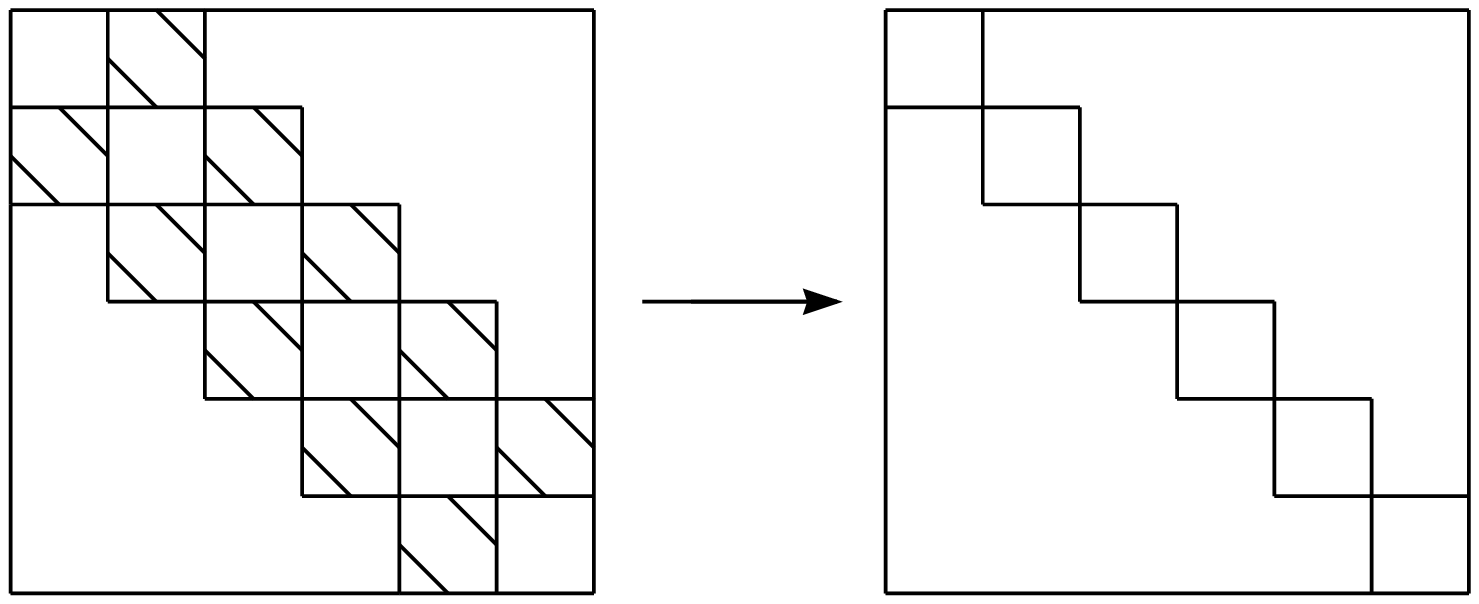}
\setlength{\unitlength}{0.240900pt}
\begin{picture}(0,0)
\put(-1500,780){\makebox(0,0){particle number}}
\put(-900,420){\makebox(0,0){$U(\La,\la)$}}
$$
\end{picture}
\vspace{0.5cm}
$$
\label{fig1}
\end{figure}
\begin{center}
Figure 1
\end{center}

\newpage
\addcontentsline{toc}{section}  
{Table 1: Effective light-front QED Hamiltonian}
\begin{table}
\vspace{2cm}
\begin{tabular}{|r|c|c|c|c|c|} \hline
 & $|\ga>$ & $|e\bar{e}>$ & $|\ga\ga>$ & $|e\bar{e}\ga>$ 
& $|e\bar{e}e\bar{e}>$ \\ \hline
$|\ga>$ & \floadeps{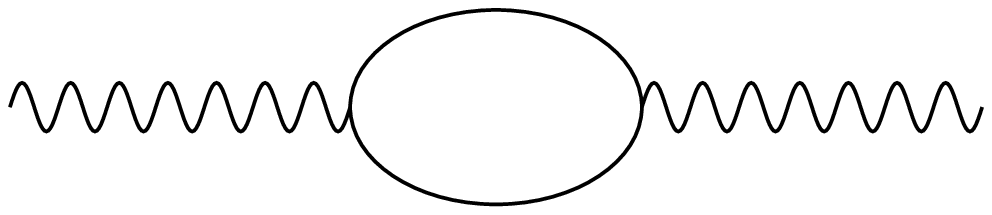} & \floadeps{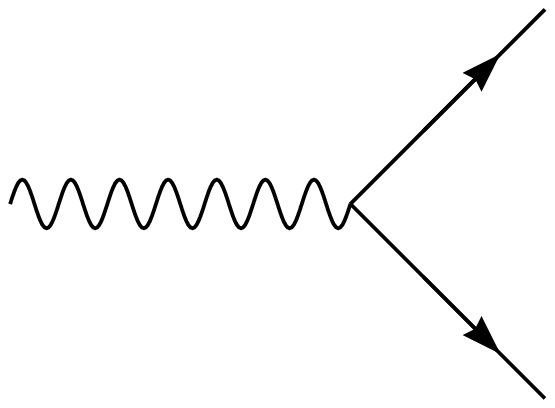} 
& \floadeps{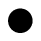} & \floadeps{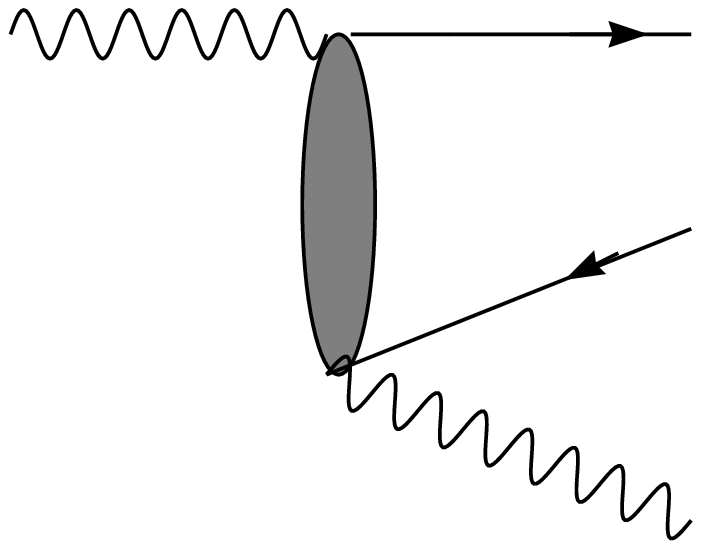} & \floadeps{table15} \\ \hline
$|e\bar{e}>$ & \floadeps{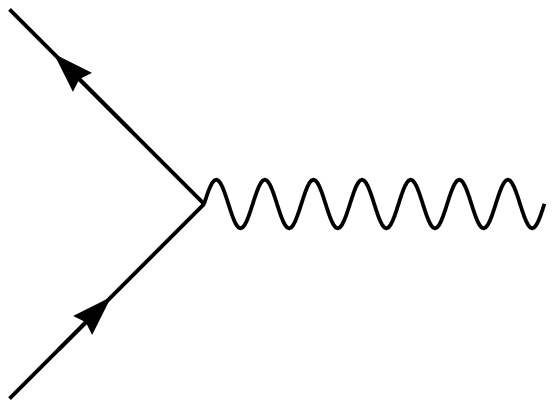} & \floadeps{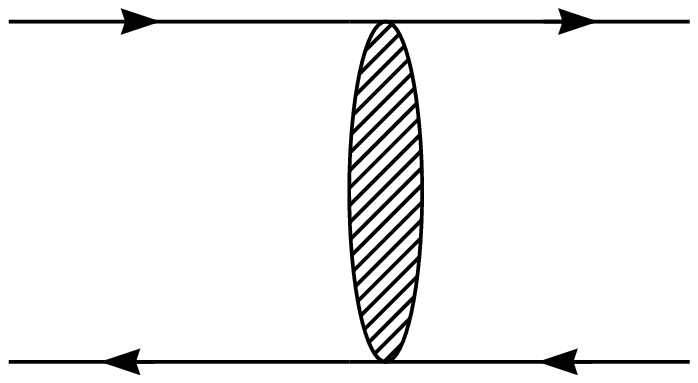} 
& \floadeps{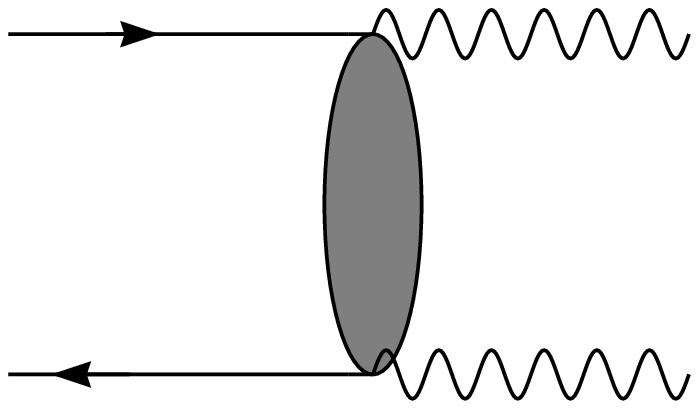} & \floadeps{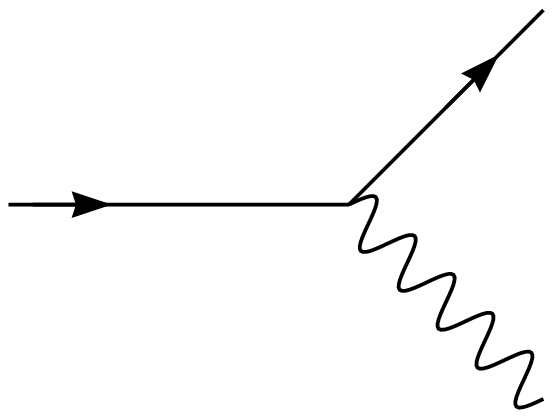} & \floadeps{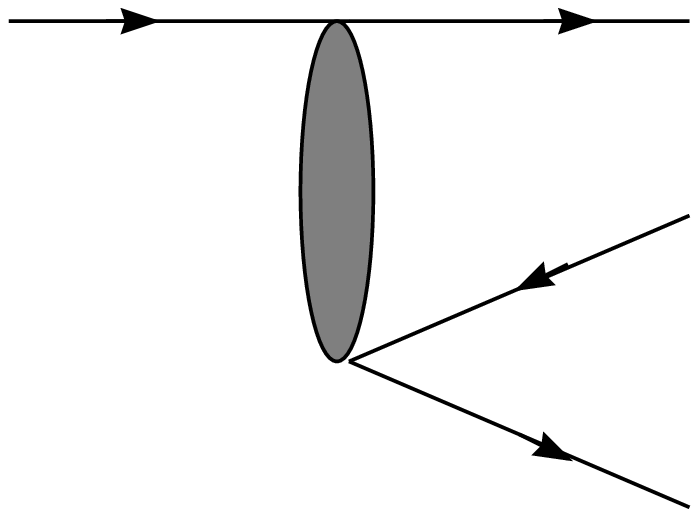} \\ \hline
$|\ga\ga>$ & \floadeps{table15} & \floadeps{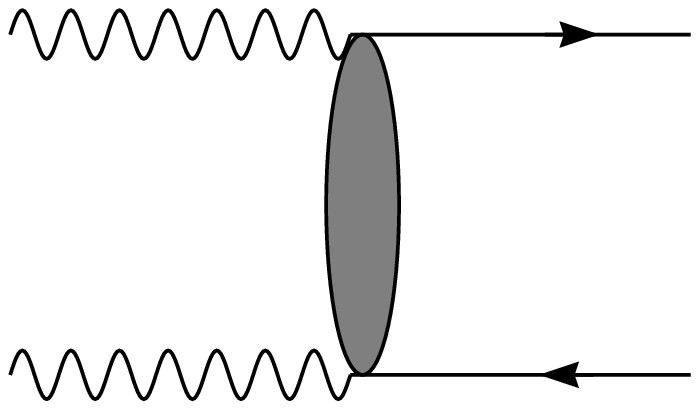} 
& \floadeps{table15} & \floadeps{tab12} & \floadeps{table15} \\ \hline
$|e\bar{e}\ga>$ & \floadeps{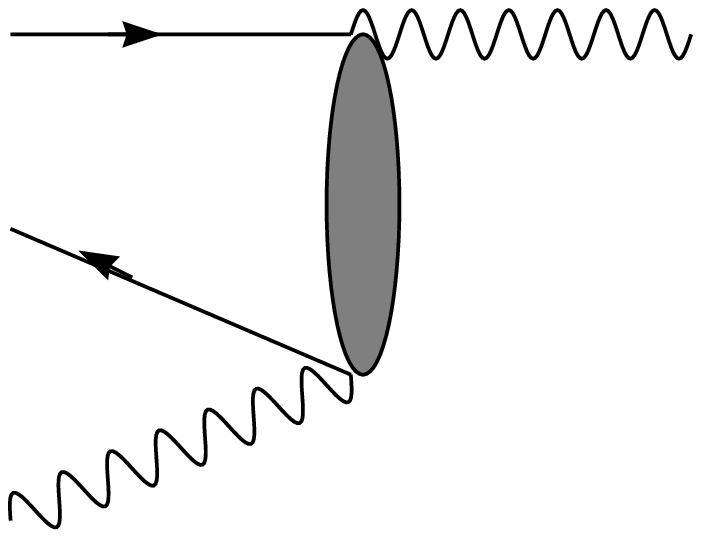} & \floadeps{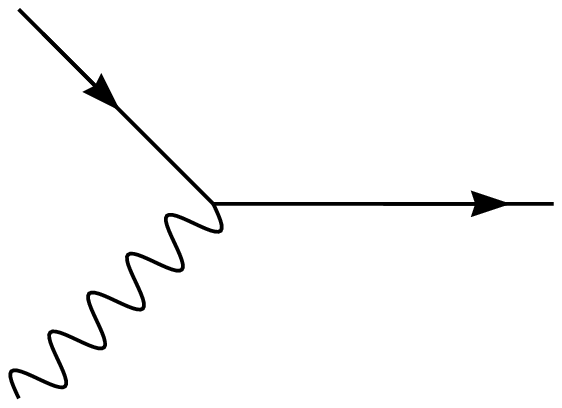} 
& \floadeps{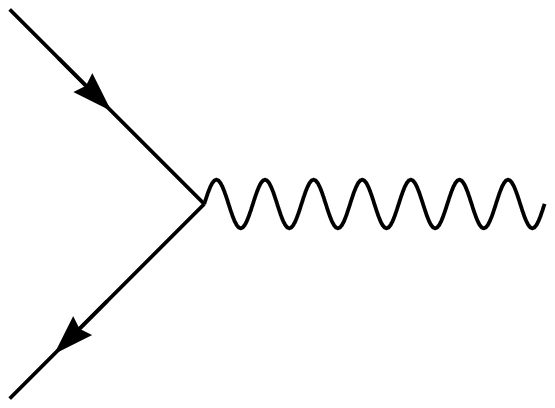} & \floadeps{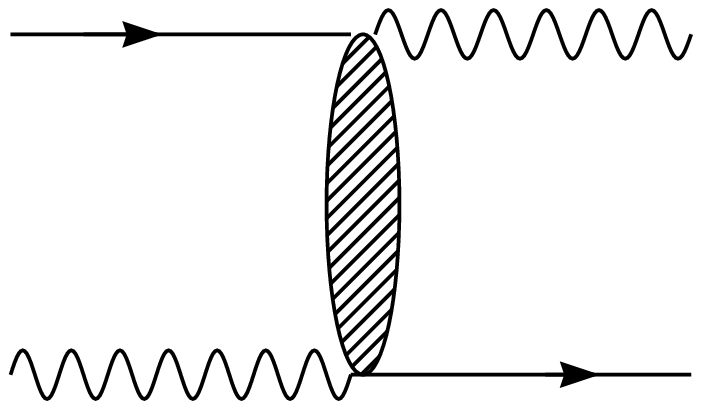} & \floadeps{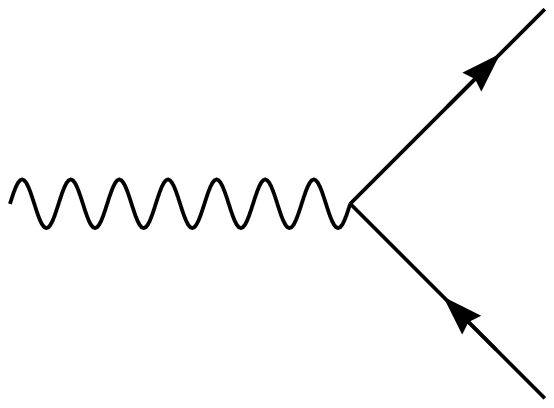} \\ \hline
$|e\bar{e}e\bar{e}>$ & \floadeps{table15} & \floadeps{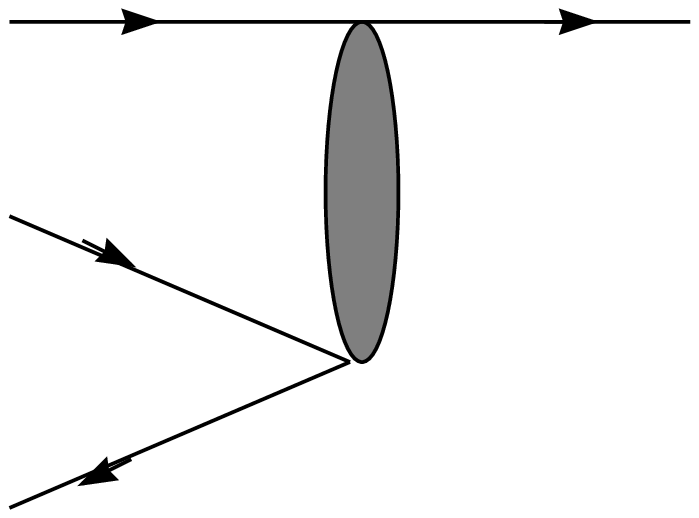} 
& \floadeps{table15} & \floadeps{tab43} & \floadeps{table22} \\ \hline
\end{tabular}
\vspace{1cm}
\label{table}
\end{table}
\begin{center}
Table 1
\end{center}

\newpage
\addcontentsline{toc}{section}
{Figure 2: Matrix elements of the effective Hamitonian}
\begin{figure}[!htb]
\setlength{\unitlength}{1mm}
\begin{picture}(170,219)
\put(0,185){\makebox(56,34.61){ \loadeps{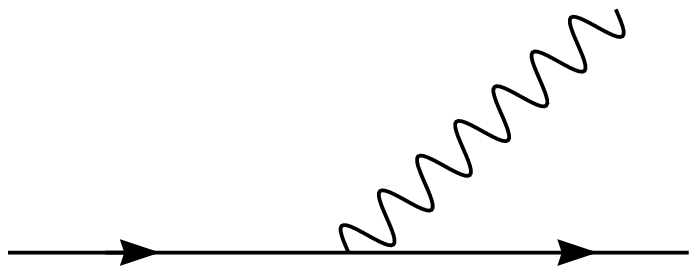} }}
  \diagform{180}{
    \bea
    &&\hspace{-15mm} -e_{\la}\exp{\left\{-\frac{\De_{p_1p_2}^2}{\la^2}\right\}}
      \ch^+_2\Ga^i_\la(p_1,p_2,k) \ch_1 \varep^{i \ast} \nn \\
    && \nn \\
    &&\hspace{-5mm} \Ga^i_\la(p_1,p_2,k) =   
      2 \frac{k^i}{[k^+]} - \frac{\si \cdot p_2^\bot - im_{\la}}{[p_2^+]} \si^i
       - \si^i \frac{\si \cdot p_1^\bot + im_{\la}}{[p_1^+]} \nn \\
    &&\hspace{-3mm} i=1,2 \nn                           
    \eea 
    }
\put(10,190){$p_1$}
\put(42,190){$p_2$}
\put(34,210){$k \; (i)$}
\put(0,140){\makebox(56,34.61){ \loadeps{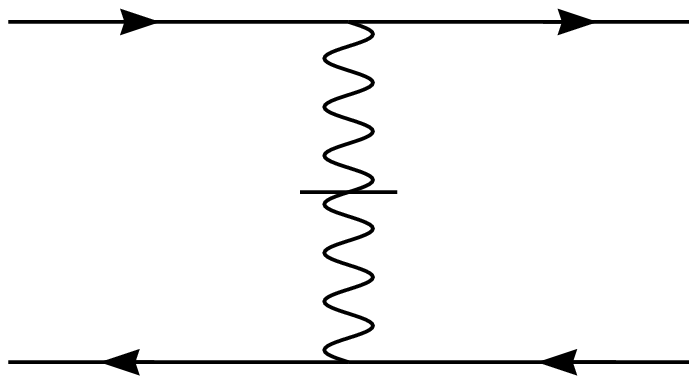} }}
  \diagform{135}{
    \bea
     e_{\la}^2\ \ch^+_3 \ch^+_{\bar{4}}
       \frac{4}{[p_1^+ - p_3^+]^2} \ch_1 \ch_{\bar{2}} \nn
    \eea
    }
\put(10,171){$p_1$}
\put(42,171){$p_3$}
\put(10,142){$p_2$}
\put(42,142){$p_4$}
\put(0,95){\makebox(56,34.61)[r]{ \loadeps{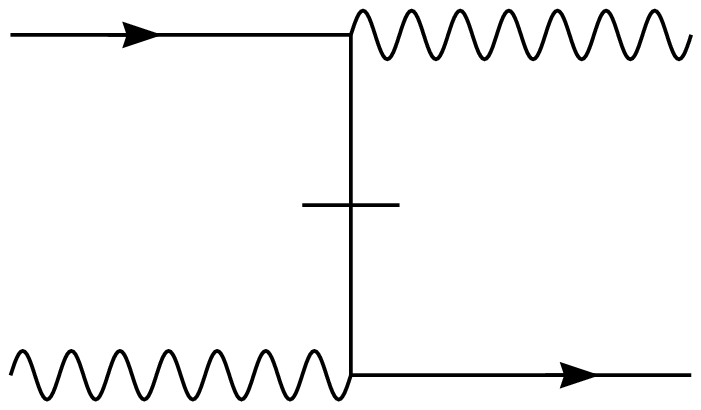} }}
  \diagform{90}{
    \bea
    e_{\la}^2 \ch^+_2
       \frac{\si^j \si^i}{[p_1^+ - k_1^+]} \ch_1 \varep^{i \ast} \varep^j \nn
    \eea
    }
\put(10,126){$p_1$}
\put(42,126){$k_1 \; (i)$}
\put(10,97){$k_2 \; (j)$}
\put(42,97){$p_2$}
\put(0,50){\makebox(56,34.61){ \loadeps{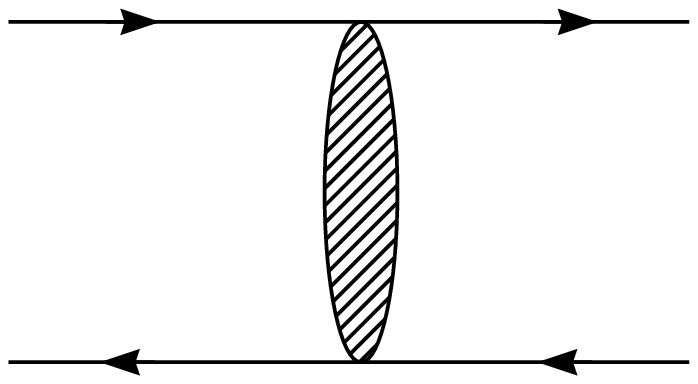} }}
  \diagform{45}{
    \bea
    &&\hspace{-17mm} -e_{\la}^2 M_{2ij,\la} \, \de^{ij} 
      \frac{1}{[p_1^+ - p_3^+]} \nn \\
    &&\hspace{-12mm} 
      \times \left(\frac{\De_{p_1p_3\la}+\De_{p_4p_2\la}}
      {\De_{p_1p_3\la}^2+\De_{p_4p_2\la}^2}\right)
      \left( 1 - \exp{\left\{-\frac{\De_{p_1p_3\la}^2+\De_{p_4p_2\la}^2}{\la^2}
      \right\}}\right) \nn \\
    && \nn \\
    &&\hspace{-7mm} M_{2ij,\la} = 
      \biggl(\ch^+_3 \Ga^i_\la(p_1,p_3,p_1 \!-\! p_3)\ch_1\biggr) \nn \\
    &&\hspace{13mm} \times \biggl(\ch^+_{\bar{2}} 
      \Ga^j_\la(-p_4,-p_2,-(p_1 \!-\! p_3)) \ch_{\bar{4}}\biggr) \nn
    \eea 
    }
\put(10,81){$p_1$}
\put(42,81){$p_3$}
\put(10,52){$p_2$}
\put(42,52){$p_4$}
\put(0,5){\makebox(56,34.61){ \loadeps{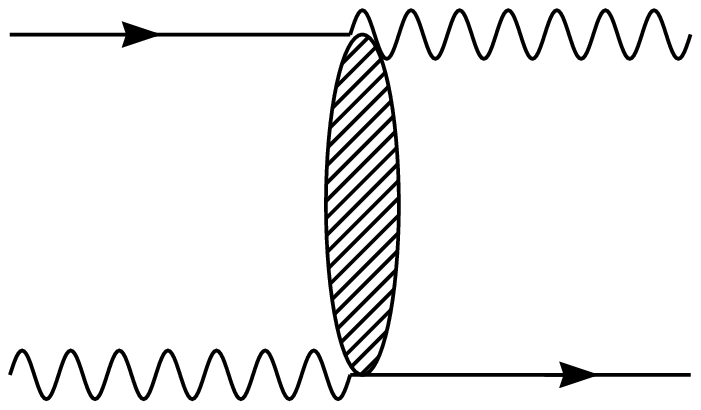} }}
  \diagform{0}{
    \bea
    &&\hspace{-17mm} e_{\la}^2\widetilde{M}_{2ij,\la} \,
      \varep^{i \ast} \varep^j \nn \\
    &&\hspace{-12mm} 
      \times \left(\frac{\De_{p_1k_1\la}+\De_{p_2k_2\la}}
      {\De_{p_1k_1\la}^2+\De_{p_2k_2\la}^2}\right)
      \left( 1 - \exp{\left\{
      -\frac{\De_{p_1k_1\la}^2+\De_{p_2k_2\la}^2}{\la^2}
      \right\}}\right) \nn \\
    && \nn \\
    &&\hspace{-7mm} \widetilde{M}_{2ij,\la} = 
       \ch^+_2 \Ga^i_\la(p_1,p_1 \!-\! k_1,k_1) \: 
       \Ga^j_\la(p_1 \!-\! k_1,p_2,k_2) \ch_1 \nn
    \eea 
    }
\put(10,36){$p_1$}
\put(42,36){$k_1 \; (i)$}
\put(10,7){$k_2 \; (j)$}
\put(42,7){$p_2$}
\end{picture}
\end{figure}

\newpage
\begin{figure}[!htb]
\setlength{\unitlength}{1mm}
\begin{picture}(170,189)
\put(0,155){\makebox(56,34.61){ \loadeps{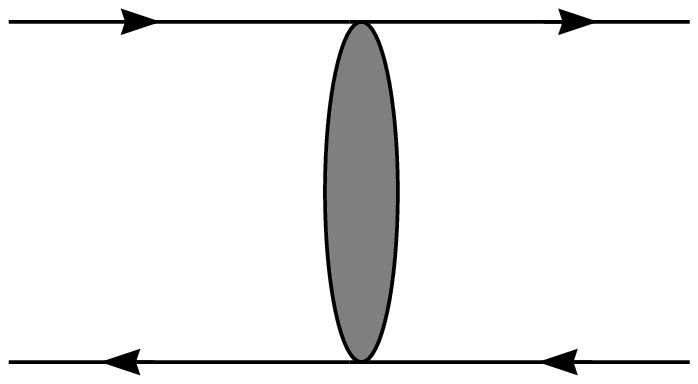} }}
  \diagform{150}{
    \bea
    &&\hspace{-17mm} -e_{\la}^2 \exp{\left\{-\frac{\De_{p_ip_f}^2}{\la^2}
      \right\}} 
      M_{2ij,\la} \, \de^{ij} 
      \frac{1}{[p_1^+ - p_3^+]} \nn \\
    &&\hspace{-12mm} 
      \times \frac{1}{2} \left( \frac{1}{\De_{p_1p_3\la}} + 
      \frac{1}{\De_{p_4p_2\la}} \right) \,
      \left( 1 - \exp{\left\{ -2 \, \frac{\De_{p_1p_3\la} \cdot 
      \De_{p_4p_2\la}}{\la^2} \right\} }
       \right) \nn \\
    && \nn \\
    &&\hspace{-7mm} M_{2ij,\la} = \biggl(\ch^+_2 \Ga^i_\la(p_1,p_2,p_1 \!-\! p_2)
      \ch_1\biggr) \nn \\
    &&\hspace{13mm} \times \biggl(\ch^+_4 \Ga^j_\la(p_3,p_4,-(p_1 \!-\! p_2)) 
      \ch_3\biggr) \nn
    \eea 
    }
\put(10,186){$p_1$}
\put(42,186){$p_3$}
\put(10,157){$p_2$}
\put(42,157){$p_4$}
\put(0,110){\makebox(56,34.61){ \loadeps{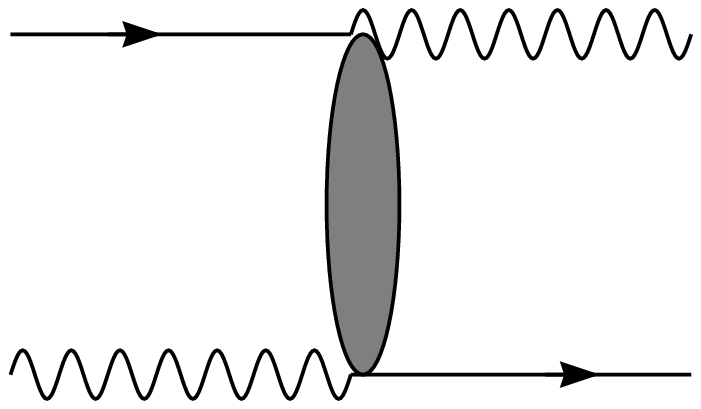} }}
  \diagform{105}{
    \bea
    &&\hspace{-17mm} e_{\la}^2\exp{\left\{-\frac{\De_{p_ip_f}^2}{\la^2}
      \right\}}
      \widetilde{M}_{2ij,\la} \,\varep^{i \ast} \varep^j \nn \\
    &&\hspace{-12mm} 
      \times \frac{1}{2} \left( \frac{1}{\De_{p_1k_1\la}} + 
      \frac{1}{\De_{p_2k_2\la}} \right) \,
      \left( 1 - \exp{\left\{ -2 \, \frac{\De_{p_1k_1\la} \cdot 
      \De_{p_2k_2\la}}{\la^2} \right\} }
       \right) \nn \\
    && \nn \\ 
    &&\hspace{-7mm} \widetilde{M}_{2ij,\la} = 
       \ch^+_2 \Ga^i_\la(p_1,p_1 \!-\! k_1,k_1) \: 
    \Ga^j_\la(p_1 \!-\! k_1,p_2,k_2) \ch_1 \nn
    \eea 
    }
\put(10,141){$p_1$}
\put(42,141){$k_1 \; (i)$}
\put(10,112){$k_2 \; (j)$}
\put(42,112){$p_2$}
\end{picture}
\vspace{0.5cm}
\begin{center}
Figure 2
\end{center}
\label{feynrules}
\end{figure}

\newpage
\addcontentsline{toc}{section}
{Figure 3: Effective electron-positron interaction}
\begin{figure}[!htb]
\setlength{\unitlength}{1mm}
\begin{picture}(170,71)
\put(19,36){\makebox(56,34.61){ \loadeps{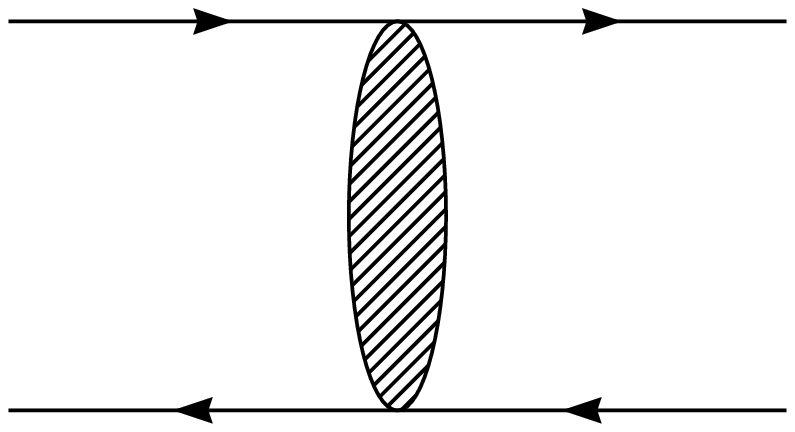} }}
\put(18,67){$p_1\;(x,\kappa^\bot)$}
\put(50,67){$p_3\;(x',\kappa'^\bot)$}
\put(18,38){$p_2\;(1\!-\!x,-\kappa^\bot)$}
\put(50,38){$p_4\;(1\!-\!x',-\kappa'^\bot)$}
  \put(75,36){\makebox(56,34.61){ \loadeps{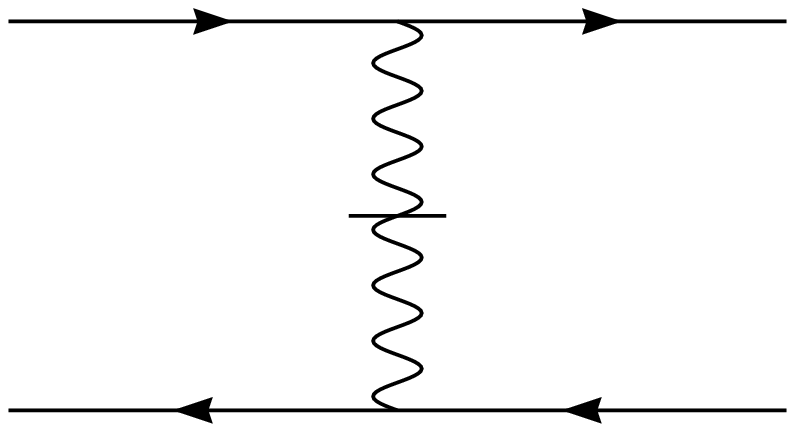} }}
\put(73,53.305){$+$}
\end{picture}
\vspace{0.5cm}
\label{reneebarint}
\end{figure}
\begin{center}
Figure 3
\end{center}

\newpage
\addcontentsline{toc}{section}
{Figure 4: Effective electron-positron interaction (another choice
of coordinates)}
\begin{figure}[!htb]
\setlength{\unitlength}{1mm}
\begin{picture}(170,71)
\put(19,36){\makebox(56,34.61){ \loadeps{FigureIII1} }}
\put(18,67){$p'_1\;(x',\vec{k'}_{\perp})$}
\put(50,67){$p_1\;(x,\vec{k}_{\perp})$}
\put(18,38){$p'_2\;(1\!-\!x',-\vec{k'}_{\perp})$}
\put(50,38){$p_2\;(1\!-\!x,-\vec{k}_{\perp})$}
  \put(75,36){\makebox(56,34.61){ \loadeps{FigureIII2} }}
\put(73,53.305){$+$}
\end{picture}
\vspace{0.5cm}
\label{reneebarint2}
\end{figure}
\begin{center}
Figure 4
\end{center}

\newpage
\addcontentsline{toc}{section}
{Figure 5: Positronium spectrum}
\begin{figure}
\centerline{
\psfig{figure=\graphpath yrast.epsi,width=16cm,angle=0}}
\end{figure}
\vspace{0.5cm} 
\begin{center}
Figure 5
\end{center}

\newpage
\addcontentsline{toc}{section}  
{Figure 6: Stability of positronium spectrum}
\begin{figure}
\centerline{
\psfig{figure=\graphpath spectrum_J0.epsi,width=16cm,angle=0}}
\end{figure}
\vspace{0.5cm} 
\begin{center}
Figure 6
\end{center}

\newpage
\addcontentsline{toc}{section}  
{Figure 7: Deviation of corresponding eigenvalues
for $J_z=0$ and $J_z=1$}
\begin{figure}
\centerline{
\psfig{figure=\graphpath difference.epsi,width=16cm,angle=0}}
\end{figure}
\vspace{0.5cm} 
\begin{center}
Figure 7
\end{center}

\newpage
\addcontentsline{toc}{section}  
{Figure 8: Electron self energy}  
\begin{figure}
\setlength{\unitlength}{1mm}
\begin{picture}(170,34.61)
\put(0,0){\makebox(56,34.61){ \loadeps{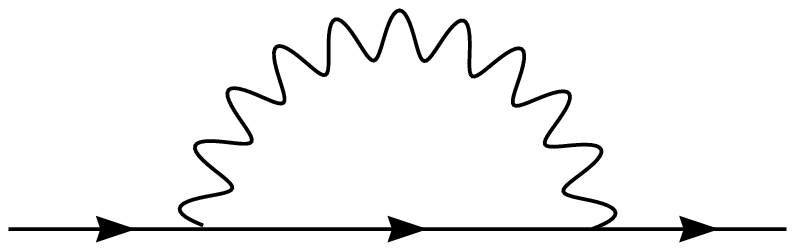} }}
\put(10,14){$p$}
\put(42,14){$p$}
\put(26,32){$k$}
  \put(57,0){\makebox(56,34.61){ \loadeps{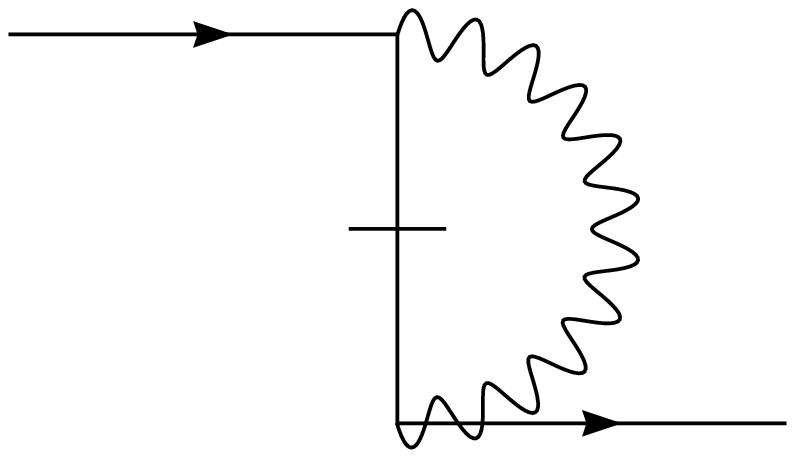} }}
  \put(67,30){$p$}
  \put(100,16){$k$}
  \put(100,3){$p$}
    \put(114,0){\makebox(56,34.61){ \loadeps{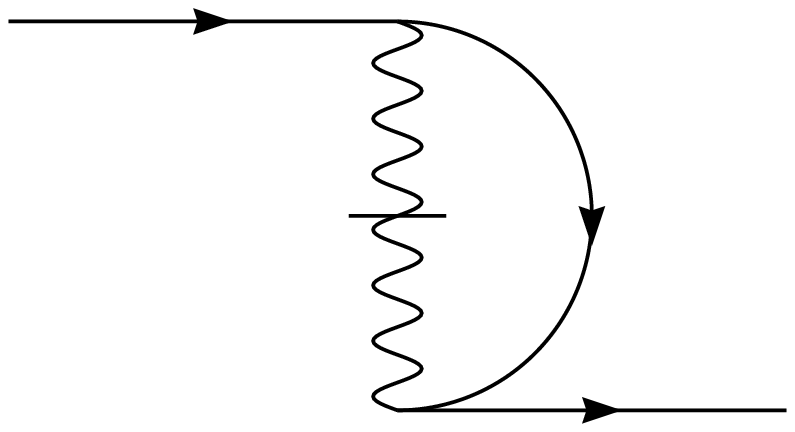} }}
    \put(124,30){$p$}
    \put(155,16){$k$}
    \put(157,3){$p$}
\put(55,17.305){$+$}
\put(112,17.305){$+$}
\end{picture}
\vspace{0.5cm}
\label{eselfen}
\end{figure}

\begin{center}
Figure 8
\end{center}

\newpage
\addcontentsline{toc}{section}  
{Figure 9: Photon self energy}
\begin{figure}
\setlength{\unitlength}{1mm}
\begin{picture}(170,34.61)
\put(28,0){\makebox(56,34.61){ \loadeps{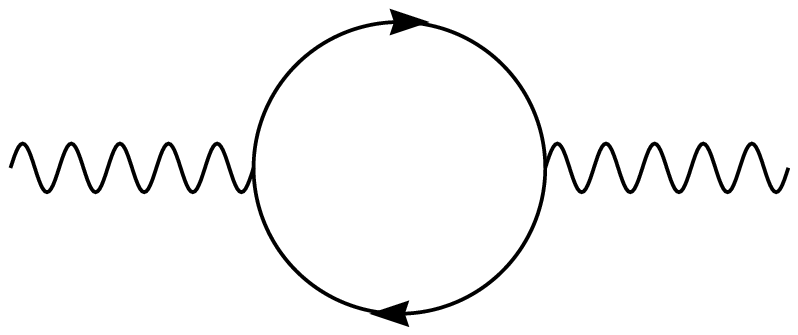} }}
\put(38,12){$p$}
\put(70,12){$p$}
\put(54,28){$k$}
  \put(86,0){\makebox(56,34.61){ \loadeps{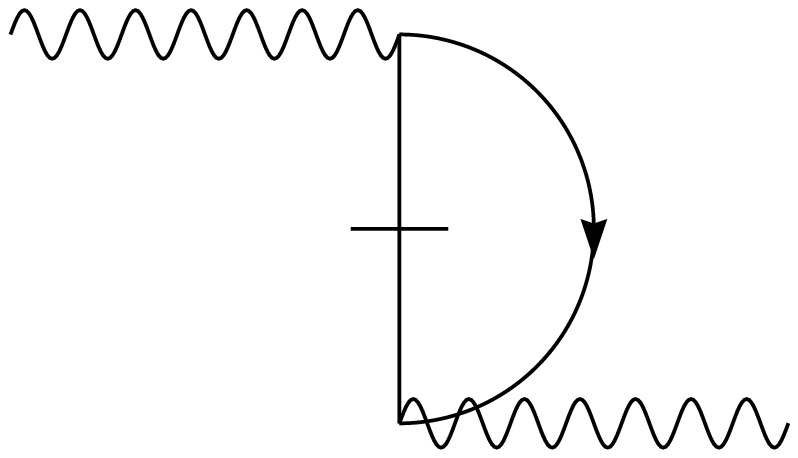} }}
  \put(96,31){$p$}
  \put(126,16){$k$}
  \put(129,3){$p$}
\put(84,17.305){$+$}
\end{picture}
\vspace{0.5cm}
\label{photselfen}
\end{figure}

\begin{center}
Figure 9
\end{center}

\thispagestyle{empty}
\addcontentsline{toc}{chapter}{Acknowledgments}
\begin{center}
{\bf Acknowledgments}
\end{center}

I want to thank my adviser Prof. Franz Wegner for
providing me the chance to work on this project
under his guidance. I thank him for the guidance
and for the help in the work, and also for his warm
and considerate attitude to me during the whole time. 

I extend warm thanks to Prof. H-C. Pauli
for his constant interest in the work
and many useful discussions that improved 
my understanding of the light-front physics.
I thank him for his patience and help.

I thank Prof. H.G. Dosch and Prof. H.J. Pirner
for the opportunity to perform
my Ph.D. study in Heidelberg.

I would like to thank Prof. Yu.A. Simonov 
for his  support and help.

A special thanks to Dr. Brett van de Sande
and Dr. Koji Harada
for introducing me many ideas.

I would like to thank Dr. Uwe Trittmann
for a work in collaboration and for performing
the numerical calculations. 

I would like to thank Prof. Stan Glazek, Prof. Robert Perry,
Prof. John Hiller and also Prof. Matthias Burkardt,
Dr. Billy Jones, Dr. Martina Brisudova,
Dr. Alex Kalloniatis, Dr. Thomas Heinzl 
for useful discussions.

I thank Dr. Andreas Mielke and Dr. Stephan Kehrein
for the fruitful discussions on the method of flow equations. 

I thank Dr. Andreas Metz and Dr. Hilmar Forkel
for correction and improving the manuscript,
and also Dr. Gabor Papp, Dr. Yong-Bin He and Dr. Umbert D'Alesio
for the help with the computer.

\end{document}